\documentclass[reprint,showkeys,showpacs,preprintnumbers,nofootinbib,amsmath,amssymb,aps,prd,longbibliography,10pt]{revtex4-1}
\usepackage{array,multirow,graphicx}
\usepackage{dcolumn}
\usepackage{txfonts}
\usepackage{newlfont}
\usepackage{times}
\usepackage{bm}
\usepackage[colorlinks,citecolor=blue,urlcolor=blue,linkcolor=blue]{hyperref}
\usepackage[figtopcap]{subfigure}
\begin{document}
\preprint{FU-PCG-14}
\title{Bounce inflation in $f(T)$ Cosmology: A unified inflaton-quintessence field}
\author{Kazuharu Bamba$^{1}$}%
\email{bamba@sss.fukushima-u.ac.jp}
\author{G.G.L. Nashed$^{2,3,4}$}
\email{nashed@bue.edu.eg}
\author{W. El Hanafy$^{2,4}$}%
\email{waleed.elhanafy@bue.edu.eg}
\author{Sh.K. Ibraheem$^{2,3}$}%
\email{shymaa{\_}77@sci.asu.edu.eg}
\affiliation{$^{1}$Division of Human Support System, Faculty of Symbiotic Systems Science, Fukushima University, Fukushima 960-1296, Japan}
\affiliation{$^{2}$Centre for Theoretical Physics, The British University in Egypt, P.O. Box 43, Sherouk City, Cairo 11837, Egypt}
\affiliation{$^{3}$Mathematics Department, Faculty of Science, Ain Shams University, Cairo 11566, Egypt}
\affiliation{$^{4}$Egyptian Relativity Group (ERG), Cairo University, Giza 12613, Egypt}
\begin{abstract}
We investigate a bounce inflation model with a graceful exit into the Friedmann-Robertson-Walker (FRW) decelerated Universe within $f(T)$ gravity framework, where $T$ is the torsion scalar in the teleparallelism. We study the cosmic thermal evolution, the model predicts a supercold Universe during the precontraction phase, which is consistent with the requirements of the slow-roll models, while it performs a reheating period by the end of the contraction with a maximum temperature just below the grand unified theory (GUT) temperature. However, it matches the radiation temperature of the hot big bang at later stages. The equation-of-state due to the effective gravitational sector suggests that our Universe is self-accelerated by teleparallel gravity. We assume the matter component to be a canonical scalar field. We obtain the scalar field potential that is induced by the $f(T)$ theory. The power spectrum of the model is nearly scale invariant. In addition, we show that the model unifies inflaton and quintessence fields in a single model. Also, we revisited the primordial fluctuations in $f(T)$ bounce cosmology, to study the fluctuations that are produced at the precontraction phase.
\end{abstract}
\maketitle
\section{Introduction} \label{Sec1}
The standard model of cosmology (big bang) has succeeded to trace the cosmic thermal evolution in an elegant way by comparing the particles interactions rate with the expansion rate. At very hot stages, the rate of interactions is much larger than the expansion rate and the local thermal equilibrium could be achieved, while at later stages, when the Universe cools down, the interaction rate decreases faster than the expansion allowing the particles to decouple from the thermal path at the equality of the rates. However, the big bang suffers many problems, e.g., \textit{causal connected}, \textit{flatness}, \textit{horizon}, etc. This requires a superfast accelerated expansion phase at some early time, i.e., cosmic inflation~\cite{Starobinsky:1980te, Sato:1980yn, Guth:1980zm, Linde:1981mu, Albrecht:1982wi}, that is usually represented by an exponential expansion at $\sim 10^{-35}$ s after the big bang. As a result, the Universe becomes isotropic, homogeneous and approximately flat. Standard inflation models assume the existence of a self-coupled scalar field (inflaton) minimally coupled to gravity, whose potential governs the inflation model. During this stage, when the initial quantum fluctuations cross the horizon and transform into classical fluctuations producing a nearly scale-invariant scalar perturbations spectrum. Although inflation solves the above mentioned problems, one of the fundamental problems still exists, that is the \textit{initial singularity} which arises when tracing the Universe back in time as divergences of the cosmic temperature and density. Since the initial singularity is before inflation raids, the problem can not be solved within inflation framework. Another serious problem is the \textit{trans-Planckian} problem which also appears in inflationary cosmology where the cosmological scales that we observe at present time correspond to length scales smaller than the Planck length at the onset of inflation \cite{transPl1,transPl2}.\\

One of the suggested alternatives is by assuming that the scale factor initially shrinks down to a nonzero minimal value then \textit{bounce} to an expanding phase. In this case a singular or nonsingular bounce Universe can be obtained \cite{Oikonomou:2015PRD, bounce4}. This idea has been extended to recognize nonsingular cyclic Universe models, e.g., pre-big-bang \cite{preBB}, ekpyrotic model \cite{Ekpy}. Other than the nonsingular issue, bounce cosmologies have many interesting features such as solving the horizon and flatness problems even in the initial shrinking phase; also, these models can generate scale-invariant scalar perturbations as supported by observations. However, bounce models are usually faced by two main problems \cite{bounce2,bounce3}: The first is called the \textit{anisotropy} problem, that is in the contraction phase the anisotropies grow faster than the background so that the contraction ends with a complete anisotropic Universe which violates the cosmological principle and bouncing to an expanding phase will not occur. The second is called the \textit{ghost instability} problem, that is the bounce cosmology violates the null energy condition (NEC), which gives rise to ghost degrees-of-freedom. However, both two issues have been successfully resolved within a nonsingular bounce cosmology \cite{Cai:2012JCAP,Cai:2013JCAP,Cai:2014SCPMA}.

As a matter of fact the above mentioned anisotropy problem can be deluded if the equation-of-state is larger than unity during contraction, then the background dominates the anisotropies. Indeed, a large equation-of-state constrains the potential to be \textit{negative} in scalar field models. On the other hand, the ghost degrees-of-freedom is an outcome of using the GR theory, while other modified gravity theories could alter the situation (for reviews on modified gravity theories, see, for instance,~\cite{Nojiri:2010wj, Nojiri:2006ri, Book-Capozziello-Faraoni, Capozziello:2011et, delaCruzDombriz:2012xy, Bamba:2012cp, Joyce:2014kja, Koyama:2015vza, Bamba:2015uma}). In $f(T)$ gravity, where $T$ is the torsion scalar described by the Weitzenb\"{o}ck connection in the teleparallelism~\cite{Hehl:1976kj, Hayashi:1979qx, Flanagan:2007dc, Garecki:2010jj, Bamba:2015jqa}, it has been shown that nonsingular bounce solutions can be constructed in a straightforward way \cite{Cai:2011tc,CQSW14,bounce4}. Also, it has been shown that $f(T)$ gravity combined with holonomy corrected loop
quantum cosmology supports the bounce Universe model \cite{bounce3,Haro:2013bea,Haro:2014wha,Haro:2015zta}\\

In this sense, we organize the work as follows. In Sec. \ref{Sec2}, we review the general relativistic cosmology showing its limited wilingness in cosmological applications. In Sec. \ref{Sec3}, we discuss a possible choice of a scale factor capable to perform a reliable cosmological model. We show that two possible scenarios could be obtained according to the values of the model parameter: a graceful exit inflation or a bounce graceful exit inflation. Also, we use the nice feature of $f(T)$ cosmology to represent the modified Friedmann equation as a one-dimensional autonomous differential equation. This enables us to construct the corresponding ($\dot{H}-H$) phase space, where the dynamical evolution of the model can be shown clearly. In Sec. \ref{Sec4}, we construct an $f(T)$ theory corresponding to the bounce inflation model. Also, we evaluate the equation-of-state of the torsion gravity showing its role to describe a healthy bounce Universe. In Sec. \ref{Sec5}, we discuss the thermal evolution of the Universe showing that its maximum reheating temperature is at the bounce point. We show how the slow-roll condition can arise naturally in this model as a consequence of the thermal evolution. We assume the matter to be a canonical scalar field, then we obtain the potential corresponding to the $f(T)$ theory. The slow-roll potential provides a nearly scale invariant spectrum consistent with observations. So the model does not suffer from a large tensor-to-scalar ratio that is usually obtained in bounce scenarios. In addition, we show that for a particular case, the model can unify inflaton-quintessence fields in a single model. We also show that the NEC is not generally violated, which makes the model safe from the ghost instability problem. In Sec. \ref{Sec6}, we extend our analysis to investigate the $f(T)$ theory at the perturbation level to study the primordial fluctuations during the precontraction phase. The work has been summarized in Sec. \ref{Sec7}.
\section{Einstein's Cosmology}\label{Sec2}
The Copernican (or cosmological) principle is believed to be a good approximation to construct a reliable cosmological model. Standard cosmology today is a manifestation of the Copernican principle and Einstein's field equations,
\begin{equation}\label{GR}
\mathfrak{G}_{\mu \nu}=\kappa^2 \mathfrak{T}_{\mu\nu},
\end{equation}
when they have been applied to the whole Universe. Where $\mathfrak{G}_{\mu\nu}$ is the Einstein tensor, $\kappa^2=8\pi G/c^4$, $G$ is the Newtonian's gravitational constant and $c$ is the speed of light in vacuum. We assume the natural unit system $c=\hbar=k_{B}=1$, and the stress-energy tensor $\mathfrak{T}_{\mu\nu}$ is taken as for a perfect fluid
\begin{equation}\label{stress-energy}
\mathfrak{T}_{\mu\nu}=\rho u_{\mu}u_{\nu}+p(u_{\mu}u_{\nu}-g_{\mu\nu}),
\end{equation}
where $u_{\mu}=\delta^{0}_{\mu}$ is the 4 velocity of the fluid in comoving coordinates, and $\rho$ and $p$ are the density and pressure of the fluid, respectively. We also assume the Universe is FRW spatially flat, that gives rise to the metric
\begin{equation}\label{FRW-metric}
ds^2=-dt^2+a(t)\delta_{ij}dx^{i}dx^{j},
\end{equation}
where $a(t)$ is the scale factor. Applying the Einstein's field equations to the FRW Universe leads to Friedmann's equations
\begin{equation}
  H^2 = \frac{\kappa^2}{3} \rho, \quad \dot{H}+H^2 = -\frac{\kappa^2}{6}(\rho+3p),
\end{equation}
where $H\equiv \dot{a}/a$ is the Hubble parameter and the dot denotes the derivative with respect to time. Constraining Friedmann's equations by the linear equation-of-state $p=\omega\rho$, then solve for the scale factor \cite{Early2004}
\begin{equation}\label{FRWscf}
a_{\textmd{FRW}}=\left\{
  \begin{array}{ll}
    a_{k}\left[(1+\omega)(t-t_{i})\right]^{\frac{2}{3(1+\omega)}}, & \hbox{$\omega \neq -1$;} \\[3pt]
    e^{H_{0}(t-t_{i})}, & \hbox{$\omega = -1$,}
  \end{array}
\right.
\end{equation}
where $a_{k}$, $H_{0}$ and $t_{i}$ are constants. The former is the usual power-law scale factor, for $\omega>-1/3$ the Universe is expanding with deceleration, while it is accelerated when $\omega<-1/3$. The later gives a de Sitter universe, where $\omega=-1$, which does not allow the Universe to evolve, that could be considered in late phases rather inflation.

We next discuss necessary consequences of using the above power-law scale factor. Since, the classical laws of physics breakdown beyond Planck time, we usually assume our description is valid at an initial time at Planck's time $t=t_{p}\sim 10^{-44}$~s, where the temperature $\Theta_{p}\sim 10^{32}$~K is the Planck temperature, and the length $\ell_{p}\sim 10^{-33}$~cm is Planck length. At a present time $t_{0}$, the rough estimation of the horizon scale is $ct_{0}\sim 10^{28}$ cm. So the ratio of the present value of the scale factor $a_{0}$ to its initial one at Planck's time $a_{i}$ is given as $a_{0}/a_{i}=\Theta_{p}/\Theta_{k}\sim 10^{32}$. However, the initial size of the present Universe is $L_{i}\sim ct_{0}\frac{a_{i}}{a_{0}}$. If we assume that nothing is faster than speed of light, the casual region size is $L_{c}\sim c t_{p}$. Thus, at Planck's limit the ratio of the expected initial to the casual region size of the Universe is $L_{i}/L_{c}\sim 10^{28}$.

Here the need to an early accelerated expansion episode, \textit{inflation}, becomes clear. As a matter of fact this inflation requires the Universe to grow up by a factor $> 10^{28}\sim 64$ $e$-folds to be causally connected at a time $\sim 10^{-35}$ s from the big bang. But the power-law scale factor (\ref{FRWscf}) for $\omega>-1/3$ gives $\dot{a}\sim a/t$ such that $L_{i}/L_{c}\sim \dot{a}_{i}/\dot{a}_{0}\gg 1$, which is consistent with the standard idea of gravity as an attractive force. However, the causal connected Universe condition implies that $L_{i}/L_{c}\sim \dot{a}_{i}/\dot{a}_{0}< 1$ at some early time. In this sense, gravity should act as a repulsive force during inflation. In addition, we need this early accelerated expansion phase to end at $\sim 10^{-32}$ s with a smooth transition to the standard FRW model in order to gain the benefits of the big bang nucleosynthesis successes.

In GR theory, the cosmic evolution is constrained by the scale factor only so that any modification needed must be through the choice of the equation-of-state. As a matter of fact, in GR theory, any choice of the scale factor different from equation (\ref{FRWscf}) leads to \textit{inconservative} Universe as long as the equation-of-state is chosen $p=\omega \rho$. In an alternative to GR theory, e.g., the teleparallel gravity, we may see gravity in a different way \cite{martin1,Waleed3,Chen2,Nashed1,Martin3,Nashed3,Nashed4}. However, in cosmological applications the teleparallel equivalent to general relativity (TEGR) suffers the same problem. One of the interesting modifications in gravity is $f(T)$-theories, this extended version of the teleparallel gravity has received a wide echo in the literature in cosmology \cite{Saridakis2,Waleed1,Karami,Chen1,Waleed2,cosm1,cosm2,cosm3,cosm4,cosm5,cosm6} and in astrophysics as well \cite{Iorio,Ruggiero,Rod1,Rod2,astro1,astro2,astro3}.
\section{A Modified scale factor}\label{Sec3}
Fortunately, the $f(T)$ gravity shows more flexibility with the FRW model, it allows two unknowns $a(t)$ and $f(T)$ in the field equations so that we have two possible ways to well identify the universe: either by introducing a specific $f(T)$ in addition to the equation-of-state then solving for $a(t)$; or by introducing a scale factor in addition to an equation-of-state then solving for $f(T)$. In these two cases, the Universe is conservative, and the gravitational sector is expected to play an important role in the cosmic dynamics. We take the second path to obtain a possible $f(T)$ theory describing how does teleparallel gravity can perform an early acceleration episode with a smooth transition to the usual decelerated FRW epoch, with no need to the slow-roll approximation.\\

\begin{figure}[t]
\centering
\includegraphics[scale=.3]{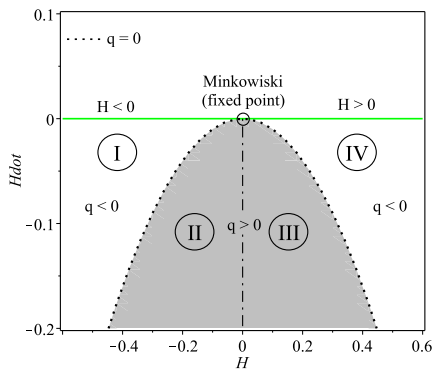}
\caption{($\dot{H}\textmd{-}H$) phase space diagram. The dot curve represents the zero acceleration boundary, it divided the phase space into two regions. The shaded region is the deceleration region, while the unshaded is the acceleration one. The labels (I)$-$(IV) give four possible behaviors.}
\label{Fig1}
\end{figure}
We summarize a useful tool to qualitatively describe the dynamical behavior of a flat FRW model by constructing its ($\dot{H}\textmd{-}H$) phase space diagram \cite{awad2013}. This method uses the pressure properties such as asymptotic behavior and fixed points to analyze cosmological solutions. We first identify the zero acceleration curve by the deceleration parameter $q\equiv -\frac{a\ddot{a}}{\dot{a}^{2}}=0$, i.e., $\dot{H}=-H^2$, which divides the phase space into two regions. The inner region characterizes the usual decelerated FRW models, this is shown by the shaded region in Fig. \ref{Fig1}. However, the unshaded region represents the accelerated phases. We classify different phases in Fig. \ref{Fig1} as follows: (I) region represents an accelerated contracting Universe as $q<0$ and $H<0$, (II) region represents a decelerated contacting Universe as $q>0$ and $H<0$, (III) region represents a decelerated expanding Universe as $q>0$ and $H>0$ which characterizes the usual FRW models, and (IV) region represents an accelerated expanding Universe as $q<0$ and $H>0$ which characterizes the so-called inflation or dark energy phases according to the dynamical evolution of the model. It is worth to mention that one can engineer a Universe using a particular scale factor to fulfill the observational requirements.
\subsection{A possible choice}\label{Sec3.1}  
As a result of the above discussion, we showed that how the GR theory limits the choices to perform accelerated-to-decelerated expansion transition, unless we change the equation-of-state by hand from $\omega<-1/3$ to $\omega>-1/3$, respectively. However, the $f(T)$ gravity can perform this task with no need to change the equation-of-state manually. This can be done by plugging a suitable scale factor into the $f(T)$ equations of motion. As a matter of fact, we need the scale factor to construct an ($\dot{H}-H$) phase space able to cross the zero acceleration curve from the (IV) region into (III) region. For this reason, let us reintroduce the power-law scale factor to the game with a correction term
\begin{equation}\label{scale-factor}
    a(t)=\underbrace{a_{k}\left[(1+\omega)(t-t_{i})\right]^{\frac{2}{3(1+\omega)}}}_{a_{\textmd{FRW}}}
\underbrace{e^{\frac{-\alpha}{(1+\omega)(t-t_{i})}}}_{a_{\textmd{corr}}},
\end{equation}
where $\alpha$ is a parameter with units of time; the usual FRW model is recovered by setting $\alpha=0$. Also, one finds that the radiation and matter dominant epochs are achievable at a late time when $t-t_{i}\gg |\alpha|$. In order to make our terminology clear as far as we can, it is worth mentioning that we take the equation-of-state parameter of ultrarelativistic (e.g., radiation) matter $\omega=1/3$ as $t_{p}<t<t_{eq}$, where $t_{eq}$ denotes the time of the matter-radiation equality [i.e., $\rho_{r}(t_{eq})=\rho_{d}(t_{eq})$] where the subscripts $r$ and $d$ denote , respectively, the radiation and dust phases. At time $t>t_{eq}$, the equation-of-state parameter will be taken as $\omega=0$ of cold matter (e.g., dust). It is convenient now to fix the values of the constants $t_{i}$ and $a_{k}$ in (\ref{scale-factor}) in addition to the parameter $\alpha$. So we take the three conditions: $a(t=0)=0$ at the initial singularity with the equation-of-state parameter $\omega=1/3$, the acceleration $\ddot{a}(t_{\textmd{end}})=0$, where $t_{\textmd{end}}=10^{-32}$ s denotes the time at the end of inflation, and $a(t_{0})=1$ with an equation-of-state parameter $\omega=0$ at the present time $t_{0}\sim 10^{17}$ s. Using (\ref{scale-factor}) with the just mentioned conditions, we fix $t_{i}=0$, $a_{k}=4.6\times10^{-12}$, while the parameter $\alpha$ may have the values $1.61\times 10^{-32}$ s, or $-2.76\times 10^{-33}$ s.
\begin{figure}[t]
\centering
\subfigure[~Graceful inflation]{\label{fig2a}\includegraphics[scale=.25]{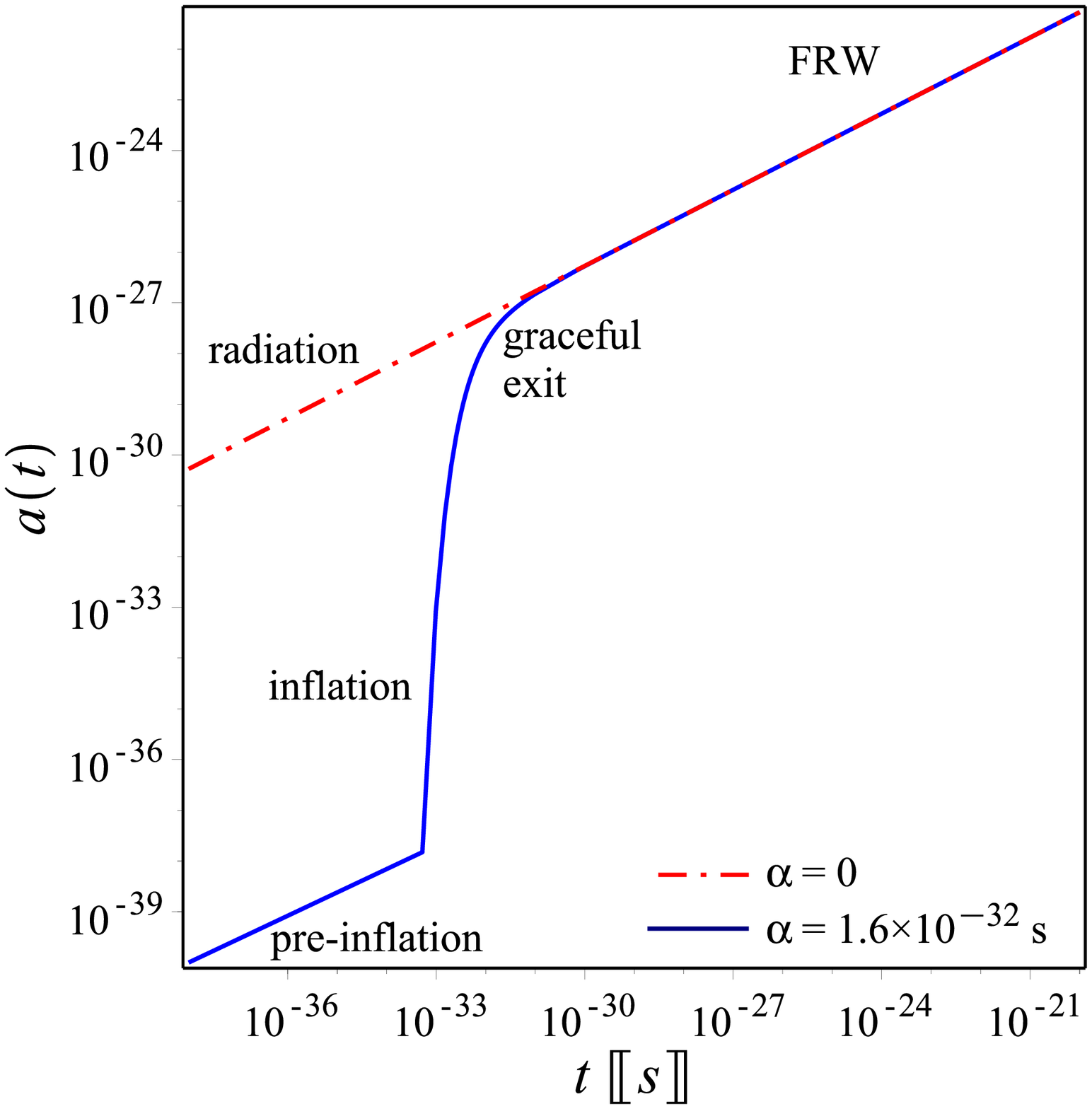}}
\subfigure[~Bounce universe]{\label{fig2b}\includegraphics[scale=.25]{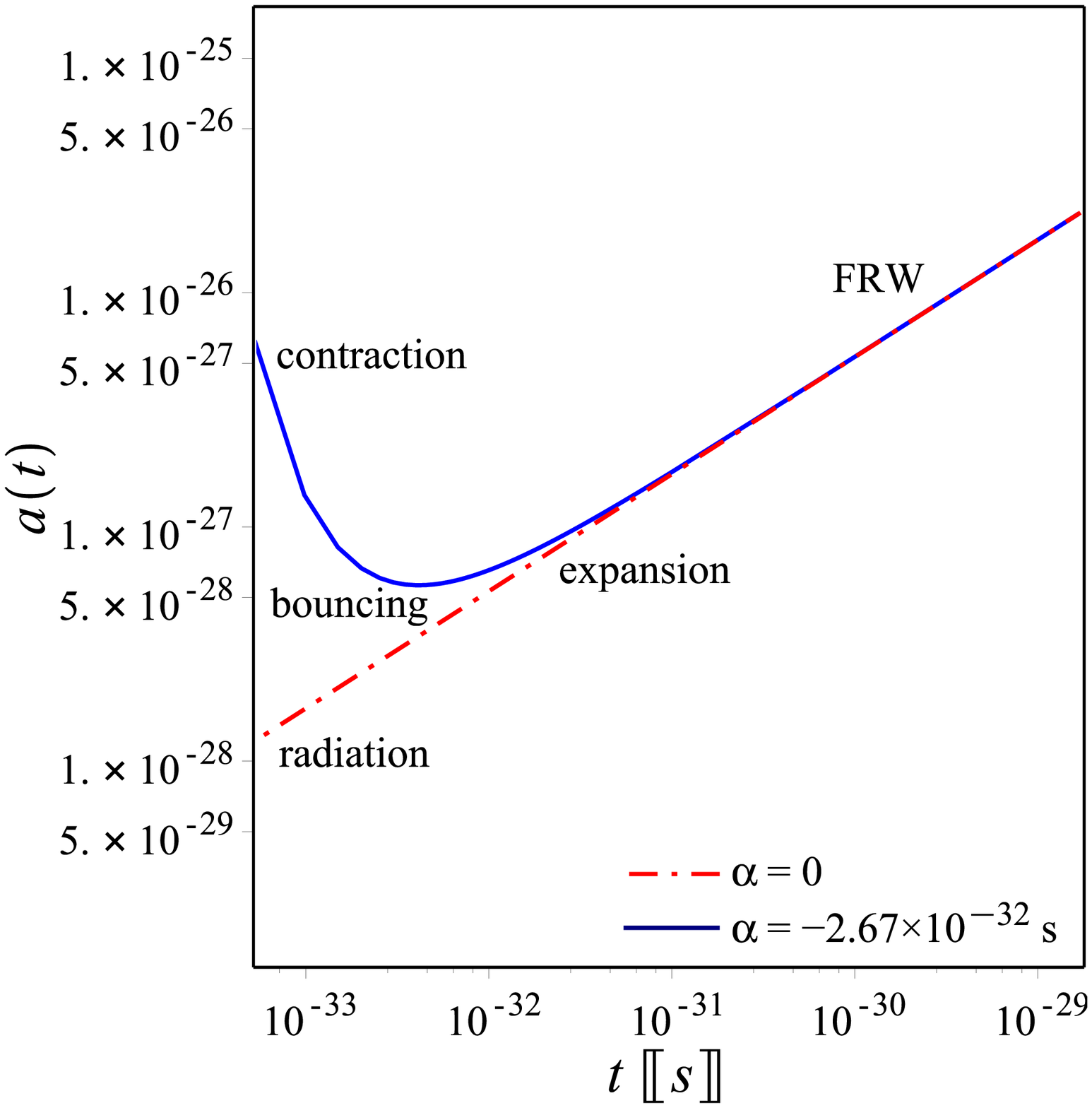}}
\subfigure[~Bounce inflation]{\label{fig2c}\includegraphics[scale=.25]{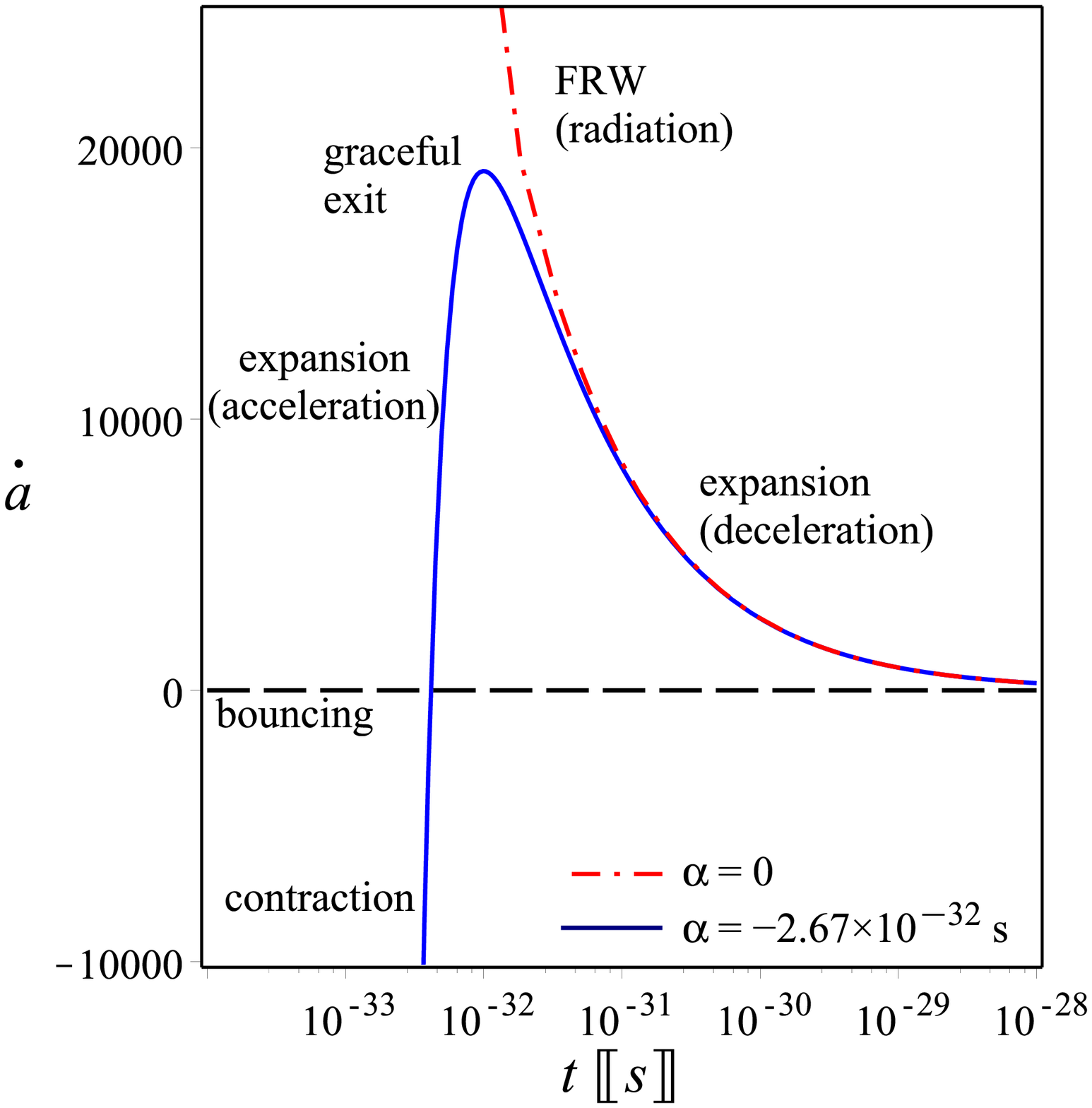}}
\caption[figtopcap]{The models: \subref{fig2a} For $\alpha>0$, we have an initial big bang singularity followed by an inflation period capable to evolve to FRW phase;
\subref{fig2b} For $\alpha<0$, we have a bouncing behavior which avoids the trans-Planckian problems of inflationary models; \subref{fig2c} For $\alpha<0$, the velocity curve shows clearly a bouncing behavior. However, after the bouncing time $t_{B}\sim 4.14\times 10^{-33}$ s, the model can also perform an early accelerated expansion period with a smooth transition (graceful exit) into a FRW model.}
\label{Fig2}
\end{figure}

In a nonphantom regime, and when $t>0$, we discuss qualitatively two possible cases:
\begin{itemize}
  \item [(i)] \textit{For} $\alpha>0$. As $t\rightarrow 0$, the scale factor is initially $a_{i}\rightarrow 0$, then we expect $\rho_{i}(t)=\infty$; that is the initial big bang singularity. At $t\ll\alpha$, we have $a_{\textmd{FRW}}\gg a_{\text{corr}}$, while at $t\approx \alpha$, we have $a_{\textmd{FRW}}< a_{\text{corr}}$; also at $t\gg\alpha$, we get $a_{\textmd{FRW}}\gg a_{\text{corr}}\sim 1$. This case gives rise to a typical graceful inflation model, see Fig. \ref{Fig2}\subref{fig2a}.
  \item [(ii)] \textit{For} $\alpha<0$. As $t\rightarrow 0^{+}$, the scale factor is initially $a_{i}\rightarrow \infty$. At $t\ll|\alpha|$, we have $a_{\textmd{FRW}}\ll a_{\text{corr}}$, while at $t\sim |\alpha|$ we have $a_{\textmd{FRW}}< a_{\text{corr}}$, also at $t\gg |\alpha|$ we get $a_{\textmd{FRW}}\gg a_{\text{corr}}\sim 1$. This case gives rise to a bouncing universe, see Fig. \ref{Fig2}\subref{fig2b}.
\end{itemize}
In both cases, we find that $a_{\textmd{corr}}\rightarrow 1$ asymptotically, which matches perfectly the FRW phase at the late time, see Fig. \ref{Fig2}. However, we are interested to study case (ii) in the above. So we take the negative parameter model, the bouncing behavior could avoid the big bang singularity so that regular problems of inflationary cosmology, e.g., trans-Planckian problems, would not have been addressed here. Interestingly, the (ii) model can perform an early accelerating expansion phase with a smooth transition to a FRW decelerated expansion later. We determine the bouncing time $t_{B}$ at which the velocity $\dot{a}=0$, where $\dot{a}<0$ (contraction phase) at $t<t_{B}$,
 while $\dot{a}>0$ (expansion phase) at $t>t_{B}$. This determines the bouncing time $t_{B}=-\frac{3\alpha}{2}\approx 4.14\times 10^{-33}$ s. The plot in Fig. \ref{Fig2}\subref{fig2c}, regardless of the initial contraction phase ($t \leq t_{B}$), shows that the velocity $\dot{a}$ experiences an increasing phase between the bounce point $\dot{a}=0$ and the inflation end $\ddot{a}=0$ (i.e., $\dot{a}=$maximum) as $t\in(t_{B},t_{\textmd{end}})$, then the velocity curve matches the power-law scale factor (FRW) decreasing phase. This indicates that the (ii) model shares same features required for a successful graceful exit inflation model after the bouncing at $t> t_{\textmd{end}}$.
\subsection{($\dot{H}-H$) phase space analysis}\label{Sec3.2} 
In the following, we construct the ($\dot{H}-H$) phase space corresponds to the modified scale factor (\ref{scale-factor}), then we track its phase portrait to extract information about the model at hand in a clear and transparent way.

\paragraph{Autonomous system.} For the scale factor (\ref{scale-factor}), we obtain the useful relation
\begin{equation}\label{phase space}
    \dot{H}_{\pm}=\frac{6\sqrt{1+9(1+\omega)\alpha H}(1+\omega)H^{2}}{-2\sqrt{1+9(1+\omega)\alpha H}\pm(2+9(1+\omega)\alpha H)},
\end{equation}
which represents one-dimensional autonomous system. Here, $\dot{H}(H)$ is a double valued function as it should be in bounce cosmology. Such a double valued function often appears when there is a first-order phase transition \cite{Nojiri:2005prd}. We take the plus sign to represent the $\dot{H}>0$ branch, while the minus sign represents the $\dot{H}<0$ branch.

\begin{figure}[t]
\centering
\subfigure[~Bounce universe]{\label{fig3a}\includegraphics[scale=.38]{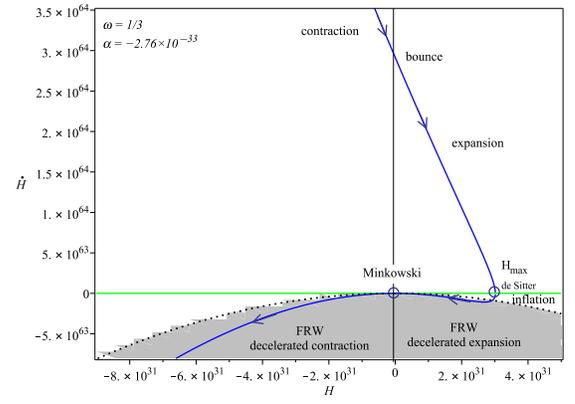}}
\subfigure[~Graceful inflation]{\label{fig3b}\includegraphics[scale=.42]{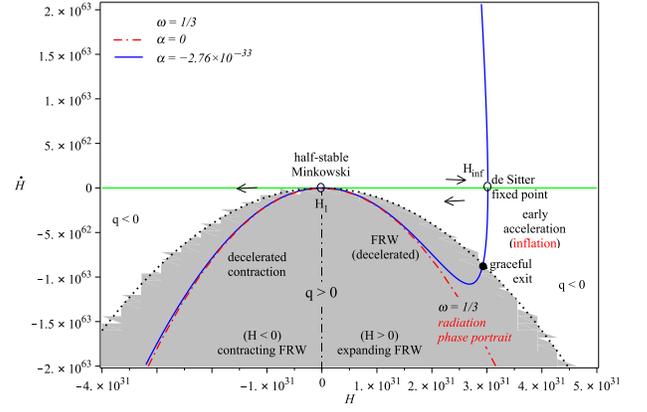}}
\caption[figtopcap]{($\dot{H}\textmd{-}H$) phase space diagram corresponds to Eq. (\ref{phase space}): \subref{fig3a} In the $\dot{H}>0$, the bounce is at $H=0$, where the transition from $H<0$ to $H>0$ is at $\dot{H}>0$;
\subref{fig2b} In the $\dot{H}<0$ and $H>0$ region, it shows an early accelerated cosmic expansion (inflation) as it appears in the (IV) region, then it is followed by a decelerated expansion phase after crossing the zero acceleration curve (FRW) as it appears in (III) region.}
\label{Fig3}
\end{figure}

\paragraph{Bounce cosmology.} In Fig. \ref{Fig3} we draw the phase space diagram corresponding to (\ref{phase space}), where the bounce point is clearly shown on Fig. \ref{Fig3}\subref{fig3a} at the point ($H_{B}=0$, $\dot{H}>0$). Before this point, the contraction phase can be shown as $H<0$ and $\dot{H}>0$, while after this point the expansion phase is determined as $H>0$ and $\dot{H}>0$. The contraction period can be evaluated as
\begin{equation}\label{bounce-contraction}
t=\int_{-\infty}^{H_{B}}\frac{dH}{\dot{H}_{+}}=-3\alpha/2\approx 4.14\times 10^{-33}~ \textmd{s},
\end{equation}
which is in agreement with the previous calculations.

\paragraph{Phantom crossing.} We first determine that the fixed points (i.e., $dH/dt=0$) are at the minimal Hubble $H_{\textmd{min}}=H_{1}=0$ (i.e., Minkowski space) and at the maximal Hubble $H_{\textmd{inf}}=H_{2}=\frac{-1}{9\alpha(1+\omega)}\sim 2.07\times 10^{7}$ GeV, which represents an inflationary Universe with $\omega_{eff}=-1$ (i.e., de Sitter space). So the period after bounce to reach de Sitter $H_{\textmd{inf}}$ can be evaluated as
\begin{equation}\label{bounce-expansion}
    t=\int_{H_{B}}^{H_{\textmd{inf}}}\frac{dH}{\dot H_{+}}=-3\alpha/2\approx 4.14\times 10^{-33}~ \textmd{s}.
\end{equation}
This makes the Universe to stay in the $\dot{H}>0$ branch a period of $-3\alpha\approx 8.28\times 10^{-33}$ s. Since the point $H_{2}$ is a fixed point, the above result seems to be unconventional. We expect that the time to reach any fixed point is an infinite, this is true when the trajectories are forced to increase or decrease monotonically. However, in our case the double valued function could alter the picture. We next investigate the possibility to cross from $\dot{H}>0$ branch to $\dot{H}<0$ through the de Sitter fixed point $H_{2}$. The former branch goes effectively as a phantomlike ($\omega_{eff}<-1$), while the latter is a nonphantom ($\omega_{eff}>-1$). The conditions for this transition to occur are listed as follows \cite{awad2013}:
\begin{itemize}
  \item [(i)] $\lim_{H\rightarrow H_{\textmd{inf}}} \dot{H}_{+}=0$,
  \item [(ii)] $\lim_{H\rightarrow H_{\textmd{inf}}} d\dot{H}_{+}/dH=\pm \infty$,
  \item [(iii)] $t=\int_{H}^{H_{\textmd{inf}}}dH/\dot{H}_{+}<\infty$.
\end{itemize}
The first condition is to ensure that the crossing point be at the fixed point $H_{\textmd{inf}}$. The second condition indicates that the pressure satisfying $dp(H)/dH$ has an infinite discontinuity at $H_{\textmd{inf}}$ so that the Universe reaches $\omega_{eff}=-1$ in a finite time, but in the general relativistic framework, the solution is not causal. The time to reach the crossing can be determined from the third condition. In addition to these conditions, it has been shown that the crossing is possible only when $\dot{H}(H)$ is a double valued function \cite{Nojiri:2010wj, awad2013}. Since the above mentioned conditions are fulfilled in this model, then the de Sitter fixed point is accessible and the transition to the standard inflationary era is valid.

\paragraph{Inflationary Universe.} In the following discussion one should deal with the $\dot{H}_{-}$ branch. The phase portrait of (\ref{phase space}) in Fig. \ref{Fig3}\subref{fig3b} shows clearly a short inflationary period after the de Sitter stage to the intersection with the zero acceleration curve $\dot{H}_{-}=-H^{2}$, which is required to go from (IV) to-(III) regions as indicated by Fig. \ref{Fig1}. So we determine possible transitions from acceleration-to-deceleration by identifying the intersections with the zero acceleration curve $\dot{H}_{-}=-H^{2}$. So we obtain these transitions at
\begin{equation}\label{transitions}
    H=\frac{2(1+3\omega)}{3\alpha}\pm \frac{2(2+3\omega)}{3\alpha}\sqrt{\frac{(1+3\omega)}{3(1+\omega)}}.
\end{equation}
This shows that possible transitions are, only, when the matter fluid is a phantom $\omega<-1$ or when $\omega\geq -1/3$. We are interested in the more physical case $\omega>-1/3$. Now if we restrict ourselves to the radiation case by taking $\omega=1/3$, it predicts the transition from acceleration-to-deceleration at $H_{\textmd{exit}}\sim 2.01\times 10^{7}$ GeV just below the $H_{\textmd{inf}}$ at the de Sitter Universe. Furthermore, we determine the inflation period by evaluating the following integral
\begin{equation}\label{inflation-phase}
    t=\int_{H_{\textmd{inf}}}^{H_{\textmd{exit}}}\frac{dH}{\dot H_{-}}\approx 1.71\times 10^{-33}~ \textmd{s}.
\end{equation}
This makes the Universe to stay in the accelerating expansion phase $\sim 5.85\times 10^{-33}$ s.

\paragraph{Graceful exit inflation.} Moreover, the model is capable to end the inflationary phase gracefully to a decelerated expansion phase, which characterizes the standard FRW cosmology at $t\sim 10^{-32}$ s. This can be shown easily by summing up (\ref{bounce-contraction}), (\ref{bounce-expansion}) and (\ref{inflation-phase}).

\paragraph{Standard decelerated FRW cosmology.} As discussed before, when the cosmic time $t>|\alpha|$, the model approaches the standard decelerated FRW cosmology. This is shown clearly on Fig. \ref{Fig3}\subref{fig3b} when the phase portrait model goes from region (IV) to (III). Also, it shows that the model matches the radiation phase portrait of the standard cosmology at late time. This is an important feature to match the thermal history of the Universe. This point will be revisited in detail in Sec. \ref{Sec5}. Finally, we show that the model has no future singularity, since the time required to approach the next fixed point, i.e., Minkowski space, at ($H=0,\dot{H}=0$) is infinite as
\begin{equation}\label{future-singularity}
    t=\int_{H_{\textmd{exit}}}^{H=0}\frac{dH}{\dot H_{-}}=\infty.
\end{equation}

\paragraph{Conclusion.} In this sense, we find the scale factor (\ref{scale-factor}) with these constraints is not only a good candidate to describe a reliable graceful exit inflation model but also its bouncing behavior avoids the trans-Planckian problems of the standard inflationary models. However, using this version of scale factors can not work properly with the linear equation-of-state, if one insists to use the standard GR. This is because of the breaking of the continuity equation. On the contrary, we can keep using the linear equation-of-state along with the scale factor (\ref{scale-factor}), if we switch to modified gravity theories.\\

One of the modified gravity theories which has been used widely in cosmology is the $f(T)$ theory. Although, this can be applied generally in modified gravity, the modified Friedmann equations of any $f(T)$ theory can be viewed as \textit{a one-dimensional autonomous system} \cite{HNS:2016}, i.e., $\dot{H} = \mathcal{F}(H)$. This feature is not available for other modified gravity theories, e.g., $f(R)$, which contain higher derivatives of $H$. In this sense, we find that the phase space analysis is more consistent with $f(T)$ cosmology. However, similar models have been investigated, without using the phase space, in Gauss-Bonnet modified gravity \cite{Oikonomou:2015PRD, Oikonomou:2016ass}.
\section{The model}\label{Sec4}
\subsection{Teleparallel space}\label{Sec4.1} 
In this section, we give a brief account of the absolute parallelism (AP) space. This space is denoted in the literature by many names teleparallel, distant parallelism, Weitzenb\"{o}ck, absolute parallelism, vielbein, parallelizable space. An AP-space is a pair $(M,\,h_{a})$, where $M$ is an $n$-dimensional smooth manifold and $h_{a}$ ($a=1,\cdots, n$) are $n$ independent vector fields defined globally on $M$. The vector fields $h_{a}$ are called the parallelization vector fields. Recent versions of vielbein space with a Finslerian flavor may have an important impact on physical applications \cite{Wanas2,Nabil1,Nabil2,Nabil3}.

Let\, $h_{a}{^{\mu}}$ $(\mu = 1, ..., n)$ be the coordinate components of the $a$ th vector field $h_{a}$, where Greek and Latin indices are constrained by the Einstein summation convention. The covariant components $h_{a \mu}$ of $h_{a}$ are given via the relations
\begin{equation}\label{orthonormality}
h_{a}{^{\mu}}h^{a}{_{\nu}}=\delta^{\mu}_{\nu}\quad \textmd{and}\quad h_{a}{^{\mu}}h^{b}{_{\mu}}=\delta^{b}_{a},
\end{equation}
where $\delta$ is the Kronecker tensor. Because of the independence of $h_{a}$, the determinant $h\equiv \det (h_{a}{^{\mu}})$ is nonzero. However, the vielbein space is equipped with many connections \cite{Wanas1,Wanas3,Wanas4,Waleed4}; on a teleparallel space $(M,\,h_{a})$, there exists a unique linear connection, namely the Weitzenb\"{o}ck connection, with respect to which the parallelization vector fields $h_{a}$ are parallel. This connection is given by
\begin{equation}\label{W_connection}
\Gamma^{\alpha}{_{\mu\nu}}\equiv h_{a}{^{\alpha}}\partial_{\nu}h^{a}{_{\mu}}=-h^{a}{_{\mu}}\partial_{\nu}h_{a}{^{\alpha}},
\end{equation}
and is characterized by the property that
\begin{equation}\label{AP_condition}
\nabla^{(\Gamma)}_{\nu}h_{a}{^{\mu}}\equiv\partial_{\nu}
{h_a}^\mu+{\Gamma^\mu}_{\lambda \nu} {h_a}^\lambda\equiv 0,
\end{equation}
where the operator $\nabla^{(\Gamma)}_{\nu}$ is the covariant derivative with respect to the Weitzenb\"{o}ck connection. The connection (\ref{W_connection}) is referred to as the canonical connection. The relation (\ref{AP_condition}) is known in the literature as the AP condition.

The noncommutation of an arbitrary vector fields $V_{a}$ is given by
$$\nabla^{(\Gamma)}_{\nu}\nabla^{(\Gamma)}_{\mu}V_{a}{^{\alpha}} - \nabla^{(\Gamma)}_{\mu}\nabla^{(\Gamma)}_{\nu}V_{a}{^{\alpha}} = R^{\alpha}{_{\epsilon\mu\nu}}
V_{a}{^{\epsilon}} + T^{\epsilon}{_{\nu\mu}} \nabla^{(\Gamma)}_{\epsilon} V_{a}{^{\alpha}},$$
where $R^{\alpha}{_{\epsilon\mu\nu}}$ and $T^{\epsilon}{_{\nu\mu}}$ are the curvature and the torsion tensors of the canonical connection, respectively. The AP condition (\ref{AP_condition}) together with the above noncommutation formula force the curvature tensor $R^{\alpha}_{~~\mu\nu\sigma}$ of the canonical connection $\Gamma^{\alpha}_{~\mu\nu}$ to vanish identically. Moreover, the parallelization vector fields define a metric tensor on $M$ by
\begin{equation}\label{metric}
g_{\mu \nu} \equiv \eta_{ab}h^{a}{_{\mu}}h^{b}{_{\nu}},
\end{equation}
with the inverse metric
\begin{equation}\label{inverse}
g^{\mu \nu} = \eta^{ab}h_{a}{^{\mu}}h_{b}{^{\nu}}.
\end{equation}
The Levi-Civita connection associated with $g_{\mu\nu}$ is
\begin{equation}\label{Christoffel}
\overcirc{\Gamma}{^{\alpha}}{_{\mu\nu}}= \frac{1}{2} g^{\alpha \sigma}\left(\partial_{\nu}g_{\mu \sigma}+\partial_{\mu}g_{\nu \sigma}-\partial_{\sigma}g_{\mu \nu}\right).
\end{equation}
In view of (\ref{AP_condition}), the canonical connection $\Gamma{^\alpha}{_{\mu\nu}}$ (\ref{W_connection}) is metric:
$$\nabla^{(\Gamma)}_{\sigma}g_{\mu\nu}\equiv 0.$$
The torsion tensor of the canonical connection (\ref{W_connection}) is defined as
\begin{equation}
T^\alpha{_{\mu\nu}}\equiv{\Gamma^\alpha}_{\nu\mu}-{\Gamma^\alpha}_{\mu\nu}={h_a}^\alpha\left(\partial_\mu{h^a}_\nu
-\partial_\nu{h^a}_\mu\right).\label{Torsion}\\
\end{equation}
The contortion tensor $K^{\alpha}_{~\mu\nu}$ is defined by
\begin{equation}
K^{\alpha}{_{\mu\nu}} \equiv \Gamma^{\alpha}_{~\mu\nu} - \overcirc{\Gamma}{^{\alpha}}_{\mu\nu}=h_{a}{^{\alpha}}~ \nabla^{(\overcirc{\Gamma})}_{\nu}h^{a}{_{\mu}}. \label{contortion}
\end{equation}
where the covariant derivative $\nabla^{(\overcirc{\Gamma})}_{\sigma}$ is with respect to the Levi-Civita connection. Since $\overcirc{\Gamma}{^{\alpha}}{_{\mu\nu}}$ is symmetric, it
follows that [using (\ref{contortion})] one can also show the following useful relations:
\begin{equation}\label{torsion03}
        T_{\alpha \mu \nu}=K_{\alpha \mu \nu}-K_{\alpha \nu \mu},
\end{equation}
\begin{equation}\label{contortion03}
        K_{\alpha \mu \nu}=\frac{1}{2}\left(T_{\nu\alpha\mu}+T_{\alpha\mu\nu}-T_{\mu\alpha\nu}\right),
\end{equation}
where $T_{\mu\nu\sigma} =
g_{\epsilon\mu}\,T^{\epsilon}_{~\nu\sigma}$\, and \,$K_{\mu\nu\sigma} =
g_{\epsilon\mu}\,K^{\epsilon}_{~\nu\sigma}$. It is to be noted that $T_{\mu\nu\sigma}$ is skew symmetric in the last pair of indices whereas $K_{\mu\nu\sigma}$ is skew symmetric in the first pair of indices. Moreover, it follows from (\ref{torsion03}) and (\ref{contortion03}) that the torsion tensor vanishes if and only if the contortion tensor vanishes. In the teleparallel space there are three Weitzenb\"{o}ck invariants: $I_{1}=T^{\alpha \mu \nu}T_{\alpha \mu \nu}$, $I_{2}=T^{\alpha \mu \nu}T_{\mu \alpha \nu}$ and $I_{3}=T^{\alpha}T_{\alpha}$, where $T^{\alpha}=T_{\rho}{^{\alpha \rho}}$. We next define the invariant $T=AI_{1}+BI_{2}+CI_{3}$, where $A$, $B$ and $C$ are arbitrary constants \cite{M2013}. For the values: $A=1/4$, $B=1/2$ and $C=-1$  the invariant $T$ is just the Ricci scalar up to a total derivative term; then a teleparallel version of gravity equivalent to GR can be achieved. The teleparallel torsion scalar is given in the compact form
\begin{equation}
T \equiv {T^\alpha}_{\mu \nu}{S_\alpha}^{\mu \nu},\label{Tor_sc}
\end{equation}
where the superpotential tensor
\begin{equation}
{S_\alpha}^{\mu\nu}=\frac{1}{2}\left({K^{\mu\nu}}_\alpha+\delta^\mu_\alpha{T^{\beta\nu}}_\beta-\delta^\nu_\alpha{T^{\beta \mu}}_\beta\right),\label{superpotential}
\end{equation}
is skew symmetric in the last pair of indices.  Also, there are different extensions of TEGR, e.g., Born-Infeld extension of the TEGR \cite{FF08,F13}, another interesting variant is the modified teleparallel equivalent of Gauss-Bonnet gravity and its applications \cite{KS114,KS214,KS314}. Another extension is the $f(T)$ gravity, it has been inspired by the $f(R)$-gravity when the Ricci scalar is replaced by an arbitrary function $f(R)$ in the Einstein-Hilbert action. But the former is by replacing the teleparallel torsion scalar by an arbitrary function $f(T)$ \cite{BF09, L10,1008.4036,1011.0508}. We consider the action of the $f(T)$ gravity
\begin{equation}\label{action}
{\cal S}=\int d^{4}x~ |h|\left[\frac{1}{2 \kappa^2}f(T)+L_{m}\right],
\end{equation}
where ${L}_{m}$ is the Lagrangian of the matter and $|h|=\sqrt{-g}=\det\left({h}_\mu{^a}\right)$. The variation of the action (\ref{action}) with respect to the tetrad gives
\begin{equation}\label{field_eqns}
\frac{1}{h} \partial_\mu \left( h S_a^{\verb| |\mu\nu} \right) f^{\prime}-h_a^\lambda  T^\rho_{\verb| |\mu \lambda} S_\rho^{\verb| |\nu\mu}f^{\prime}
+S_a^{\verb| |\mu\nu} \partial_\mu T f^{\prime\prime}
+\frac{1}{4} h_a^\nu f=\frac{{\kappa}^2}{2} h_a^\rho \mathfrak{T}_\rho^{\verb| |\nu},
\end{equation}
where $f=f(T)$, $f^{\prime}=\frac{\partial f(T)}{\partial T}$, $f^{\prime\prime}=\frac{\partial^2 f(T)}{\partial T^2}$ such that the TEGR theory is recovered by setting $f(T)=T$. Also, the stress-energy tensor is assumed to be for perfect fluid as given by (\ref{stress-energy}). The applications of the $f(T)$ gravity in cosmology show interesting results, for example, avoiding the big bang singularity by presenting a bouncing solution \cite{Cai:2011tc,CQSW14}. Also, $f(T)$ cosmology provides an alternative tool to study inflationary models \cite{FF07,FF08,BF09,FF11,Waleed5,BNO14,BO14,Waleed6,JMM14,HLOS14,Nd15,Saridakis2,Nd215}. Although $f(T)$-theories lack invariance under a local Lorentz transformation \cite{1010.1041,1012.4039,LLT1} (for the related considerations, see~\cite{Li:2011rn, Ong:2013qja, Bamba:2013jqa, Bamba:2013ooa, Izumi:2013dca, Chen:2014qtl, Bahamonde:2015zma}), a recent modification by considering nontrivial spin connections may solve the problem \cite{Martin2}. For more details of $f(T)$ gravity, see the recent review \cite{Saridakis1}.
\subsection{Constructing an $f(T)$ theory}\label{Sec4.2} 
As presumed that the Copernican principle is valid, we take the flat FRW metric (\ref{FRW-metric}), which may give rise to the vierbein
\begin{equation}\label{tetrad}
{h_{\mu}}^{a}=\textmd{diag}\left(1,a(t),a(t),a(t)\right).
\end{equation}
For the vierbein (\ref{tetrad}) and by using (\ref{scale-factor}), the teleparallel torsion scalar (\ref{Tor_sc}) gives rise to the useful relation
 \begin{equation}\label{Tor_sc1}
T(t)=-6H(t)^{2}=-\frac{2\left[3\alpha+2(t-t_{i})\right]^{2}}{3(1+\omega)^{2}(t-t_{i})^{4}}.
\end{equation}
As we mentioned earlier, in the $f(T)$ framework, one needs to enter a particular scale factor or viable $f(T)$ in addition to a specific equation-of-state. In this model, we are interested to construct an $f(T)$ theory corresponding to the modified scale factor (\ref{scale-factor}), where the equation-of-state is chosen to be linear $p=\omega\rho$. We apply the $f(T)$ field equations (\ref{field_eqns}) to the vierbein (\ref{tetrad}), then the modified Friedmann equations read
\begin{eqnarray}
\rho&=&\frac{1}{2 \kappa^{2}}(f+12 H^2 f'),\label{dens1}\\
p&=&-\frac{1}{2 \kappa^{2}}\left[(f+12 H^2 f') + 4\dot{H}(f'-12 H^2 f'')\right].\label{press1}
\end{eqnarray}
It is convenient to write the $f(T)$ in terms of time $t$. One easily can show that
\begin{equation}\label{Fd2T}
f^{\prime} = \dot{f}/\dot{T},~ f^{\prime\prime} = \left(\dot{T} \ddot{f}-\ddot{T} \dot{f}\right)/\dot{T}^{3}.
\end{equation}
Substitute (\ref{Tor_sc1}) and (\ref{Fd2T}) into (\ref{dens1}) and (\ref{press1}), then matter density
\begin{equation}\label{dens}
    \rho(t)= \frac{1}{4\kappa^2}\frac{2\left[3\alpha+(t-t_{i})\right]f+\left[3\alpha+2(t-t_{i})\right](t-t_{i})\dot{f}}{3\alpha+(t-t_{i})},
\end{equation}
and the matter pressure
\begin{eqnarray}\label{press}
\nonumber    p(t)&=& \left\{-2\left[3\alpha+(t-t_{i})\right]^{2}f\right.\\
\nonumber        &-&\left[(2+\omega)(t-t_{i})\left(2(t-t_{i})+9\alpha\right)+9\alpha^{2}\right](t-t_{i})\dot{f}\\
\nonumber        &-&\left.(1+\omega)\left[3\alpha+(t-t_{i})\right](t-t_{i})^{3}\ddot{f}\right\}
                 /{4\kappa^2[3\alpha+(t-t_{i})]^{2}}.\\
\end{eqnarray}
The continuity equation can be integrated to
\begin{equation}\label{reln}
    \rho=\rho_{0}a^{-3(1+\omega)},
\end{equation}
where the integration constant $$\rho_{0}\equiv \rho(t_{0}) \approx 1.88\times 10^{-29}~\Omega h^{2}~\textmd{g.cm}^{-3},$$ the density parameter $\Omega$, and the dimensionless hubble constant $h$ are given by the observations. Combining (\ref{reln}) with (\ref{dens}), then solving for $f(t)$, we get
\begin{equation}\label{f(t)}
    f(t)=\frac{c_{1}\left[3\alpha+2(t-t_{i})\right]}{(t-t_{i})^{2}}+c_{2}\frac{e^{3\alpha/(t-t_{i})}}{(t-t_{i})},
\end{equation}
where the constant $c_2=-\frac{4\kappa^2\rho_{0}a_{k}^{-3(1+\omega)}}{3\alpha (1+\omega)^{2}}$; $\alpha \neq 0$. Using the inverse relation of (\ref{Tor_sc1}), one can rewrite the above result as $f(T)$ as usual. Thus the corresponding $f(T)$ theory which generates the scale factor (\ref{scale-factor}) is given by
\begin{equation}\label{f(T)}
    f(T)_{\pm}=c_{1}\sqrt{T}+c_{2}(-T)^{1/4}\left(\frac{G(T)\pm 6}{G(T)\mp 6}\right)^{1/2}e^{\frac{\mp G(T)}{6}},
\end{equation}
where $G(T)\equiv \left(36+54\alpha (1+\omega)\sqrt{-T}\right)^{1/2}$. Here, $f(T)_{+}$ corresponds to the branch $\dot{H}>0$, and $f(T)_{-}$ corresponds to the branch $\dot{H}<0$. One can show that the first term $\sim \sqrt{T}$ in (\ref{f(T)}) has no contribution in the field equations, so we omit this term in the following without affecting the generality of the model. It is convenient to evaluate the evolution of the density and pressure of the matter, this can be achieved by substituting from (\ref{f(t)}) into (\ref{dens}) and (\ref{press}); the density and pressure can be written as
\begin{equation}\label{dens-press}
    \rho=\rho_{0}\frac{a_{k}^{-3(1+\omega)}e^{3\alpha/(t-t_{i})}}{(1+\omega)^2(t-t_{i})^{2}};\quad p=\omega \rho_{0}\frac{a_{k}^{-3(1+\omega)}e^{3\alpha/(t-t_{i})}}{(1+\omega)^2(t-t_{i})^{2}}.
\end{equation}
It is obvious that as $t\gg |\alpha|$ the density and pressure of the standard FRW model is recovered.
\subsection{Effective equation of state}\label{Sec4.3}   
\begin{figure}[t]
\centering
\includegraphics[scale=.3]{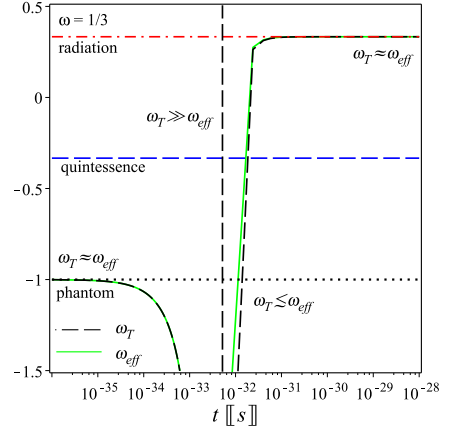}
\caption{The equation-of-state parameter of the torsion is given by the solid line, while the effective (total) equation-of-state parameter is given by the dash-doted line.}
\label{Fig5}
\end{figure}

It is convenient to transform from the \textit{matter} frame we have been using to \textit{Einstein} frame, which gives the Einstein's field equations form and additional degrees-of-freedom by $f(T)$ gravity. So we write the modified Friedmann equations in the case of $f(T)$ gravity, i.e.,
\begin{eqnarray}
{H}^2& =& \frac{\kappa^2}{3} \left( \rho+  \rho_{ T} \right), \label{MFR1}\\
2 \dot{{H}} + 3{H}^2&=& - \kappa^2 \left(p+p_{ T }\right),\label{MFR2}
\end{eqnarray}
where the standard matter energy density $\rho$ and pressure $p$ have their torsion scalar  counterpart $\rho_{ T}$ and $p_{ T}$,
\begin{equation}\label{rhoT}
    \rho_{T}=\frac{1}{2\kappa^2}[2{T}f'-f-{T}],
\end{equation}
\begin{equation}\label{pT}
    p_{T}=\frac{2}{\kappa^2}\dot{H}(2{T}f''+f'-1)-\rho_{T}.
\end{equation}
are the torsion contributions to the energy density and pressure, respectively, which satisfy the continuity
\begin{equation}\label{continuity}
    \dot{\rho}_{T}+3H(\rho_{T}+p_{T})=0.
\end{equation}
One can show that $\rho_{T}$ and $p_{T}$ vanish where $f(T)=T$ and the standard Friedmann equations are recovered. We argue here that the quantities  $\rho_{ T}$ and $p_{ T}$ can explain the early self-acceleration of the Universe. Then, by using Eqs. (\ref{rhoT}) and (\ref{pT}), we can define the effective torsion equation-of-state parameter as
\begin{eqnarray}\label{Tor_EoS}
\nonumber    \omega_{T}&\equiv &\frac{p_{ T}}{\rho_{ T}}=-1+\frac{4\dot H(2{ T}f''+f'-1)}{2{ T}f'-f-{ T}},\\[3pt]
                          &=&\frac{3\omega\kappa^2\rho_{0}t^{2}e^{\frac{3\alpha}{t}}+a_{k}^{3(1+\omega)}\left[9\alpha^2-4\omega t(3\alpha+t)\right]}{3\kappa^2\rho_{0}t^2 e^{\frac{3\alpha}{t}}-a_{k}^{3(1+\omega)}(2t+3\alpha)^2}.
\end{eqnarray}
Where the last equation has been evaluated by using (\ref{Tor_sc1}), (\ref{f(t)}), it can be shown that $\omega_{T}=\omega$ at $t\gg |\alpha|$. We plot the evolution of the equation-of-state parameter of the teleparallel torsion fluid as seen in Fig. \ref{Fig5}. Equation (\ref{Tor_EoS}) is ill defined at
$$t_{\pm}=\frac{-3\alpha}{2W(\pm \Sigma)+1},$$
where $\Sigma=\frac{\sqrt{3}}{2e}\kappa\rho_{0}^{1/2}a_{k}^{-3(1+\omega)/2}$ and $W(\Sigma)$ is the Lambert-$W$ function which is the solution of $We^{W}=x$. This defines an interval $(t_{+},t_{-})\approx(3.54\times 10^{-33}\textmd{s},5.85\times 10^{-33}\textmd{s})$. So it initially begins as a cosmological constant $\omega_{T}\rightarrow -1$, then it goes to $-\infty$ as $t\rightarrow t_{+}$, while $\omega_{T}\gg 1$, where $t_{+}<t<t_{-}$; this includes the bounce time $t_{B}$, which shows that the torsion equation of state is greater than unity at the contraction phase as required for solving the anisotropy problem. After that, $\omega_{T}$ is negative again, while it goes back to cross $\omega_{T}=-1$ to connect the observed expanding Universe \cite{bounce2}, then it crosses $\omega_{T}=-1/3$ ending the early accelerated expansion, at $t\sim t_{\textmd{end}}\approx10^{-32}$ s, to enter a new phase of a decelerated expansion. Finally, it approaches the radiation limit $\omega_{T}=\omega=1/3$ as $t\gg \alpha$ as required to match the hot big bang consistently.

As we discussed above, the torsion equation-of-state evolutions fulfills the requirements of a successful bounce cosmology. In addition, it matches precisely the results of the phase space analysis in Sec. \ref{Sec3.2}. This behavior supports our argument that the cosmic bounce is a manifestation of a higher-order teleparallel gravity. In other words, the vacuum $f(T)$ is a good candidate to describe bounce cosmology. We also define the effective (total) equation-of-state parameter
\begin{eqnarray}\label{eff_EoS}
\nonumber    \omega_{eff}&\equiv& \frac{p+p_{T}}{\rho+\rho_{T}},\\
                         &=&\omega-\frac{9\alpha^2(1+\omega)}{(2t+3\alpha)^{2}}.
\end{eqnarray}
Where the last equation has been evaluated by using (\ref{dens-press}), (\ref{rhoT}) and (\ref{pT}), it is obvious that $\omega_{eff}=\omega$ at $t\gg |\alpha|$. We plot the $\omega_{eff}$ as shown in Fig. \ref{Fig5}. Equation (\ref{eff_EoS}) shows that the Universe effectively initially is a cosmological constant with $\omega_{eff}=-1$, then it evolves to $\omega_{eff}\rightarrow -\infty$ at the bounce time $t_{B}=-\frac{3}{2}\alpha\sim 4.14\times 10^{-33}$ s, while it is $-1/3$ at the acceleration expansion ends $t_{\textmd{end}}$. Finally, it matches the radiation limit, i.e., $\omega_{eff}=\omega=1/3$. In conclusion, we find that the torsion and the effective equation-of-state parameters agree in all stages except at the bounce time. The later goes to $-\infty$, while the former is much greater than unity at that time.

As is well-known, the violation of the NEC is necessary to obtain a bouncing solution. In addition, violation of the strong energy condition (SEC) is necessary to obtain an accelerated expansion phase. The above results show clearly that the model effectively breaks these energy conditions at the early Universe, where $\omega_{eff}<-1$. However, due to the limitation of the GR, the violation of these energy conditions in the matter component is unavoidable. Even in the effective field theory, a bounce Universe is usually achieved by introducing matter fields, which violate the NEC. On the contrary, this picture could be altered if we use the $f(T)$ gravity which violates the NEC effectively. We can always use this feature to produce a healthy bounce solution, where the matter component in this scenario will be consistent with the NEC. This will be discussed in detail in Sec. \ref{Sec5.4}.
\section{Thermalization of the universe}\label{Sec5}
\subsection{Reheating in bounce universe}\label{Sec5.1}  
As mentioned before, the key of the thermal history is to compare the rate of interactions $\Gamma$ with the rate of expansion $H$. In the case of $\Gamma \gg H$, the time scale of the particle interactions is much smaller than the expansion time scale as
$$t_{c}\equiv \frac{1}{\Gamma}~ \ll~ t_{H}\equiv \frac{1}{H}.$$
Thus, a local thermal equilibrium can be reached before the effect of the expansion becomes relevant.
After, as the Universe cools down, the $\Gamma$ decreases faster than $H$ so that at $t_{c} \sim t_{H}$, the particles decouple from the thermal bath. Different particle species may have different interaction rates and so may decouple at different times. On the other hand, one of the essential ingredients of inflationary models is the reheating process of the Universe by the end of the inflation. In order to examine the capability of the model to predict a successful thermal evolution, we define the entropy $S$ of all particles in thermal equilibrium at temperature $\Theta$ in volume $V$. According to the first law of thermodynamics, in the expanding universe, we have
\begin{equation}\label{Thermodynamics2}
    \Theta dS=d(\rho V)+p dV,
\end{equation}
with the integrability condition \cite{Wberg1972}
\begin{equation*}
    \frac{\partial^2 S}{\partial \Theta \partial V}=\frac{\partial^2 S}{\partial V \partial \Theta},
\end{equation*}
the energy density and pressure satisfy
\begin{equation}\label{Temp}
    \frac{dp}{d\Theta}=\frac{\rho+p}{\Theta}.
\end{equation}
Using (\ref{dens-press}) and (\ref{Temp}), we evaluate the temperature
\begin{eqnarray}\label{Temperature}
\nonumber    \Theta(t)&=&\Theta_{0}~e^{\displaystyle{\int}\frac{\frac{d}{dt}p(t)}{\rho(t)+p(t)}dt},\\[3pt]
\nonumber             &\propto&a^{-3\omega},\\
                      &=&\Theta_{0}\left[(1+\omega)t\right]^{\frac{-2\omega}{1+\omega}}e^{\frac{3\alpha\omega}{(1+\omega)t}},
\end{eqnarray}
\begin{figure}[t]
\centering
\includegraphics[scale=.4]{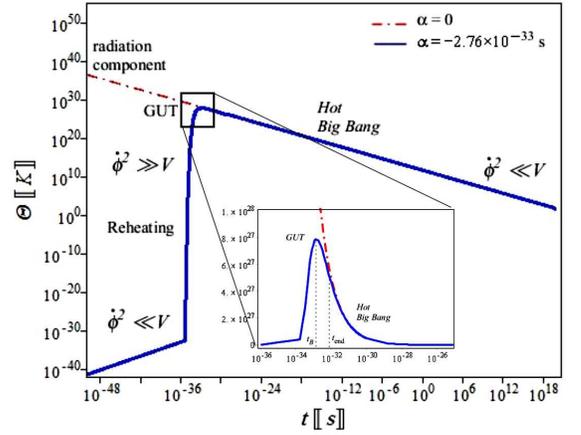}
\caption{The temperature evolution (\ref{Temperature}) shows a reheating after inflation, the maximum effective temperature by the end of reheating is just below the GUT temperature $\sim \Theta_{\textmd{eff}}=\Theta_{\text{rh,max}}\sim 10^{26}~K$. Then the effective temperature evolves similar to the radiation temperature $\Theta_{\textmd{r}}$ of the hot big bang.}
\label{Fig4}
\end{figure}
where $\Theta_{0}\equiv\Theta(t_{0})$ is an arbitrary constant, with a dimension $K$. We choose a boundary condition such that the temperature $\Theta\sim 2.73~K$ at the present time $t_{0}\sim 10^{17}~\textmd{s}>t_{eq}$, with an equation-of-state parameter $\omega=0$. This determines the value $\Theta_{0}= 2.73~K$. In standard cosmology we expect an extremely high temperature as Universe goes back towards the initial singularity, i.e., $\Theta_{i} \rightarrow \infty$ as $a_{i}\rightarrow 0$. However, in the present model, equation (\ref{Temperature}) indicates that the temperature initially is extremely small, i.e., $\Theta_{i} \rightarrow 0~K$ as $a_{i}\rightarrow \infty$, during the precontraction phase. Then the temperature increases as $a$ decreases during contraction to its maximal value at the bounce time $t_{B}$. From the temperature (\ref{Temperature}), it can be shown that the maximum temperature by the end of the reheating, at $t_{B}=-\frac{3\alpha}{2}\sim 4.14\times 10^{-33}$ s, is $\Theta_{\textmd{rh,max}}\sim 4.8\times 10^{26}~K$, see Fig. \ref{Fig4}. Also, it is clear from (\ref{Temperature}) that the temperature evolves just as the standard cosmology at $t \gg |\alpha|$. In conclusion, the model predicts an initial low temperature, then a reheating of the Universe occurs during the contraction phase. At bounce time, the Universe reaches its maximum temperature $10^{26}~K$, which is just below the GUT temperature $\Theta_{\textmd{GUT}}\sim 10^{27}~K$. So the model is safe from reproducing unacceptable amount of monopoles after inflation. This is followed by a very short period of accelerated expansion to cool the Universe down to match exactly the standard model thermal evolution. So we gain the successes of the hot big bang scenario as well.
\subsection{Unified inflaton-quintessence field}\label{Sec5.2}            
The above result seems unfamiliar at first impression. One may expect the temperature to start with $\Theta_{p}$ at Planck's time, not $\Theta\sim ~0~K$. As a matter of fact, this model predicts a more physical scenario when dealing with a scalar field component. We will discuss this point in detail in the following section. In order to investigate the scalar field induced by the theory at hand, we take the matter component to be a canonical scalar field with density $\rho_{\phi}$ and pressure $p_{\phi}$ to be defined as
\begin{eqnarray}
  \rho_{\phi} &=& \frac{\dot{\phi}^2}{2}+V(\phi), \label{rho_phi}\\
  p_{\phi} &=& \frac{\dot{\phi}^2}{2}-V(\phi), \label{p_phi}
\end{eqnarray}
where $\dot{\phi}^2$ represents a kinetic term of the scalar field and $V(\phi)$ is its potential. Combining (\ref{rho_phi}), (\ref{p_phi}), (\ref{MFR1}) and (\ref{MFR2}), we write the kinetic and the potential of the scalar field
\begin{equation}\label{kinetic}
    \dot{\phi}^{2}(t)=\frac{1}{6\kappa^2}\left(\frac{\dot{H}\ddot{f}-\dot{f}\ddot{H}}{\dot{H}^2}\right),
\end{equation}
\begin{equation}\label{potential}
    V(t)=\frac{1}{2\kappa^2}\left[f+\left(\frac{\ddot{H}}{6\dot{H}^2}-\frac{H}{\dot{H}}\right)\dot{f}-\frac{\ddot{f}}{6\dot{H}}\right].
\end{equation}
In order to be consistent with the literature, we may use $\kappa^2=1/M_{p}^2$ with $M_{p}=1.22\times 10^{19}$ GeV. The above equations are consistent with the scalar field background (Klein-Gordon) equation of a homogeneous scalar field in the expanding FRW Universe \begin{equation}\label{Klein-Gordon}
    \ddot{\phi}+3H\dot{\phi}+\frac{dV}{d\phi}=0.
\end{equation}
Inserting (\ref{Tor_sc1}) and (\ref{f(t)}) into (\ref{kinetic}) and (\ref{potential}), we get
\begin{equation}\label{kinetic2}
    \dot{\phi}^{2}(t)=\frac{\rho_{0}a_{k}^{-3(1+\omega)}e^{3\alpha/t}}{(1+\omega)t^2},
\end{equation}
which can be integrated exactly to
\begin{eqnarray}
    \phi(t)&=&\phi_{0}+\xi~\textmd{Ei}\left(1,-\frac{3\alpha}{2~t}\right),\label{phi1}\\
              &\approx&\phi_{0}-\xi\left[\gamma+\ln\left(-\frac{3\alpha}{2t}\right)
              +\frac{3\alpha}{2t}+O\left(\frac{\alpha^2}{t^2}\right)\right],\label{phi2}
\end{eqnarray}
where $\phi_{0}$ is a constant of integration, $\xi\equiv \sqrt{\frac{\rho_{0}}{1+\omega}}a_{k}^{-\frac{3}{2}(1+\omega)}$, Ei($i$,$x$) is the exponential integral, and $\gamma$ is Euler's constant which is approximately $0.5772\ldots$. Since $\alpha<0$, the Ei function in (\ref{phi1}) is real and consequently, the scalar field. Integrating the Klein-Gordon equation (\ref{Klein-Gordon}), we have
\begin{equation}\label{potential2}
V(t)=V_{0}+\frac{(1-\omega)\rho_{0}a_{k}^{-3(1+\omega)}e^{3\alpha/t}}{2(1+\omega)^{2}t^{2}},
\end{equation}
where $V_{0}$ is the constant of integration. One can reobtain the above solution from (\ref{potential}) without the $V_{0}$ term. However, the presence of $V_{0}$ can be recovered by considering a cosmological constant in the $f(T)$. We shall discuss this issue in more concretely later on in this section. Using the inverse relation of (\ref{phi2}), we can eliminate $t$ in (\ref{potential2}) and rewrite the potential as $V(\phi)$ as usual. Substituting from (\ref{kinetic2}) and (\ref{potential2}) into (\ref{rho_phi}) and (\ref{p_phi}), we evaluate the scalar field density energy and pressure
\begin{eqnarray}
  \rho_{\phi} &=& \rho_{0}a_{k}^{-3(1+\omega)}\frac{e^{3\alpha/t}}{(1+\omega)^{2}t^{2}}+V_{0},\label{rho_phi1} \\
  p_{\phi} &=& \omega \rho_{0}a_{k}^{-3(1+\omega)}\frac{e^{3\alpha/t}}{(1+\omega)^{2}t^{2}}-V_{0}. \label{p_phi1}
\end{eqnarray}
Then the equation-of-state parameter $\omega_{\phi}=p_{\phi}/\rho_{\phi}$ of the scalar field is
\begin{equation}\label{omega-phi}
\omega_{\phi}(t) \equiv \frac{p_{\phi}}{\rho_{\phi}}=\frac{\omega \rho_{0}a_{k}^{-3(1+\omega)}e^{3\alpha/t}-(1+\omega)^{2}t^{2}V_{0}}
{\rho_{0}a_{k}^{-3(1+\omega)}e^{3\alpha/t}+(1+\omega)^{2}t^{2}V_{0}}.
\end{equation}
\begin{figure}[t]
\centering
\includegraphics[scale=.4]{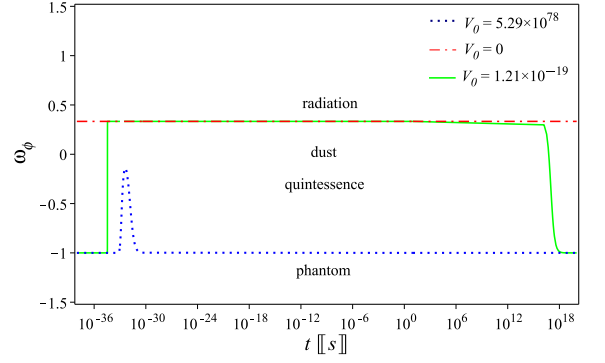}
\caption{The equation-of-state parameter of the scalar field.}
\label{Fig6}
\end{figure}
It is important to investigate a possible crossing of an equation of state to the phantom divide line $\omega_{\phi}=-1$ or the quintessence limits at $\omega_{\phi}=-1/3$. Independent of the $V_{0}$ value, Eq. (\ref{omega-phi}) indicates no crossing to the phantom phase, so the scalar field bounce model is always connected with the observed expanding Universe. We determine the value of $V_{0}$ by requiring that $\omega_{\phi}=-1/3$ at time $t_{s}$ chosen according to cosmological constraints. Thus we have
\begin{equation}\label{V0}
    V_{0}=\frac{\rho_{0}a_{k}^{-3(1+\omega)}(1+3\omega)e^{3\alpha/t_{s}}}{2(1+\omega)^{2}t_{s}^{2}}.
\end{equation}
If we assume that $\omega=1/3$ along with $t_{s}=10^{-32}$ s at the graceful exit time, we have $V_{0}\sim 5.29\times 10^{78}$. If we choose $t_{s}=10^{17}$ s at late time acceleration, we have $V_{0}\sim 1.21 \times 10^{-19}$. However, inserting (\ref{V0}) into (\ref{omega-phi}) implies that the scalar field crossing to the quintessence limit has a three patterns according to the value of $V_{0}$:
\paragraph*{a. Case ($V_{0}=0$).} For a vanishing value of $V_{0}$, we get $\omega_{\phi}=\omega$. So the scalar field has a fixed equation of state.\\

\paragraph*{b. Case ($V_{0}\neq 0$).} For nonvanishing values of $V_{0}$, we have two possible scenarios:

(\textbf{i}) for large $V_{0}$, the Universe is trapped in an inflation phase. It begins with $\omega_{\phi}=-1$, then it goes to higher values. In order to make the acceleration-to-deceleration transition at the end of the inflation, i.e., $t_{s}=10^{-32}$ s, we choose a large value $V_{0}\sim 5.29\times 10^{78}$. This makes the equation-of-state parameter just above the quintessence limit $\omega_{\phi}\gtrsim -1/3$ for a very short period $t\in (10^{-33},~10^{-32})$ s, then goes back towards $\omega_{\phi}\rightarrow -1$ or an eternal de Sitter and will never match the radiation limit.

(\textbf{ii}) for small $V_{0}$, similar to the previous case, the Universe begins with $\omega_{\phi}=-1$; however in this case, it has a chance to end its early accelerating expansion phase entering a deceleration one for a reasonable long period, with a later transition to a de Sitter Universe just as $\Lambda$CDM cosmology. We found that for the smaller $V_{0}$ values, the later transition is towards de Sitter. In order to make the late transition at $t_{s}= 10^{17}$ s, we choose $V_{0}\sim 1.21\times 10^{-19}$. Besides, the early transition from acceleration to deceleration can be obtained at $t\sim 10^{-35}$ s. Interestingly, the radiation limit in this case is allowed at $t\in (\sim 10^{-35}\textmd{s},t_{eq})$. So we find that the scalar field unifies the inflaton and the quintessence fields in a single model. The three patterns of the scalar field equation-of-state parameter are shown in the plots of Fig. \ref{Fig6}.\\

In order to comprehend the results of an induced scalar field in the frame of the phase space analysis in Sec. \ref{Sec3.2}, we see that the presence of the $V_{0}$ term representing a ground state or background of a cosmological constant. In large $V_{0}$ regimes, the Universe has a high potential to remain in a de Sitter universe, while for small $V_{0}$ regimes, the Universe has being outside de Sitter for a longer period before pulling to de Sitter once again as a final fate. This can be compared to the situation when large or small values of cosmological constant are adopted in the theory. This conclusion can be easily seen on the phase space diagram, where the presence of a positive cosmological constant shifts the phase portrait vertically upwards such that the larger value of $\Lambda$, the more shifts of the phase portrait. In this model the portrait, Fig. \ref{fig3b}, generally cuts the zero acceleration curve at two points. When the cosmological constant is large, the period between these two cuttings is short just as in the large $V_{0}$ regime of the scalar field. On the other hand, when the cosmological constant is small, the phase portrait will be allowed to remain in the decelerated FRW cosmology for a longer period. However, in both cases, the Universe evolves towards a de Sitter fixed point instead of Minkowski in an infinite time, which represents a similar scenario of a $\Lambda$CDM Universe.
\subsection{Slow-roll validity}\label{Sec5.3}            
As discussed in Sec. \ref{Sec5.1}, the cosmic temperature begins with very low temperature $\Theta\sim 0~K$ as predicted by the model. This result is compatible with the slow-roll condition $V(\phi) \gg \dot{\phi}^{2}$ so that the inflation epoch is dominated by the scalar field potential only. Accordingly, its equation-of-state parameter $\omega_{\phi}\equiv p_{\phi}/\rho_{\phi}\approx -1$, where $V(\phi) \gg \dot{\phi}^2$. The later assumption is called the slow-roll condition. In fact this condition can not be justified unless the temperature at that episode is very small. According to Eq. (\ref{Temperature}), the slow roll condition can be valid at the precontraction phase as well as at a late Universe phase when the temperature is low as shown in Fig. \ref{Fig4}. In order to examine the viability of the model at hand, we write the slow-roll parameters:
\begin{equation}\label{slow_roll}
    \epsilon_{V}=\frac{1}{2\kappa^{2}}\left(\frac{V_{\phi}}{V}\right)^{2},\qquad \eta_{V}=\frac{1}{\kappa^{2}}\left(\frac{V_{\phi\phi}}{V}\right),
\end{equation}
where $V_{\phi}={dV}/{d\phi}$ and $V_{\phi\phi}={d^{2}V}/{d\phi^{2}}$. Using (\ref{slow_roll}), we evaluate the two observable parameters: For large $V_{0}$, the tensor-to-scalar ratio $r=16\epsilon_{V} \sim 9.7\times 10^{-4}$, and scalar tilt (spectral index) $n_{s}=1-6\epsilon_{V}+2\eta_{V} \sim 1.0004$. Although the spectrum is nearly scale invariant, the spectral index in this case is slightly blue tilted which is disfavored by the observations. However, for a vanishing or small $V_{0}$, we evaluate the two observables $r \sim 1.56\times 10^{-2}$ and $n_{s} \sim 0.997$, which are in agreement with the recent observations by the Planck satellite and BICEP2 and Keck Array~\cite{Planck:2015xua, Ade:2015lrj, Ade:2015tva, Array:2015xqh}. In the above calculations, we assumed that $\omega=1/3$, for different choices of $\omega$ the corresponding values of $V_{0}$ will be different but the qualitative behavior is the same.
\begin{figure}
\centering
\subfigure[~$V_{0} \gg 1$]{\label{fig7a}\includegraphics[scale=.32]{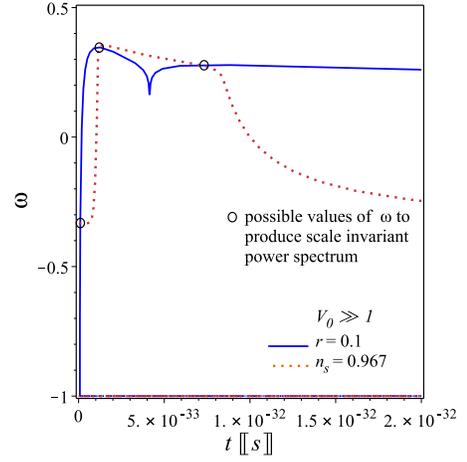}}
\subfigure[~$V_{0}\ll 1$]{\label{fig7b}\includegraphics[scale=.32]{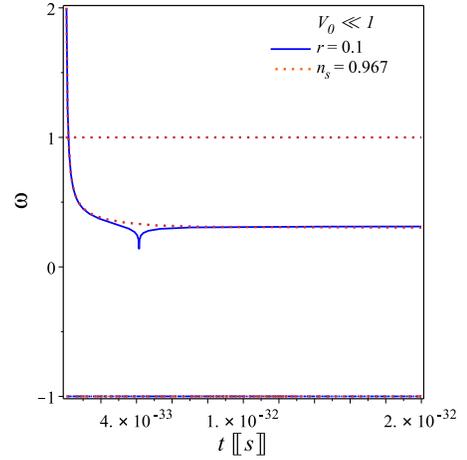}}
\caption[figtopcap]{Matter equation of state that produces the observed power spectrum;
\subref{fig7a} For $V_{0}=5.3\times 10^{78}$,
\subref{fig7b} For $V_{0}=1.21\times 10^{-19}$.
}
\label{Fig7}
\end{figure} However, for $V_{0}\neq 0$ models and the choice $\omega=1$, we find that the scalar power spectrum is the scale invariant Harrison-Zel'dovich spectrum with $r=0$ and $n_{s}=1$, that it is ruled out by the Planck 2015 results. We develop a novel technique to trace the matter equation of state in order to produce nearly scale invariant power spectrum at different time near the bounce to the end of inflation. This can be done by substituting from (\ref{V0}) into (\ref{potential2}), which can be rewritten as
\begin{equation}\label{potential3}
    V(t)=\frac{3}{2}\frac{\rho_{0}\left(t^2 (\omega+\frac{1}{3})e^{3\alpha/t_{s}}+\frac{1}{3}t_{s}^{2}(1-\omega)e^{3\alpha/t}\right)}{a_{k}^{3(1+\omega)(1+\omega)^2 t^2 t_{s}^2}}.
\end{equation}
Then the slow roll parameters (\ref{slow_roll}) read
\begin{eqnarray}
  \epsilon_{V} &=& \frac{(1-\omega)^2 (1+\omega)t_{s}^4 a_{k}^{3(1+\omega)}(2t+3\alpha)^2 e^{3\alpha/t}}{18\kappa^2 \rho_{0}t^{2}\left[t^2 (\omega+\frac{1}{3})e^{\frac{3\alpha}{t_{s}}}+\frac{1}{3}t_{s}^{2}(1-\omega)e^{\frac{3\alpha}{t}}\right]^{2}}, \\
  \eta_{V} &=& \frac{a_{k}^{3(1+\omega)}t_{s}^{2}(1+\omega^2)(9\alpha^2+24\alpha t+8 t^2)}{6\kappa^2 \rho_{0}t^{2}\left[t^2 (\omega+\frac{1}{3})e^{\frac{3\alpha}{t_{s}}}+\frac{1}{3}t_{s}^{2}(1-\omega)e^{\frac{3\alpha}{t}}\right]} .
\end{eqnarray}

Now, we obtain the observable parameters $r$ and $n_{s}$ as functions of $\omega$ and $t$, since all the constants appear in the above equations are known. By requiring the reasonable values of $r$ and $n_{s}$ from observations, one can get an explicit relation between $\omega$ and $t$ that produces the observed power spectrum. We give the results in the plots of Fig. (\ref{Fig7}), which is  represented in the ($\omega-t$) plane, where the small (large) $V_{0}$ is due to the choice $t_{s}=10^{-32}$ s ($t=10^{17}$ s) just as previously identified. The intersections of the curves determine which choice of the matter equation-of-state at different times verify the desired values of $r$ and $n_{s}$. As shown by Fig. \ref{Fig7}\subref{fig7b}, the small $V_{0}$ model is more flexible with observations. In addition, it shows clearly that $\omega>1$ in the contraction phase is the natural choice allowing the contraction phase not only to dominate over the anisotropy evolution but also to produce a scale invariant power spectrum.\\

In conclusion, we find that the slow-roll condition is natural in this model. In addition, by applying the slow-roll approximation, we find that the scalar field induced by $f(T)$ of this bounce model does not suffer from the problem of a large tensor-to-scalar ratio, which usually faces bouncing models. However, it produces a nearly scale invariant spectrum of the scalar field in a good agreement with observations. Moreover, it provides a unified field representing inflaton and quintessence in a single model.
\subsection{Energy conditions}\label{Sec5.4}             
In the previous section, we have shown that the large tensor-to-scalar ratio in bouncing models is avoidable in our model. Here, we investigate another major problem that usually faces bouncing models, that is the violation of the NEC, which gives rise to the ghost instability problem. As we have shown in Sec. \ref{Sec4.3}, the $f(T)$ gravity breaks the NEC effectively at the early time as required to obtain a bounce solution. However, in this section we argue that this feature can help to induce a scalar field model free from ghosts. Next, we briefly present the necessary background of the energy conditions. In the context of a geometric field theory, e.g., the GR theory, it is enough to apply the Bianchi II identity to guarantee the matter conservation. This leaves us with a huge amount of arbitrariness of the choice of the matter source. This implies to impose a particular kind of matter, e.g., dust, radiation, scalar field, electromagnetism, $\ldots$. However, in modified gravity theories, e.g., $f(T)$ gravity, the dark sector of the Universe arises as an effective gravity in the field equations. Energy conditions strategy can be used to limit the arbitrariness of the $\mathfrak{T}_{\mu\nu}$ for a variety of different sources.\\

In order to describe the interaction between any two nearby bits of matter, we should remember the Raychaudhuri equation. This equation represents the fundamental lemma of the Penrose-Hawking singularity theorems. Raychaudhuri equation for a congruence of timelike (or null) geodesics, respectively, in spacetime can be written as
\begin{eqnarray}
&& \frac{d\vartheta}{d\tau}=-\frac{1}{3}\vartheta^2-\sigma_{\mu \nu}\sigma^{\mu \nu}+\omega_{\mu \nu}\omega^{\mu \nu}-R_{\mu \nu} u^\mu u^\nu,\nonumber\\
&& \frac{d\vartheta}{d\tau}=-\frac{1}{3}\vartheta^2-\sigma_{\mu \nu}\sigma^{\mu \nu}+\omega_{\mu \nu}\omega^{\mu \nu}-R_{\mu \nu} k^\mu k^\nu.
\end{eqnarray}
The lhs of Raychaudhuri equation identifies the temporal evolution of the expansion of scalar $\vartheta$, while the rhs contains two classifications: the first promotes a collapsing configuration due to a nonzero initial expansion scalar, shearing $\sigma^{\mu \nu}$, and the second opposes the collapsing configuration due to a nonzero vorticity $\omega^{\mu \nu}$. However, the contribution of the last term $R_{\mu \nu} u^\mu u^\nu$, where $u^{\mu}$ is an arbitrary timelike vector and $k^{\mu}$ is an arbitrary null vector, is restricted by the energy conditions. There are four forms of energy conditions namely: weak energy condition (WEC), NEC, SEC and dominant energy condition (DEC).\\

As a result of the attraction of gravity, the focusing theorem states that $\frac{d\vartheta}{d\tau} < 0$, which implies the positivity of the trace of the tidal tensor, i.e.,
\[  R_{\mu \nu} u^\mu u^\nu \geq 0, \] \[R_{\mu \nu} k^\mu k^\nu\geq 0.\]
This precisely gives the SEC and the NEC, respectively. These in terms of a stress-energy tensor and its trace can be written as
\begin{equation}
R_{\mu \nu}=\mathfrak{T}_{\mu \nu}-\frac{\mathfrak{T}}{2}g_{\mu \nu}.
\end{equation}
As a result, the inequalities of SEC and NEC having the form
\begin{eqnarray}\label{c1}
& & R_{\mu \nu} u^\mu u^\nu=\left(\mathfrak{T}_{\mu \nu}-\frac{\mathfrak{T}}{2}g_{\mu \nu}\right)u^\mu u^\nu \geq 0, \nonumber\\
 & &  R_{\mu \nu} k^\mu k^\nu=\left(\mathfrak{T}_{\mu \nu}-\frac{\mathfrak{T}}{2}g_{\mu \nu}\right)k^\mu k^\nu\geq 0.\end{eqnarray}
In the case of a perfect fluid, these energy conditions SEC and NEC, namely (\ref{c1}), are reduced to $\rho+p\geq 0$ and $\rho+3p\geq 0$, while the WEC and DEC demand the following constrains $\rho\geq 0$ and $\rho\pm p\geq 0$. We summarize the energy conditions of the perfect fluid as
\begin{equation}\label{energy_cond}
\begin{array}{lc}
\text{Name}&\text{For perfect fluid}\\
\text{Weak}\phantom{\Big|}&  \rho\geq 0,\quad \rho+p\geq 0;\\[0.2cm]
\text{Null}& \rho+p\geq 0;\\[0.2cm]
\text{Strong}& \quad\rho+p\geq 0,\quad \rho+3p\geq 0;\\[0.2cm]
\text{Dominant}& \rho\geq |p\,|.
\end{array}
\end{equation}

The Raychaudhuri equation is a pure geometrical relation, so that using different geometries gives rise to different descriptions of the Raychaudhuri equation. Also, the idea of energy conditions can be generalized to the modified theories of gravity. All ordinary matter, even the vacuum expectation value (vev) of a scalar field obey the DEC, while during inflation this condition should be relaxed.\\
\begin{figure}
\centering
\subfigure[~$V_{0} \gg 1$]{\label{Fig8a}\includegraphics[scale=.3]{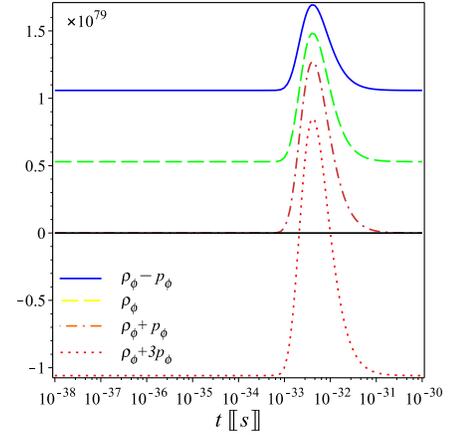}}
\subfigure[~$V_{0}=0$]{\label{Fig8b}\includegraphics[scale=.3]{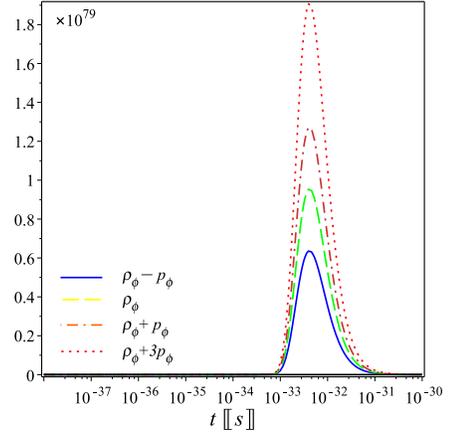}}
\subfigure[~$0 <V_{0} \ll 1$]{\label{Fig8c}\includegraphics[scale=.3]{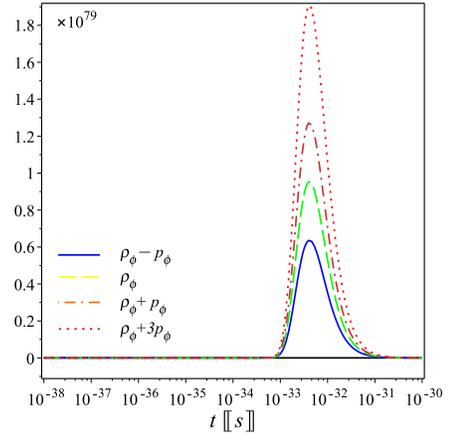}}
\caption[figtopcap]{The above plots represent the four ingredients ($\rho_{\phi}$, $\rho_{\phi}\pm p_{\phi}$, and $\rho_{'phi}+3p_{\phi}$), which are necessary to study the energy conditions (\ref{energy_cond}); The three models are as follows:
\subref{Fig8a} For $V_{0}=5.3\times 10^{78}$, the SEC is generally violated except during $t \in (2.1\times 10^{-33},10^{-32}) s$;
\subref{Fig8b} For $V_{0}=0$, the energy conditions (\ref{energy_cond}) are fulfilled;
\subref{Fig8c} For $V_{0}=1.2\times 10^{-19}$, only the SEC is violated during $t \in \left(\right. 0,~3.4\times 10^{-35}~s\left.\right] \cup (10^{17}~s,~\infty)$. All the above plots are in agreement with Fig. \ref{Fig6}, also a summary of the final results is given in Table \ref{Table1}. The numerical values $10^{79}$ which appear on the vertical axis represent the scale of the axis.}
\label{Fig8}
\end{figure}

For the scalar matter (\ref{rho_phi1}) and (\ref{p_phi1}), it is convenient to visualize the evolution of four ingredients ($\rho_{\phi}$, $\rho_{\phi}\pm p_{\phi}$, and $\rho_{\phi}+3p_{\phi}$) which are necessary to study any of the energy conditions (\ref{energy_cond}). This is shown on the plots of Fig. \ref{Fig8}.

(\textbf{i}) The case of the large value of $V_{0}$ is shown in Fig. \ref{Fig8}\subref{Fig8a}; we find that only the SEC is violated except during the very short time interval $2.1 \times 10^{-33}\lesssim t\lesssim 10^{-32}$ s it is not verified. This result is consistent with the previous result of Sec. \ref{Sec5.3} where the equation-of-state $\omega_{\phi}$ is allowed to exceed the quintessence limit $-1/3$ only during this short period. This has been explained before as a result of the large background potential $V_{0}$ which drags the Universe back to its de Sitter phase very shortly. Also, we can see that other energy conditions are verified for the large $V_{0}$ model.

(\textbf{ii}) The case of a vanishing $V_{0}$ is shown in Fig. \ref{Fig8}\subref{Fig8b}; we find that all the energy conditions are verified. This is also in agreement with the previous results where $\omega_{\phi}>-1/3$ for this model.

(\textbf{iii}) The case of the small value of $V_{0}$ is shown in Fig. \ref{Fig8}\subref{Fig8c}; we find the the SEC is verified during the time interval $3.4 \times 10^{-35} < t < 10^{17}$ s, while it is violated elsewhere. This result can be also verified by Fig. \ref{Fig6}, where $\omega_{\phi}>-1/3$ during this interval. However, at very early and very late times, $-1< \omega_{\phi}<-1/3$, which explains these accelerated expansion phases.

We summarize these results in Table \ref{Table1}. As is well-known, the bounce models require a temporary violation of the NEC about the bounce time. Interestingly, we find that the NEC is fulfilled for the matter field during the bounce phase which makes the model free from a ghost instability. However, the NEC violation is due to the effective torsion field not in the matter field. This nice feature of $f(T)$ gravity could provide a better environment to obtain a healthy bounce model.
\begin{table}
\caption{Verification of the energy conditions}
\begin{tabular*}{\columnwidth}{@{\extracolsep{\fill}}ccccc@{}}
\hline\hline
\multicolumn{2}{c}{Energy condition}& $V_{0}=0$ & $V_{0}=1.2\times 10^{-19}$ & $V_{0}=5.3\times 10^{78}$\\
\hline
\multirow{2}{*}{WEC} & $\rho_{\phi}\geq 0$         & always & always & always\\[3pt]
                     & $\rho_{\phi}+p_{\phi}\geq 0$       & always & always & always\\
  \hline
 NEC                 & $\rho_{\phi}+p_{\phi}\geq 0$       & always & always & always\\
  \hline
\multirow{2}{*}{SEC} & $\rho_{\phi}+p_{\phi}\geq 0$  & always & always & always\\[3pt]
                     & $\rho_{\phi}+3p_{\phi}\geq 0$      & always & $10^{-35}\lesssim t \lesssim 10^{17}$ s& $10^{-33}\lesssim t \lesssim 10^{-32}$ s\\
  \hline
 DEC                 & $\rho_{\phi}\geq |p_{\phi}\,|$         & always & always& always\\
  \hline
\hline
\end{tabular*}\label{Table1}
\end{table}
\section{Primordial fluctuations in $f(T)$ cosmology}\label{Sec6}
In the Sec. \ref{Sec4}, we discussed the cosmological bounce in the $f(T)$ cosmology at the background level. In this section, we extend our analysis to investigate the theory at the perturbation level. In order to identify the true perturbations, which do not change under gauge coordinate transformation, is to fix the gauge freedom. We choose the longitudinal (conformal Newtonian) gauge which fixes the gauge completely and only involves two scalars metric fluctuation as
\begin{equation}\label{pert-metric}
    ds^2=(1+2\Phi)dt^2-a^2(1-2\Psi)dx^2.
\end{equation}
By comparison with the weak field limit of GR about Minkowski space, one can realize that the metric fluctuation $\Phi$ plays the role of the gravitational potential. Assuming that the anisotropic stress vanishes, one can obtain that $\Psi=\Phi$. The authors of \cite{Cai:2011tc} have shown that the gravitational potential $\Phi$ can be completely determined by the scalar field fluctuation $\delta \phi$. Therefore, there exists only a single degree of freedom in the scenario of $f(T)$ gravity minimally coupled to a canonical scalar field. In order to understand the evolution of scalar-sector metric perturbations, we use the perturbed equation of motion for the gravitational potential $\Phi$ instead of a scalar field fluctuation $\delta \phi$. Then the complete form of the equation of motion for one Fourier mode $\Phi_k$ with the comoving wave number $k$ is given by \cite{Cai:2011tc}
\begin{equation}\label{Fourier-mode}
    \ddot{\Phi}_{k}+\tilde{\alpha}\dot{\Phi}_{k}+\left(\mu^2 +c_s^2\frac{k^2}{a^2}\right)\Phi_k=0.
\end{equation}
Here, the functions $\tilde{\alpha}$, $\mu$, and $c_s$ represent, respectively, the frictional term, the effective mass and the sound speed parameter for the gravitational potential $\Phi$. These functions are defined by \cite{Cai:2011tc}
\begin{eqnarray}
\label{alpha}
\tilde{\alpha}  &=& 7H + \frac{{2{V_{\phi }}}}{{\dot \phi }} - \frac{{36H\dot H\left( {{f_{TT}} - 4{H^2}{f_{TTT}}} \right)}}{{{f_{T}} - 12{H^2}{f_{TT}}}},
\\
\label{mu}
{\mu ^2} &=& 6{H^2} + 2\dot H + \frac{{2H{V_{\phi }}}}{{\dot \phi }} - \frac{{36H\dot H\left( {{f_{TT}} - 4{H^2}{f_{TTT}}} \right)}}{{{f_{T}} - 12{H^2}{f_{TT}}}},
\\
\label{cs}
c_s^2 &=& \frac{{{f_{T}}}}{{{f_{T}} - 12{H^2}{f_{TT}}}}.
\end{eqnarray}
Moreover, if we use the Friedmann equation (\ref{kinetic}) and the evolution equation (\ref{Klein-Gordon}) for the scalar field, then we can rewrite (\ref{Fourier-mode}) as
\begin{equation}\label{Fourier-mode1}
{\ddot \Phi _k} + \left( {H - \frac{{\ddot H}}{{\dot H}}} \right){\dot \Phi _k} + \left( {2\dot H - \frac{{H\ddot H}}{{\dot H}}} \right){\Phi _k} + c_s^2\frac{{{k^2}}}{{{a^2}}}{\Phi _k} = 0.
\end{equation}
This is the equation of motion for the gravitational potential $\Phi$ in $f(T)$ gravity in the presence of a canonical scalar field. We see that this equation is identical with the one in the standard Einstein gravity, except the new sound speed parameter $c_s$ has been introduced.
\subsection{Mukhanov-Sasaki equations}\label{Sec6.1} 
We assume that the curvature fluctuation in comoving coordinates, which characterizes the cosmological inhomogeneities, to be a gauge-invariant variable $\zeta$ as the same as in the standard cosmological perturbation theory\footnote{Another important gauge-invariant variable is the inflaton fluctuation $\delta \phi(t; x)$ in the uniform curvature gauge. The gauge-invariant field fluctuation is directly related to the comoving curvature perturbation $\mathcal{R}=-H \frac{\delta \phi}{\dot{\phi}}$. At large scales, $\mathcal{R}\thickapprox \zeta$.}
\begin{equation}\label{zeta}
\zeta  = \Phi  - \frac{H}{{\dot H}}\big( {\dot \Phi  + H\Phi } \big).
\end{equation}
Using the above equation and (\ref{Fourier-mode1}), we get
\begin{equation}\label{zetakdot}
{\dot \zeta _k} = \frac{H}{{\dot H}}\frac{{c_s^2{k^2}}}{{{a^2}}}{\Phi _k}.
\end{equation}
For the case of a generic expanding Universe, ${\dot \zeta _k}$ approaches zero at large length scales, $k \to 0$, because the dominant mode of ${\dot \Phi _k}$ is approximately constant.
\subsubsection{The scalar power spectrum}         
In order to simplify the calculations, we change to the conformal time $\tau  \equiv \int {dt/a} $, and rescale the field (\ref{zeta}) by introducing a canonically normalized field variable
\begin{equation}\label{v}
v = z_s \zeta,
\end{equation}
where
\begin{equation}\label{zs}
z_s = \frac{a}{\kappa}\sqrt {2\epsilon_{1}},
\end{equation}
and $\epsilon_{1} \equiv -\frac{d \ln H/dt}{H}$ is the first slow roll parameter, where $0\leq \epsilon_{1}<1$ during inflation. Using (\ref{zeta}) - (\ref{zs}), we rewrite the equation of motion (\ref{Fourier-mode1}) for the scalar perturbations as
\begin{equation}\label{d2vk}
v_k'' + \left( {c_s^2{k^2} - \frac{{z_s''}}{z_s}} \right){v_k} = 0,
\end{equation}
where the prime denotes the derivative with respect to the conformal time. It is clear that the \textit{Mukhanov-Sasaki equation} of Einstein gravity is recovered, if the sound speed is equal to the light speed, i.e., $c_s=1$ \cite{Mukhanov:1992}. The above equation can be solved in two limits
\begin{itemize}
  \item [(i)] At small scales when $c_{s}k \gg a H$ (sub-Hubble): Eq. (\ref{d2vk}) becomes $v_k'' + c_s^2{k^2}{v_k} \approx 0$. We assume the quantum fluctuations around the initial vacuum state to be much earlier than the bounce time. The initial conditions of the vacuum state is the Bunch-Davies vacuum \cite{Bunch:1978yq}, then the solution of (\ref{d2vk}) will be
      \begin{equation}\label{vk1}
{v_k} \simeq \frac{{{{\mathop{ e}\nolimits} ^{ - i c_s k\tau }}}}{{\sqrt {2c_s k} }}, \quad (\textmd{free oscilations}).
\end{equation}
We mention that the imaginary phase disappears when the norm of the comoving curvature $|\zeta|^{2}$ is used.
  \item [(ii)] At large scales when $c_{s}k \ll a H$ (super-Hubble): When the fluctuations exit the sound horizon, we have $z_s''/{z_s} \gg c_s^2 k^2$ in (\ref{d2vk}). Consequently, its solution gives rise to
\begin{equation}\label{vk2}
{v_k} = {C_k}z_{s},\quad (\textmd{frozen fluctuations})
\end{equation}
where $C_k$ is a constant, which can be determined by matching the solutions (\ref{vk1}) and (\ref{vk2}) at the sound horizon exit (i.e., $c_s k = aH$). Consequently, we have
\begin{equation}\label{Ck}
{\left| {{C_k}} \right|^2} = \frac{1}{{2{c_s}kz_s^2}}.
\end{equation}
\end{itemize}
Using (\ref{v}) and (\ref{zs}), we obtain the power spectrum for the scalar perturbations in the framework of $f(T)$ gravity as \cite{Karami}
\begin{equation}\label{Ps}
{{\cal P}_s} \equiv \frac{{{k^3}}}{{2{\pi ^2}}}{\left| \zeta  \right|^2} = {\left. {\frac{{{k^3}}}{{2{\pi ^2}}}\frac{{{{\left| {{v_k}} \right|}^2}}}{{z_s^2}}} \right|_{ {c_s}k=aH}} = {\left. {\frac{{{H^2}}}{{8{\pi ^2}M_P^2c_s^3{\epsilon_{1}}}}} \right|_{ {c_s}k=aH}},
\end{equation}
which should be evaluated at the sound horizon exit specified by $c_s k=a H$. This equation reduces to the standard result for slow-roll inflation if the sound speed is equal to the light speed ($c_s=1$).

The scalar spectral index is defined as
\begin{equation}\label{ns}
{n_s} - 1 \equiv \frac{{d\ln {{\cal P}_s}}}{{d\ln k}}.
\end{equation}
Since during slow-roll inflation, the Hubble parameter $H$ and the sound speed $c_s$ are almost constant, therefore using the relation $c_s k=a H$ that is valid at the sound horizon exit, we can obtain the relation
\begin{equation}\label{dlnk}
d\ln k \approx Hdt.
\end{equation}
\subsubsection{The tensor power spectrum}                    
Now we turn to study the tensor perturbations in $f(T)$ gravity. According to \cite{Chen:2010va,Cai:2011tc}, the equation governing the tensor perturbation $h_{ij}$ can be obtained as
\begin{equation}\label{d2hij}
{\ddot h_{ij}} + 3H{\dot h_{ij}} - \frac{{{\nabla^2}}}{{{a^2}}}{h_{ij}} + \gamma {\dot h_{ij}} = 0,
\end{equation}
where the definition of the parameter $\gamma$ is given by \cite{Karami}
\begin{equation}\label{gamma}
\gamma  \equiv \frac{{\dot T{f_{,TT}}}}{{{f_{,T}}}}.
\end{equation}
When the tensor perturbation $h_{ij}$ is constrained to be symmetric (${h_{ij}} = {h_{ji}}$), transverse ($\partial^ih_{ij} = 0$), and traceless ($h_{ii} = 0$), then it will have only two degrees of freedom corresponding to two polarization modes of the gravitational waves. The Fourier transformations of the tensor perturbation is given by
\begin{equation}\label{hijtx}
{h_{ij}}(t,x) = \sum\limits_{r = 1}^2 {\int {\frac{{{d^3}k}}{{{{(2\pi )}^{3/2}}}}} } {h^r}(t,k)\,\xi _{ij}^r\,{{\mathop{ e}\nolimits} ^{ikx}},
\end{equation}
where the index $r$ is to identify the polarization state. Therefore, each state of ${h_{ij}}(t,x)$ can be written as a scalar field $h^r(t, x)$ multiplied by a polarization tensor $\xi _{ij}^r$, which is constant in space and time. Using the above in (\ref{d2hij}), we get
\begin{equation}\label{d2hr}
{\ddot h^r} + \left( {3H + \gamma } \right){\dot h^r} + \frac{{{k^2}}}{{{a^2}}}{h^r} = 0.
\end{equation}
Change to the conformal time and rescale the field by introducing the canonically normalized field variable
\begin{equation}\label{vrk}
v_k^r = \frac{{{z_t}}}{2}h^r{M_P},
\end{equation}
where
\begin{equation}\label{zt}
{z_t} = a\exp \left( { \int {\frac{\gamma }{2}\,dt} } \right).
\end{equation}
Then, the equation of motion of the tensor fluctuations (\ref{d2hr}) gives rise to
\begin{equation}\label{d2vrk}
{v_k^r}'' + \left( {{k^2} - \frac{{z_t ''}}{{{z_t}}}} \right)v_k^r = 0.
\end{equation}
Following the same procedure used for the scalar perturbations, we can find the two solutions of (\ref{d2vrk}) at sub-Hubble and super-Hubble scales. Therefore, the tensor power spectrum can be obtained by matching the solutions at the horizon exit $k=a H$,
\begin{equation}\label{Pt0}
{{\cal P}_t} = 2{{\cal P}_h} = {\left. {\frac{{2{a^2}{H^2}}}{{{\pi ^2}M_P^2z_t^2}}} \right|_{k = aH}},
\end{equation}
which is sum of the power spectra ${{\cal P}_h}$ for two polarization modes of $h_{ij}$.
\subsubsection{Tensor-to-scalar ratio}        
One of the most important inflationary observable is the tensor-to-scalar ratio. It is completely constrained by the Planck satellite and BICEP2 and Keck Array~\cite{Planck:2015xua, Ade:2015lrj, Ade:2015tva, Array:2015xqh}. Therefore, it can be used to exclude unviable inflationary models. This observable is defined by
\begin{equation}\label{r}
r \equiv \frac{{{{\cal P}_t}}}{{{{\cal P}_s}}}.
\end{equation}
Another inflationary observable, but not accurately tied yet, is the tensor spectral index defined as
\begin{equation}\label{nt}
{n_t} \equiv \frac{{d\ln {{\cal P}_t}}}{{d\ln k}}.
\end{equation}
In the inflationary scenario, the Hubble parameter $H$ is nearly constant, i.e., $\dot{T}\simeq 0$, then from (\ref{gamma}), we have $\gamma \simeq 0$ as well. Consequently, Eq. (\ref{zt}) reduces to the standard solution $z_t=a$, and (\ref{Pt0}) reduces to the standard expression,
\begin{equation}\label{Pt}
{{\cal P}_t} = {\left. {\frac{{2{H^2}}}{{{\pi ^2}M_P^2}}} \right|_{k = aH}},
\end{equation}
which matches the tensor power spectrum in Einstein gravity. This relation must be calculated at the time of horizon crossing for which $k=a H$. This time is not exactly the same as the time of the sound horizon crossing for which $c_s k=a H$, but to the lowest order in the slow-roll parameters this difference is negligible. Therefore, using Eqs. (\ref{Ps}), (\ref{r}) and (\ref{Pt}), the tensor-to-scalar ratio is obtained as
\begin{equation}\label{r2}
r = 16c_s^3{\epsilon_{1}}.
\end{equation}
Using (\ref{dlnk}), (\ref{nt}), and (\ref{Pt}), we can obtain the tensor spectral index as
\begin{equation}\label{nt2}
{n_t}=-2\epsilon_{1}.
\end{equation}
From (\ref{r2}) and (\ref{nt2}), we write the \textit{consistency relation} as
\begin{equation}\label{rnt}
r =  - 8c_s^3{n_t}.
\end{equation}
We conclude that the inflation in $f(T)$ gravity differs from Einstein gravity by introducing the sound speed. On the other hand, the standard inflationary of Einstein's gravity is recovered when $c_s=1$, which is valid during the contraction phase of the bounce Universe models.
\subsection{Conservation of the comoving curvature perturbations in the precontraction phase}\label{Sec6.2} 
It is convenient to present briefly some key features of the inflationary theory, then we discuss the alternative scenario of the bounce model at hand. In inflationary models, the Universe begins with an initial singularity just as in the standard cosmology, at the initial singularity, the hubble radius\footnote{The hubble radius is usually referred to as the horizon.} $r_{H}=\frac{1}{a(t) H(t)}$ is infinite as $a(t_{i})=0$. Consequently, we expect that all primordial quantum fluctuations to be subhorizon where their wavelength $\lambda\ll r_{H}$ or equivalently, the comoving wave number is at the subhorizon scales $k \gg a(t)H(t)=1/r_{H}$. Since the scale factor grows exponentially and the Hubble parameter is almost constant during inflation, the horizon shrinks exponentially allowing some modes to exit the horizon and become classical (freeze-out). By the end of inflation, the horizon expands allowing the freezed modes to reenter the horizon at a superhorizon scale and propagating as particles. The conservation of the comoving primordial fluctuations is an essential feature in inflationary models, as a matter of fact, it enables us to relate the observable quantities at the superhorizon (law energy) scale to the subhorizon (high energy) scale.\\

\begin{figure}
\centering
\includegraphics[scale=.4]{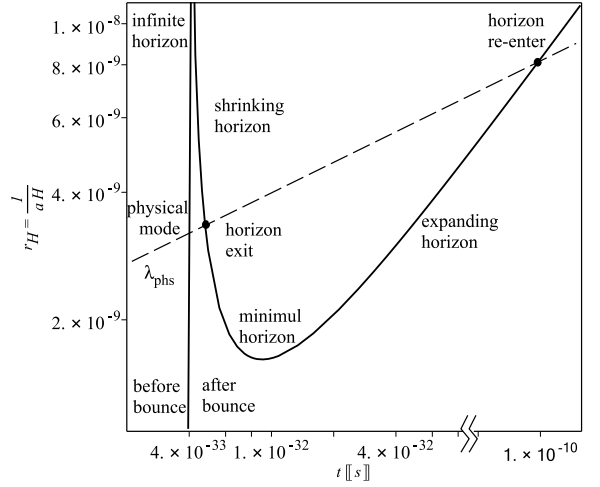}
\caption{Schematic diagram to show the evolution of the hubble radius (\ref{hubble-radius}). Immediately after bounce, it shrinks allowing the relevant physical modes to exit the horizon, then it expands allowing these physical modes to reenter the horizon at later time.}
\label{Fig9}
\end{figure}

In the present bounce model, we discuss the alternative scenario. Using (\ref{scale-factor}) and (\ref{Tor_sc1}), the hubble radius is given by
\begin{equation}\label{hubble-radius}
    r_{H}=\frac{(1+\omega)^{\frac{1+3\omega}{3(1+\omega)}}t^{\frac{2(2+3\omega)}{3(1+3\omega)}}e^{\frac{\alpha}{(1+\omega)t}}}{a_{k}(2t+3\alpha)}.
\end{equation}
At the bounce point, we expect the hubble radius (horizon) to be infinite as $H(t_{B})=0$ (i.e., $t=-\frac{3}{2}\alpha$). This means that all modes are sub-Hubble as $\lambda\ll r_{H}$. Immediately after bounce, the horizon suddenly shrinks to a minimal value, see Fig. \ref{Fig9}, during this stage when the initial quantum fluctuations cross the horizon, they transform into classical fluctuations. So it is worth to investigate if the comoving curvature fluctuations will be conserved at the super-Hubble or not. This feature is important to examine the validity of the bounce scenario.
The Universe around the bounce point is governed by quantum gravity which is beyond our reach so far. On the other hand, the perturbation during the contraction period suffers from producing a high tensor-to-scalar ratio, which is disfavored by observations \cite{Cai:2011tc}. As we have shown in Section \ref{Sec5.3} that the slow roll conditions are valid during the precontraction period, so we expect that the primordial fluctuations that are produced at this phase to play an essential role in the present model. It is obvious that the precontraction phase can be identified by $t\ll |\alpha|$ so that the correction term in (\ref{scale-factor}) is much more dominant than the FRW evolution. So the scale factor and the corresponding Hubble at this phase are approximately
\begin{equation}\label{approx-scale-fac}
    a\sim a_{k} e^{-\frac{\alpha}{(1+\omega)t}}; \quad     H(t)\sim \frac{\alpha}{(1+\omega)t^2}.
\end{equation}
We also reconstruct the $f(T)$ which generates the scale factor (\ref{approx-scale-fac}) as a function of the cosmic time $t$ as
\begin{equation}\label{approx-f(t)}
    f(t)=\frac{c_{2} 9\alpha^2 \textmd{Ei}(1,-3\alpha/t)+(t+3\alpha)t e^{3\alpha/t}}{t^2}.
\end{equation}
Also, it can be rewritten in terms of $T$ as
\begin{equation}\label{approx-f(T)}
    f_{+}(T)=c_{2}\textmd{Ei}\left(3,\frac{6^{3/4}}{2}\sqrt{\alpha(1+\omega)\sqrt{-T}}\right).
\end{equation}
We are interested in the precontraction phase before bounce, so we take only the $f(T)$ form corresponding to the $\dot H>0$ branch. Therefore, the speed of sound (\ref{cs}) reads
\begin{equation}\label{sound-speed}
    c_{s}^{2}=-2-\frac{6\alpha}{t}\textmd{Ei}(1,-3\alpha/t)e^{-3\alpha/t}.
\end{equation}
At the precontraction phase $t \ll t_B$, we have $c_{s}^{2}\simeq 0$. Consequently the evolution equation (\ref{Fourier-mode1}) takes the form
\begin{equation}\label{Fourier-mode2}
    \ddot{\Phi}+\left(\frac{3}{t}+\frac{\alpha}{(1+\omega)t^2}\right)\dot{\Phi}-\frac{\alpha}{(1+\omega)t^3}\Phi=0,
\end{equation}
which can be solved to
\begin{eqnarray}\label{scalar-prop}
\nonumber    \Phi(t)&=&(1+\omega)C_{2}\left[(1+\omega) \frac{\alpha}{t}\right]\\
    &+&\frac{e^{\frac{\alpha}{(1+\omega)t}}}{t^2}\left[C_{1}+C_{2}\alpha^2 \textmd{Ei}\left(1,\frac{-\alpha}{(1+\omega)t}\right)\right].
\end{eqnarray}
The above relation describes the perturbations that are produced at the precontraction phase. These can cross the bounce time as the horizon is infinite at that time, then they exit the horizon and evolve as classical perturbations. Later on, they reenter the horizon to produce an almost scale invariant, Harrison-Zel'dovich spectrum; this may explain its origin. So we study the evolution of the comoving curvature perturbations between horizon exiting to reentering. Since the perturbations could grow making an instability or even vanish which cannot explain the presence of our Universe in its structure, we find that this point should be investigated for the present model. In (\ref{scalar-prop}), the exponential function in the last term evolves to a constant value as time increases, while $1/t^{2}$ contribution makes the last term approaching zero. Consequently, Eq. (\ref{scalar-prop}) reads $\Phi(t) \simeq (1+\omega)C_{2}\left[(1+\omega) \frac{\alpha}{t}\right]$.  Also, we realize that all the fluctuations due to the term $\alpha/t$ are decaying as time increases. In this case we have
\begin{equation}\label{conservative-fluctuations}
\Phi(t)\simeq (1+\omega)^{2} C_{2}=\textmd{constant}.
\end{equation}
We conclude that in this bounce scenario, the comoving curvature fluctuations at the precontraction phase are conserved after the modes exit the horizon. This is an important feature to ensure that the quantum fluctuations transform into classical fluctuations holding all of the information, which are characterizing the Universe at the sub-Hubble energy scale. Finally, we note that the slow roll conditions are valid at the precontraction phase so the calculations of Sec. \ref{Sec5.3} are valid in this early stage. Therefore, the power spectrum at this stage matches the observations.
\section{Summary}\label{Sec7}
In this work, we have shown how the GR applications in cosmology are very limited. However, in $f(T)$ theories, we have more flexibility to start with a designed scale factor holding the well behaved linear equation of state of the matter content $p=\omega \rho$ without breaking the conservation principle. We have modified the usual FRW scale factor as in (\ref{scale-factor}), by introducing a new parameter $\alpha$ with units of time. The scale factor shows a good agreement with inflation-to-deceleration transition. The new parameter has been shown to control the time of transition, so the $\alpha$ parameter can have values $\alpha=1.61\times 10^{-32}$ s or $-2.76\times 10^{-33}$ s. The positive value gives a graceful exit inflation model, while the negative value gives a bounce model.\\

We have studied the negative parameter model, i.e., $\alpha=-2.76\times 10^{-33}$ s, extensively. The model shows interesting features, it performs a bouncing behavior from contraction to expansion at bouncing time $t_{B}=-\frac{3\alpha}{2}\sim 4.14\times 10^{-33}$ s with a minimal scale factor $a_{B}\neq 0$ so that the trans-Plankian problems of inflation models can be smoothed away. On the other hand, it performs an early inflation phase with a graceful exit into FRW decelerated. We have also used the useful ($\dot{H}-H$) phase space analysis to examine the capability of the model to practice the above mentioned behavior. Unlike the usual behavior in standard cosmology the model can cross the phantom divide line (de Sitter universe) safely in a finite time; then it interpolates smoothly between de Sitter and Minkowski spaces in an infinite time, i.e., the Universe is free from a future singularity.\\

We have constructed an $f(T)$ theory corresponding to the modified scale factor. Consequently, we determined the density and pressure of the matter as a function of time. In addition, we have evaluated the temperature as a function of time; the model predicts that the temperature evolves as $\Theta \propto a^{-3\omega}$. So we expect a very low initial temperature $\Theta \sim 0~K$ as expected by the bouncing behavior where $a\rightarrow \infty$ at $t\ll t_{B}$. So we have argued that this result, however, seems unfamiliar, it is indeed providing a natural environment for the slow-roll inflation condition (i.e., $V(\phi) \gg \dot{\phi}^{2}$). On the other hand, the temperature evolves to its maximum at the GUT energy scale $\Theta \sim 10^{27}~K$ by the end of inflation providing a graceful exit at $t_{\textmd{end}}\sim 10^{-32}$ s into a FRW decelerated phase as required to initiate the standard hot big bang thermal scenario.\\

We also have reexpressed the Friedmann equations in the Einstein's frame to identify the torsion gravity as a degree of freedom. The torsion equation of state suggests that the gravitational sector may provide a good candidate to describe the bounce behavior at an early time. Whereas the torsion equation of state begins with $\omega_{T}=-1$, then it goes in the phantom regime as $\omega \ll -1$. After that it evolves to $\omega_{T}\gg 1$ before the bounce time, this is required in contraction for solving the anisotropy problem. By crossing the bounce point, $\omega_{T}$ is in phantom energy phase again, while it goes back to cross $\omega_{T}=-1$ smoothly to connect the observed expanding Universe. Then it crosses $\omega_{T}=-1/3$ ending the early accelerated expansion, at $t\sim t_{\textmd{end}}\approx10^{-32}$ s, to enter a new phase of a decelerated expansion. Finally, it approaches the radiation limit $\omega_{T}=1/3$ as $t\gg \alpha$ as required to match the hot big bang consistently.\\

We have considered the case when the matter component is a canonical scalar field $\phi$. For a particular choice of the background potential $V_{0}$, the small $V_{0}$ regime, the equation-of-state $\omega_{\phi}$ of the induced scalar field begins with a pure vacuum energy $\omega_{\phi}=-1$ of a de Sitter universe, then crosses $\omega_{\phi}=-1/3$ at $t\sim 10^{-35}$ s to match the radiation limit. However, it crosses $\omega=-1/3$ once more at a late time $t\sim 10^{17}$ s entering a late accelerated expansion with a de Sitter fate just as $\Lambda$CDM cosmology. In this sense, the model provides a unified field representing inflaton and quintessence in a single model. In the slow-roll regime, we find that the scalar field induced by the $f(T)$ of this bounce model does not suffer from the problem of a large tensor-to-scalar ratio which usually faces bouncing models. In addition, we have shown the NEC is not violated so that it is free from ghost instabilities. We have developed a technique to trace the ordinary matter components that are consistent with the observed scale invariant power spectrum.\\

Finally, we have extended the investigation of the model to the perturbation level to study scalar and tensor primordial fluctuations at the early Universe when the Newtonian gauge is assumed. We have shown that the slow roll conditions are fulfilled at the precontraction phase. In addition, the comoving curvature fluctuations are conserved at the super-Hubble energy scale which allows the fluctuations to match the observable Universe.
\subsection*{Acknowledgments}
This work is partially supported by the Egyptian Ministry of Scientific Research under Project No. 24-2-12 (G.G.L.N., W.E. and Sh.K.I.), and the JSPS KAKENHI Grant Number JP 25800136 and the research-funds provided by Fukushima University (K.B.)

\begin{thebibliography}{113}%
\makeatletter
\providecommand \@ifxundefined [1]{%
 \@ifx{#1\undefined}
}%
\providecommand \@ifnum [1]{%
 \ifnum #1\expandafter \@firstoftwo
 \else \expandafter \@secondoftwo
 \fi
}%
\providecommand \@ifx [1]{%
 \ifx #1\expandafter \@firstoftwo
 \else \expandafter \@secondoftwo
 \fi
}%
\providecommand \natexlab [1]{#1}%
\providecommand \enquote  [1]{``#1''}%
\providecommand \bibnamefont  [1]{#1}%
\providecommand \bibfnamefont [1]{#1}%
\providecommand \citenamefont [1]{#1}%
\providecommand \href@noop [0]{\@secondoftwo}%
\providecommand \href [0]{\begingroup \@sanitize@url \@href}%
\providecommand \@href[1]{\@@startlink{#1}\@@href}%
\providecommand \@@href[1]{\endgroup#1\@@endlink}%
\providecommand \@sanitize@url [0]{\catcode `\\12\catcode `\$12\catcode
  `\&12\catcode `\#12\catcode `\^12\catcode `\_12\catcode `\%12\relax}%
\providecommand \@@startlink[1]{}%
\providecommand \@@endlink[0]{}%
\providecommand \url  [0]{\begingroup\@sanitize@url \@url }%
\providecommand \@url [1]{\endgroup\@href {#1}{\urlprefix }}%
\providecommand \urlprefix  [0]{URL }%
\providecommand \Eprint [0]{\href }%
\providecommand \doibase [0]{http://dx.doi.org/}%
\providecommand \selectlanguage [0]{\@gobble}%
\providecommand \bibinfo  [0]{\@secondoftwo}%
\providecommand \bibfield  [0]{\@secondoftwo}%
\providecommand \translation [1]{[#1]}%
\providecommand \BibitemOpen [0]{}%
\providecommand \bibitemStop [0]{}%
\providecommand \bibitemNoStop [0]{.\EOS\space}%
\providecommand \EOS [0]{\spacefactor3000\relax}%
\providecommand \BibitemShut  [1]{\csname bibitem#1\endcsname}%
\let\auto@bib@innerbib\@empty
\bibitem [{\citenamefont {{Starobinsky}}(1980)}]{Starobinsky:1980te}%
  \BibitemOpen
  \bibfield  {author} {\bibinfo {author} {\bibfnamefont {A.~A.}\ \bibnamefont
  {{Starobinsky}}},\ }\bibfield  {title} {\enquote {\bibinfo {title} {{A new
  type of isotropic cosmological models without singularity}},}\ }\href
  {\doibase 10.1016/0370-2693(80)90670-X} {\bibfield  {journal} {\bibinfo
  {journal} {Phys. Lett.}\ }\textbf {\bibinfo {volume} {91B}},\ \bibinfo
  {pages} {99--102} (\bibinfo {year} {1980})}\BibitemShut {NoStop}%
\bibitem [{\citenamefont {{Sato}}(1981)}]{Sato:1980yn}%
  \BibitemOpen
  \bibfield  {author} {\bibinfo {author} {\bibfnamefont {K.}~\bibnamefont
  {{Sato}}},\ }\bibfield  {title} {\enquote {\bibinfo {title} {{First-order
  phase transition of a vacuum and the expansion of the Universe}},}\ }\href
  {\doibase 10.1093/mnras/195.3.467} {\bibfield  {journal} {\bibinfo  {journal}
  {Mon. Not. R. Astron. Soc.}\ }\textbf {\bibinfo {volume} {195}},\ \bibinfo
  {pages} {467--479} (\bibinfo {year} {1981})}\BibitemShut {NoStop}%
\bibitem [{\citenamefont {{Guth}}(1981)}]{Guth:1980zm}%
  \BibitemOpen
  \bibfield  {author} {\bibinfo {author} {\bibfnamefont {A.~H.}\ \bibnamefont
  {{Guth}}},\ }\bibfield  {title} {\enquote {\bibinfo {title} {{Inflationary
  universe: A possible solution to the horizon and flatness problems}},}\
  }\href {\doibase 10.1103/PhysRevD.23.347} {\bibfield  {journal} {\bibinfo
  {journal} {Phys. Rev. D}\ }\textbf {\bibinfo {volume} {23}},\ \bibinfo
  {pages} {347--356} (\bibinfo {year} {1981})}\BibitemShut {NoStop}%
\bibitem [{\citenamefont {{Linde}}(1982)}]{Linde:1981mu}%
  \BibitemOpen
  \bibfield  {author} {\bibinfo {author} {\bibfnamefont {A.~D.}\ \bibnamefont
  {{Linde}}},\ }\bibfield  {title} {\enquote {\bibinfo {title} {{A new
  inflationary universe scenario: A possible solution of the horizon, flatness,
  homogeneity, isotropy and primordial monopole problems}},}\ }\href {\doibase
  10.1016/0370-2693(82)91219-9} {\bibfield  {journal} {\bibinfo  {journal}
  {Phys. Lett.}\ }\textbf {\bibinfo {volume} {108B}},\ \bibinfo {pages}
  {389--393} (\bibinfo {year} {1982})}\BibitemShut {NoStop}%
\bibitem [{\citenamefont {{Albrecht}}\ and\ \citenamefont
  {{Steinhardt}}(1982)}]{Albrecht:1982wi}%
  \BibitemOpen
  \bibfield  {author} {\bibinfo {author} {\bibfnamefont {A.}~\bibnamefont
  {{Albrecht}}}\ and\ \bibinfo {author} {\bibfnamefont {P.~J.}\ \bibnamefont
  {{Steinhardt}}},\ }\bibfield  {title} {\enquote {\bibinfo {title} {{Cosmology
  for grand unified theories with radiatively induced symmetry breaking}},}\
  }\href {\doibase 10.1103/PhysRevLett.48.1220} {\bibfield  {journal} {\bibinfo
   {journal} {Phys. Rev. Lett.}\ }\textbf {\bibinfo {volume} {48}},\ \bibinfo
  {pages} {1220--1223} (\bibinfo {year} {1982})}\BibitemShut {NoStop}%
\bibitem [{\citenamefont {{Martin}}\ and\ \citenamefont
  {{Brandenberger}}(2001)}]{transPl1}%
  \BibitemOpen
  \bibfield  {author} {\bibinfo {author} {\bibfnamefont {J.}~\bibnamefont
  {{Martin}}}\ and\ \bibinfo {author} {\bibfnamefont {R.~H.}\ \bibnamefont
  {{Brandenberger}}},\ }\bibfield  {title} {\enquote {\bibinfo {title} {{The
  Trans-Planckian problem of inflationary cosmology}},}\ }\href {\doibase
  10.1103/PhysRevD.63.123501} {\bibfield  {journal} {\bibinfo  {journal} {Phys.
  Rev. D}\ }\textbf {\bibinfo {volume} {63}},\ \bibinfo {eid} {123501}
  (\bibinfo {year} {2001})},\ \Eprint {http://arxiv.org/abs/hep-th/0005209}
  {arXiv:hep-th/0005209 [hep-th]} \BibitemShut {NoStop}%
\bibitem [{\citenamefont {{Brandenberger}}\ and\ \citenamefont
  {{Martin}}(2013)}]{transPl2}%
  \BibitemOpen
  \bibfield  {author} {\bibinfo {author} {\bibfnamefont {R.~H.}\ \bibnamefont
  {{Brandenberger}}}\ and\ \bibinfo {author} {\bibfnamefont {J.}~\bibnamefont
  {{Martin}}},\ }\bibfield  {title} {\enquote {\bibinfo {title}
  {{Trans-Planckian issues for inflationary cosmology}},}\ }\href {\doibase
  10.1088/0264-9381/30/11/113001} {\bibfield  {journal} {\bibinfo  {journal}
  {Classical and Quantum Gravity}\ }\textbf {\bibinfo {volume} {30}},\ \bibinfo
  {eid} {113001} (\bibinfo {year} {2013})},\ \Eprint
  {http://arxiv.org/abs/1211.6753} {arXiv:1211.6753 [astro-ph]} \BibitemShut
  {NoStop}%
\bibitem [{\citenamefont {{Oikonomou}}(2015)}]{Oikonomou:2015PRD}%
  \BibitemOpen
  \bibfield  {author} {\bibinfo {author} {\bibfnamefont {V.~K.}\ \bibnamefont
  {{Oikonomou}}},\ }\bibfield  {title} {\enquote {\bibinfo {title} {{Singular
  bouncing cosmology from Gauss-Bonnet modified gravity}},}\ }\href {\doibase
  10.1103/PhysRevD.92.124027} {\bibfield  {journal} {\bibinfo  {journal} {Phys.
  Rev. D}\ }\textbf {\bibinfo {volume} {92}},\ \bibinfo {eid} {124027}
  (\bibinfo {year} {2015})},\ \Eprint {http://arxiv.org/abs/1509.05827}
  {arXiv:1509.05827 [gr-qc]} \BibitemShut {NoStop}%
\bibitem [{\citenamefont {Cai}\ \emph {et~al.}(2016{\natexlab{a}})\citenamefont
  {Cai}, \citenamefont {Duplessis}, \citenamefont {Easson},\ and\ \citenamefont
  {Wang}}]{bounce4}%
  \BibitemOpen
  \bibfield  {author} {\bibinfo {author} {\bibfnamefont {Yi-Fu}\ \bibnamefont
  {Cai}}, \bibinfo {author} {\bibfnamefont {Francis}\ \bibnamefont
  {Duplessis}}, \bibinfo {author} {\bibfnamefont {Damien~A.}\ \bibnamefont
  {Easson}}, \ and\ \bibinfo {author} {\bibfnamefont {Dong-Gang}\ \bibnamefont
  {Wang}},\ }\bibfield  {title} {\enquote {\bibinfo {title} {Searching for a
  matter bounce cosmology with low redshift observations},}\ }\href {\doibase
  10.1103/PhysRevD.93.043546} {\bibfield  {journal} {\bibinfo  {journal} {Phys.
  Rev. D}\ }\textbf {\bibinfo {volume} {93}},\ \bibinfo {pages} {043546}
  (\bibinfo {year} {2016}{\natexlab{a}})},\ \Eprint
  {http://arxiv.org/abs/1512.08979} {arXiv:1512.08979 [astro-ph]} \BibitemShut
  {NoStop}%
\bibitem [{\citenamefont {{Gasperini}}\ and\ \citenamefont
  {{Veneziano}}(1993)}]{preBB}%
  \BibitemOpen
  \bibfield  {author} {\bibinfo {author} {\bibfnamefont {M.}~\bibnamefont
  {{Gasperini}}}\ and\ \bibinfo {author} {\bibfnamefont {G.}~\bibnamefont
  {{Veneziano}}},\ }\bibfield  {title} {\enquote {\bibinfo {title}
  {{Pre-big-bang in string cosmology}},}\ }\href {\doibase
  10.1016/0927-6505(93)90017-8} {\bibfield  {journal} {\bibinfo  {journal}
  {Astropart. Phys.}\ }\textbf {\bibinfo {volume} {1}},\ \bibinfo {pages}
  {317--339} (\bibinfo {year} {1993})},\ \Eprint {http://arxiv.org/abs/9211021}
  {arXiv:9211021 [hep-th]} \BibitemShut {NoStop}%
\bibitem [{\citenamefont {{Khoury}}\ \emph {et~al.}(2001)\citenamefont
  {{Khoury}}, \citenamefont {{Ovrut}}, \citenamefont {{Steinhardt}},\ and\
  \citenamefont {{Turok}}}]{Ekpy}%
  \BibitemOpen
  \bibfield  {author} {\bibinfo {author} {\bibfnamefont {J.}~\bibnamefont
  {{Khoury}}}, \bibinfo {author} {\bibfnamefont {B.~A.}\ \bibnamefont
  {{Ovrut}}}, \bibinfo {author} {\bibfnamefont {P.~J.}\ \bibnamefont
  {{Steinhardt}}}, \ and\ \bibinfo {author} {\bibfnamefont {N.}~\bibnamefont
  {{Turok}}},\ }\bibfield  {title} {\enquote {\bibinfo {title} {{Ekpyrotic
  universe: Colliding branes and the origin of the hot big bang}},}\ }\href
  {\doibase 10.1103/PhysRevD.64.123522} {\bibfield  {journal} {\bibinfo
  {journal} {Phys. Rev. D}\ }\textbf {\bibinfo {volume} {64}},\ \bibinfo
  {pages} {123522} (\bibinfo {year} {2001})},\ \Eprint
  {http://arxiv.org/abs/hep-th/0103239} {arXiv:hep-th/0103239 [hep-th]}
  \BibitemShut {NoStop}%
\bibitem [{\citenamefont {{Wan}}\ \emph {et~al.}(2015)\citenamefont {{Wan}},
  \citenamefont {{Qiu}}, \citenamefont {{Huang}}, \citenamefont {{Cai}},
  \citenamefont {{Li}},\ and\ \citenamefont {{Zhang}}}]{bounce2}%
  \BibitemOpen
  \bibfield  {author} {\bibinfo {author} {\bibfnamefont {Y.}~\bibnamefont
  {{Wan}}}, \bibinfo {author} {\bibfnamefont {T.}~\bibnamefont {{Qiu}}},
  \bibinfo {author} {\bibfnamefont {F.~P.}\ \bibnamefont {{Huang}}}, \bibinfo
  {author} {\bibfnamefont {Y.-F.}\ \bibnamefont {{Cai}}}, \bibinfo {author}
  {\bibfnamefont {H.}~\bibnamefont {{Li}}}, \ and\ \bibinfo {author}
  {\bibfnamefont {X.}~\bibnamefont {{Zhang}}},\ }\bibfield  {title} {\enquote
  {\bibinfo {title} {{Bounce inflation cosmology with Standard Model Higgs
  boson}},}\ }\href {\doibase 10.1088/1475-7516/2015/12/019} {\bibfield
  {journal} {\bibinfo  {journal} {J. Cosmol. Astropart. Phys.}\ }\textbf
  {\bibinfo {volume} {12}},\ \bibinfo {eid} {019} (\bibinfo {year} {2015})},\
  \Eprint {http://arxiv.org/abs/1509.08772} {arXiv:1509.08772 [gr-qc]}
  \BibitemShut {NoStop}%
\bibitem [{\citenamefont {Amorós}\ \emph {et~al.}(2013)\citenamefont {Amorós},
  \citenamefont {de~Haro},\ and\ \citenamefont {Odintsov}}]{bounce3}%
  \BibitemOpen
  \bibfield  {author} {\bibinfo {author} {\bibfnamefont {Jaume}\ \bibnamefont
  {Amorós}}, \bibinfo {author} {\bibfnamefont {Jaume}\ \bibnamefont {de~Haro}},
  \ and\ \bibinfo {author} {\bibfnamefont {Sergei~D.}\ \bibnamefont
  {Odintsov}},\ }\bibfield  {title} {\enquote {\bibinfo {title} {{Bouncing loop
  quantum cosmology from $F(T)$ gravity}},}\ }\href {\doibase
  10.1103/PhysRevD.87.104037} {\bibfield  {journal} {\bibinfo  {journal} {Phys.
  Rev. D}\ }\textbf {\bibinfo {volume} {87}},\ \bibinfo {pages} {104037}
  (\bibinfo {year} {2013})},\ \Eprint {http://arxiv.org/abs/1305.2344}
  {arXiv:1305.2344 [gr-qc]} \BibitemShut {NoStop}%
\bibitem [{\citenamefont {{Cai}}\ \emph {et~al.}(2012)\citenamefont {{Cai}},
  \citenamefont {{Easson}},\ and\ \citenamefont
  {{Brandenberger}}}]{Cai:2012JCAP}%
  \BibitemOpen
  \bibfield  {author} {\bibinfo {author} {\bibfnamefont {Y.-F.}\ \bibnamefont
  {{Cai}}}, \bibinfo {author} {\bibfnamefont {D.~A.}\ \bibnamefont {{Easson}}},
  \ and\ \bibinfo {author} {\bibfnamefont {R.}~\bibnamefont
  {{Brandenberger}}},\ }\bibfield  {title} {\enquote {\bibinfo {title}
  {{Towards a nonsingular bouncing cosmology}},}\ }\href {\doibase
  10.1088/1475-7516/2012/08/020} {\bibfield  {journal} {\bibinfo  {journal} {J.
  Cosmol. Astropart. Phys.}\ }\textbf {\bibinfo {volume} {8}},\ \bibinfo {eid}
  {020} (\bibinfo {year} {2012})},\ \Eprint {http://arxiv.org/abs/1206.2382}
  {arXiv:1206.2382 [hep-th]} \BibitemShut {NoStop}%
\bibitem [{\citenamefont {{Cai}}\ \emph {et~al.}(2013)\citenamefont {{Cai}},
  \citenamefont {{McDonough}}, \citenamefont {{Duplessis}},\ and\ \citenamefont
  {{Brandenberger}}}]{Cai:2013JCAP}%
  \BibitemOpen
  \bibfield  {author} {\bibinfo {author} {\bibfnamefont {Y.-F.}\ \bibnamefont
  {{Cai}}}, \bibinfo {author} {\bibfnamefont {E.}~\bibnamefont {{McDonough}}},
  \bibinfo {author} {\bibfnamefont {F.}~\bibnamefont {{Duplessis}}}, \ and\
  \bibinfo {author} {\bibfnamefont {R.~H.}\ \bibnamefont {{Brandenberger}}},\
  }\bibfield  {title} {\enquote {\bibinfo {title} {{Two field matter bounce
  cosmology}},}\ }\href {\doibase 10.1088/1475-7516/2013/10/024} {\bibfield
  {journal} {\bibinfo  {journal} {J. Cosmol. Astropart. Phys.}\ }\textbf
  {\bibinfo {volume} {10}},\ \bibinfo {eid} {024} (\bibinfo {year} {2013})},\
  \Eprint {http://arxiv.org/abs/1305.5259} {arXiv:1305.5259 [hep-th]}
  \BibitemShut {NoStop}%
\bibitem [{\citenamefont {{Cai}}(2014)}]{Cai:2014SCPMA}%
  \BibitemOpen
  \bibfield  {author} {\bibinfo {author} {\bibfnamefont {Y.-F.}\ \bibnamefont
  {{Cai}}},\ }\bibfield  {title} {\enquote {\bibinfo {title} {{Exploring
  bouncing cosmologies with cosmological surveys}},}\ }\href {\doibase
  10.1007/s11433-014-5512-3} {\bibfield  {journal} {\bibinfo  {journal} {Sci.
  China Phys. Mech. Astron.}\ }\textbf {\bibinfo {volume} {57}},\ \bibinfo
  {pages} {1414--1430} (\bibinfo {year} {2014})},\ \Eprint
  {http://arxiv.org/abs/1405.1369} {arXiv:1405.1369 [hep-th]} \BibitemShut
  {NoStop}%
\bibitem [{\citenamefont {{Nojiri}}\ and\ \citenamefont
  {{Odintsov}}(2011)}]{Nojiri:2010wj}%
  \BibitemOpen
  \bibfield  {author} {\bibinfo {author} {\bibfnamefont {S.}~\bibnamefont
  {{Nojiri}}}\ and\ \bibinfo {author} {\bibfnamefont {S.~D.}\ \bibnamefont
  {{Odintsov}}},\ }\bibfield  {title} {\enquote {\bibinfo {title} {{Unified
  cosmic history in modified gravity: From F(R) theory to Lorentz non-invariant
  models}},}\ }\href {\doibase 10.1016/j.physrep.2011.04.001} {\bibfield
  {journal} {\bibinfo  {journal} {Phys. Rep.}\ }\textbf {\bibinfo {volume}
  {505}},\ \bibinfo {pages} {59--144} (\bibinfo {year} {2011})},\ \Eprint
  {http://arxiv.org/abs/1011.0544} {arXiv:1011.0544 [gr-qc]} \BibitemShut
  {NoStop}%
\bibitem [{\citenamefont {Nojiri}\ and\ \citenamefont
  {Odintsov}(2007)}]{Nojiri:2006ri}%
  \BibitemOpen
  \bibfield  {author} {\bibinfo {author} {\bibfnamefont {Shin'ichi}\
  \bibnamefont {Nojiri}}\ and\ \bibinfo {author} {\bibfnamefont {Sergei~D.}\
  \bibnamefont {Odintsov}},\ }\bibfield  {title} {\enquote {\bibinfo {title}
  {{Introduction to modified gravity and gravitational alternative for dark
  energy}},}\ }\href {\doibase 10.1142/S0219887807001928} {\bibfield  {journal}
  {\bibinfo  {journal} {Int. J. Geom. Meth. Mod. Phys.}\ }\textbf {\bibinfo
  {volume} {04}},\ \bibinfo {pages} {115} (\bibinfo {year} {2007})},\ \Eprint
  {http://arxiv.org/abs/hep-th/0601213} {arXiv:hep-th/0601213 [hep-th]}
  \BibitemShut {NoStop}%
\bibitem [{\citenamefont {{Capozziello}}\ and\ \citenamefont
  {{Faraoni}}(2011)}]{Book-Capozziello-Faraoni}%
  \BibitemOpen
  \bibfield  {author} {\bibinfo {author} {\bibfnamefont {Salvatore}\
  \bibnamefont {{Capozziello}}}\ and\ \bibinfo {author} {\bibfnamefont
  {Valerio}\ \bibnamefont {{Faraoni}}},\ }\href {\doibase
  10.1007/978-94-007-0165-6} {\emph {\bibinfo {title} {{Beyond Einstein
  Gravity}}}},\ \bibinfo {series} {Fundamental Theories of Physics}, Vol.\
  \bibinfo {volume} {170}\ (\bibinfo  {publisher} {Springer},\ \bibinfo
  {address} {Dordrecht},\ \bibinfo {year} {2011})\BibitemShut {NoStop}%
\bibitem [{\citenamefont {{Capozziello}}\ and\ \citenamefont {{de
  Laurentis}}(2011)}]{Capozziello:2011et}%
  \BibitemOpen
  \bibfield  {author} {\bibinfo {author} {\bibfnamefont {S.}~\bibnamefont
  {{Capozziello}}}\ and\ \bibinfo {author} {\bibfnamefont {M.}~\bibnamefont
  {{de Laurentis}}},\ }\bibfield  {title} {\enquote {\bibinfo {title}
  {{Extended Theories of Gravity}},}\ }\href {\doibase
  10.1016/j.physrep.2011.09.003} {\bibfield  {journal} {\bibinfo  {journal}
  {Phys. Rep.}\ }\textbf {\bibinfo {volume} {509}},\ \bibinfo {pages}
  {167--321} (\bibinfo {year} {2011})},\ \Eprint
  {http://arxiv.org/abs/1108.6266} {arXiv:1108.6266 [gr-qc]} \BibitemShut
  {NoStop}%
\bibitem [{\citenamefont {{de la Cruz-Dombriz}}\ and\ \citenamefont
  {{S{\'a}ez-G{\'o}mez}}(2012)}]{delaCruzDombriz:2012xy}%
  \BibitemOpen
  \bibfield  {author} {\bibinfo {author} {\bibfnamefont {A.}~\bibnamefont {{de
  la Cruz-Dombriz}}}\ and\ \bibinfo {author} {\bibfnamefont {D.}~\bibnamefont
  {{S{\'a}ez-G{\'o}mez}}},\ }\bibfield  {title} {\enquote {\bibinfo {title}
  {{Black Holes, Cosmological Solutions, Future Singularities, and Their
  Thermodynamical Properties in Modified Gravity Theories}},}\ }\href {\doibase
  10.3390/e14091717} {\bibfield  {journal} {\bibinfo  {journal} {Entropy}\
  }\textbf {\bibinfo {volume} {14}},\ \bibinfo {pages} {1717--1770} (\bibinfo
  {year} {2012})},\ \Eprint {http://arxiv.org/abs/1207.2663} {arXiv:1207.2663
  [gr-qc]} \BibitemShut {NoStop}%
\bibitem [{\citenamefont {{Bamba}}\ \emph {et~al.}(2012)\citenamefont
  {{Bamba}}, \citenamefont {{Capozziello}}, \citenamefont {{Nojiri}},\ and\
  \citenamefont {{Odintsov}}}]{Bamba:2012cp}%
  \BibitemOpen
  \bibfield  {author} {\bibinfo {author} {\bibfnamefont {K.}~\bibnamefont
  {{Bamba}}}, \bibinfo {author} {\bibfnamefont {S.}~\bibnamefont
  {{Capozziello}}}, \bibinfo {author} {\bibfnamefont {S.}~\bibnamefont
  {{Nojiri}}}, \ and\ \bibinfo {author} {\bibfnamefont {S.~D.}\ \bibnamefont
  {{Odintsov}}},\ }\bibfield  {title} {\enquote {\bibinfo {title} {{Dark energy
  cosmology: the equivalent description via different theoretical models and
  cosmography tests}},}\ }\href {\doibase 10.1007/s10509-012-1181-8} {\bibfield
   {journal} {\bibinfo  {journal} {Astrophys. Space Sci.}\ }\textbf {\bibinfo
  {volume} {342}},\ \bibinfo {pages} {155--228} (\bibinfo {year} {2012})},\
  \Eprint {http://arxiv.org/abs/1205.3421} {arXiv:1205.3421 [gr-qc]}
  \BibitemShut {NoStop}%
\bibitem [{\citenamefont {{Joyce}}\ \emph {et~al.}(2015)\citenamefont
  {{Joyce}}, \citenamefont {{Jain}}, \citenamefont {{Khoury}},\ and\
  \citenamefont {{Trodden}}}]{Joyce:2014kja}%
  \BibitemOpen
  \bibfield  {author} {\bibinfo {author} {\bibfnamefont {A.}~\bibnamefont
  {{Joyce}}}, \bibinfo {author} {\bibfnamefont {B.}~\bibnamefont {{Jain}}},
  \bibinfo {author} {\bibfnamefont {J.}~\bibnamefont {{Khoury}}}, \ and\
  \bibinfo {author} {\bibfnamefont {M.}~\bibnamefont {{Trodden}}},\ }\bibfield
  {title} {\enquote {\bibinfo {title} {{Beyond the cosmological standard
  model}},}\ }\href {\doibase 10.1016/j.physrep.2014.12.002} {\bibfield
  {journal} {\bibinfo  {journal} {Phys. Rep.}\ }\textbf {\bibinfo {volume}
  {568}},\ \bibinfo {pages} {1--98} (\bibinfo {year} {2015})},\ \Eprint
  {http://arxiv.org/abs/1407.0059} {arXiv:1407.0059 [astro-ph]} \BibitemShut
  {NoStop}%
\bibitem [{\citenamefont {{Koyama}}(2016)}]{Koyama:2015vza}%
  \BibitemOpen
  \bibfield  {author} {\bibinfo {author} {\bibfnamefont {K.}~\bibnamefont
  {{Koyama}}},\ }\bibfield  {title} {\enquote {\bibinfo {title} {{Cosmological
  tests of modified gravity}},}\ }\href {\doibase
  10.1088/0034-4885/79/4/046902} {\bibfield  {journal} {\bibinfo  {journal}
  {Rep. Prog. Phys.}\ }\textbf {\bibinfo {volume} {79}},\ \bibinfo {eid}
  {046902} (\bibinfo {year} {2016})},\ \Eprint
  {http://arxiv.org/abs/1504.04623} {arXiv:1504.04623 [astro-ph]} \BibitemShut
  {NoStop}%
\bibitem [{\citenamefont {Bamba}\ and\ \citenamefont
  {Odintsov}(2015)}]{Bamba:2015uma}%
  \BibitemOpen
  \bibfield  {author} {\bibinfo {author} {\bibfnamefont {Kazuharu}\
  \bibnamefont {Bamba}}\ and\ \bibinfo {author} {\bibfnamefont {Sergei~D.}\
  \bibnamefont {Odintsov}},\ }\bibfield  {title} {\enquote {\bibinfo {title}
  {{Inflationary cosmology in modified gravity theories}},}\ }\href {\doibase
  10.3390/sym7010220} {\bibfield  {journal} {\bibinfo  {journal} {Symmetry}\
  }\textbf {\bibinfo {volume} {7}},\ \bibinfo {pages} {220--240} (\bibinfo
  {year} {2015})},\ \Eprint {http://arxiv.org/abs/1503.00442} {arXiv:1503.00442
  [hep-th]} \BibitemShut {NoStop}%
\bibitem [{\citenamefont {{Hehl}}\ \emph {et~al.}(1976)\citenamefont {{Hehl}},
  \citenamefont {{von der Heyde}}, \citenamefont {{Kerlick}},\ and\
  \citenamefont {{Nester}}}]{Hehl:1976kj}%
  \BibitemOpen
  \bibfield  {author} {\bibinfo {author} {\bibfnamefont {F.~W.}\ \bibnamefont
  {{Hehl}}}, \bibinfo {author} {\bibfnamefont {P.}~\bibnamefont {{von der
  Heyde}}}, \bibinfo {author} {\bibfnamefont {G.~D.}\ \bibnamefont
  {{Kerlick}}}, \ and\ \bibinfo {author} {\bibfnamefont {J.~M.}\ \bibnamefont
  {{Nester}}},\ }\bibfield  {title} {\enquote {\bibinfo {title} {{General
  relativity with spin and torsion: Foundations and prospects}},}\ }\href
  {\doibase 10.1103/RevModPhys.48.393} {\bibfield  {journal} {\bibinfo
  {journal} {Rev. Mod. Phys.}\ }\textbf {\bibinfo {volume} {48}},\ \bibinfo
  {pages} {393--416} (\bibinfo {year} {1976})}\BibitemShut {NoStop}%
\bibitem [{\citenamefont {{Hayashi}}\ and\ \citenamefont
  {{Shirafuji}}(1979)}]{Hayashi:1979qx}%
  \BibitemOpen
  \bibfield  {author} {\bibinfo {author} {\bibfnamefont {K.}~\bibnamefont
  {{Hayashi}}}\ and\ \bibinfo {author} {\bibfnamefont {T.}~\bibnamefont
  {{Shirafuji}}},\ }\bibfield  {title} {\enquote {\bibinfo {title} {{New
  general relativity}},}\ }\href {\doibase 10.1103/PhysRevD.19.3524} {\bibfield
   {journal} {\bibinfo  {journal} {Phys. Rev. D}\ }\textbf {\bibinfo {volume}
  {19}},\ \bibinfo {pages} {3524--3553} (\bibinfo {year} {1979})}\BibitemShut
  {NoStop}%
\bibitem [{\citenamefont {{Flanagan}}\ and\ \citenamefont
  {{Rosenthal}}(2007)}]{Flanagan:2007dc}%
  \BibitemOpen
  \bibfield  {author} {\bibinfo {author} {\bibfnamefont {{\'E}.~{\'E}.}\
  \bibnamefont {{Flanagan}}}\ and\ \bibinfo {author} {\bibfnamefont
  {E.}~\bibnamefont {{Rosenthal}}},\ }\bibfield  {title} {\enquote {\bibinfo
  {title} {{Can Gravity Probe B usefully constrain torsion gravity
  theories?}}}\ }\href {\doibase 10.1103/PhysRevD.75.124016} {\bibfield
  {journal} {\bibinfo  {journal} {Phys. Rev. D}\ }\textbf {\bibinfo {volume}
  {75}},\ \bibinfo {eid} {124016} (\bibinfo {year} {2007})},\ \Eprint
  {http://arxiv.org/abs/0704.1447} {arXiv:0704.1447 [gr-qc]} \BibitemShut
  {NoStop}%
\bibitem [{\citenamefont {{Garecki}}()}]{Garecki:2010jj}%
  \BibitemOpen
  \bibfield  {author} {\bibinfo {author} {\bibfnamefont {J.}~\bibnamefont
  {{Garecki}}},\ }\bibfield  {title} {\enquote {\bibinfo {title} {{Teleparallel
  equivalent of general relativity: a critical review}},}\ }\href@noop {} {\
  }\Eprint {http://arxiv.org/abs/1010.2654} {arXiv:1010.2654 [gr-qc]}
  \BibitemShut {NoStop}%
\bibitem [{\citenamefont {{Bamba}}()}]{Bamba:2015jqa}%
  \BibitemOpen
  \bibfield  {author} {\bibinfo {author} {\bibfnamefont {K.}~\bibnamefont
  {{Bamba}}},\ }\bibfield  {title} {\enquote {\bibinfo {title} {{Cosmological
  Issues in $F(T)$ Gravity Theory}},}\ }\href@noop {} {\ }\Eprint
  {http://arxiv.org/abs/1504.04457} {arXiv:1504.04457 [gr-qc]} \BibitemShut
  {NoStop}%
\bibitem [{\citenamefont {{Cai}}\ \emph {et~al.}(2011)\citenamefont {{Cai}},
  \citenamefont {{Chen}}, \citenamefont {{Dent}}, \citenamefont {{Dutta}},\
  and\ \citenamefont {{Saridakis}}}]{Cai:2011tc}%
  \BibitemOpen
  \bibfield  {author} {\bibinfo {author} {\bibfnamefont {Y.-F.}\ \bibnamefont
  {{Cai}}}, \bibinfo {author} {\bibfnamefont {S.-H.}\ \bibnamefont {{Chen}}},
  \bibinfo {author} {\bibfnamefont {J.~B.}\ \bibnamefont {{Dent}}}, \bibinfo
  {author} {\bibfnamefont {S.}~\bibnamefont {{Dutta}}}, \ and\ \bibinfo
  {author} {\bibfnamefont {E.~N.}\ \bibnamefont {{Saridakis}}},\ }\bibfield
  {title} {\enquote {\bibinfo {title} {{Matter bounce cosmology with the $f(T)$
  gravity}},}\ }\href {\doibase 10.1088/0264-9381/28/21/215011} {\bibfield
  {journal} {\bibinfo  {journal} {Classical and Quantum Gravity}\ }\textbf
  {\bibinfo {volume} {28}},\ \bibinfo {eid} {215011} (\bibinfo {year}
  {2011})},\ \Eprint {http://arxiv.org/abs/1104.4349} {arXiv:1104.4349
  [astro-ph]} \BibitemShut {NoStop}%
\bibitem [{\citenamefont {{Cai}}\ \emph {et~al.}(2014)\citenamefont {{Cai}},
  \citenamefont {{Quintin}}, \citenamefont {{Saridakis}},\ and\ \citenamefont
  {{Wilson-Ewing}}}]{CQSW14}%
  \BibitemOpen
  \bibfield  {author} {\bibinfo {author} {\bibfnamefont {Y.-F.}\ \bibnamefont
  {{Cai}}}, \bibinfo {author} {\bibfnamefont {J.}~\bibnamefont {{Quintin}}},
  \bibinfo {author} {\bibfnamefont {E.~N.}\ \bibnamefont {{Saridakis}}}, \ and\
  \bibinfo {author} {\bibfnamefont {E.}~\bibnamefont {{Wilson-Ewing}}},\
  }\bibfield  {title} {\enquote {\bibinfo {title} {{Nonsingular bouncing
  cosmologies in light of BICEP2}},}\ }\href {\doibase
  10.1088/1475-7516/2014/07/033} {\bibfield  {journal} {\bibinfo  {journal} {J.
  Cosmol. Astropart. Phys.}\ }\textbf {\bibinfo {volume} {7}},\ \bibinfo {eid}
  {033} (\bibinfo {year} {2014})},\ \Eprint {http://arxiv.org/abs/1404.4364}
  {arXiv:1404.4364 [astro-ph]} \BibitemShut {NoStop}%
\bibitem [{\citenamefont {Haro}(2013)}]{Haro:2013bea}%
  \BibitemOpen
  \bibfield  {author} {\bibinfo {author} {\bibfnamefont {Jaime}\ \bibnamefont
  {Haro}},\ }\bibfield  {title} {\enquote {\bibinfo {title} {{Cosmological
  perturbations in teleparallel Loop Quantum Cosmology}},}\ }\href {\doibase
  10.1088/1475-7516/2014/05/E01, 10.1088/1475-7516/2013/11/068} {\bibfield
  {journal} {\bibinfo  {journal} {J. Cosmol. Astropart. Phys.}\ }\textbf
  {\bibinfo {volume} {1311}},\ \bibinfo {pages} {068} (\bibinfo {year}
  {2013})},\ \bibinfo {note} {[Erratum: JCAP1405,E01(2014)]},\ \Eprint
  {http://arxiv.org/abs/1309.0352} {arXiv:1309.0352 [gr-qc]} \BibitemShut
  {NoStop}%
\bibitem [{\citenamefont {Haro}\ and\ \citenamefont
  {Amoros}(2014)}]{Haro:2014wha}%
  \BibitemOpen
  \bibfield  {author} {\bibinfo {author} {\bibfnamefont {Jaume}\ \bibnamefont
  {Haro}}\ and\ \bibinfo {author} {\bibfnamefont {Jaume}\ \bibnamefont
  {Amoros}},\ }\bibfield  {title} {\enquote {\bibinfo {title} {{Viability of
  the matter bounce scenario in $F(T)$ gravity and Loop Quantum Cosmology for
  general potentials}},}\ }\href {\doibase 10.1088/1475-7516/2014/12/031}
  {\bibfield  {journal} {\bibinfo  {journal} {J. Cosmol. Astropart. Phys.}\
  }\textbf {\bibinfo {volume} {1412}},\ \bibinfo {pages} {031} (\bibinfo {year}
  {2014})},\ \Eprint {http://arxiv.org/abs/1406.0369} {arXiv:1406.0369 [gr-qc]}
  \BibitemShut {NoStop}%
\bibitem [{\citenamefont {{Haro}}\ and\ \citenamefont
  {{Amorós}}(2015)}]{Haro:2015zta}%
  \BibitemOpen
  \bibfield  {author} {\bibinfo {author} {\bibfnamefont {Jaume}\ \bibnamefont
  {{Haro}}}\ and\ \bibinfo {author} {\bibfnamefont {Jaume}\ \bibnamefont
  {{Amorós}}},\ }\bibfield  {title} {\enquote {\bibinfo {title} {{Matter Bounce
  Scenario in $F(T)$ gravity}},}\ }\href@noop {} {\bibfield  {journal}
  {\bibinfo  {journal} {Proc. Sci.}\ }\textbf {\bibinfo {volume} {FFP142015}},\
  \bibinfo {pages} {163} (\bibinfo {year} {2015})},\ \Eprint
  {http://arxiv.org/abs/1501.06270} {arXiv:1501.06270 [gr-qc]} \BibitemShut
  {NoStop}%
\bibitem [{\citenamefont {{Bret{\'o}n}}\ \emph {et~al.}(2004)\citenamefont
  {{Bret{\'o}n}}, \citenamefont {{Cervantes-Cota}},\ and\ \citenamefont
  {{Salgad}}}]{Early2004}%
  \BibitemOpen
  \bibinfo {editor} {\bibfnamefont {N.}~\bibnamefont {{Bret{\'o}n}}}, \bibinfo
  {editor} {\bibfnamefont {J.~L.}\ \bibnamefont {{Cervantes-Cota}}}, \ and\
  \bibinfo {editor} {\bibfnamefont {M.}~\bibnamefont {{Salgad}}},\ eds.,\ \href
  {\doibase 10.1007/b97189} {\emph {\bibinfo {title} {The Early Universe and
  Observational Cosmology}}},\ \bibinfo {series} {Lecture Notes in Physics,
  Berlin Springer Verlag}, Vol.\ \bibinfo {volume} {646}\ (\bibinfo {year}
  {2004})\BibitemShut {NoStop}%
\bibitem [{\citenamefont {Kr\v{s}\v{s}\'{a}k}\ and\ \citenamefont
  {Pereira}(2015)}]{martin1}%
  \BibitemOpen
  \bibfield  {author} {\bibinfo {author} {\bibfnamefont {Martin}\ \bibnamefont
  {Kr\v{s}\v{s}\'{a}k}}\ and\ \bibinfo {author} {\bibfnamefont {J.~G.}\
  \bibnamefont {Pereira}},\ }\bibfield  {title} {\enquote {\bibinfo {title}
  {{Spin Connection and Renormalization of Teleparallel Action}},}\ }\href
  {\doibase 10.1140/epjc/s10052-015-3749-2} {\bibfield  {journal} {\bibinfo
  {journal} {Eur. Phys. J. C}\ }\textbf {\bibinfo {volume} {75}},\ \bibinfo
  {pages} {519} (\bibinfo {year} {2015})},\ \Eprint
  {http://arxiv.org/abs/1504.07683} {arXiv:1504.07683 [gr-qc]} \BibitemShut
  {NoStop}%
\bibitem [{\citenamefont {{El Hanafy}}\ and\ \citenamefont
  {{Nashed}}(2016{\natexlab{a}})}]{Waleed3}%
  \BibitemOpen
  \bibfield  {author} {\bibinfo {author} {\bibfnamefont {W.}~\bibnamefont {{El
  Hanafy}}}\ and\ \bibinfo {author} {\bibfnamefont {G.~G.~L.}\ \bibnamefont
  {{Nashed}}},\ }\bibfield  {title} {\enquote {\bibinfo {title} {{Exact
  teleparallel gravity of binary black holes}},}\ }\href {\doibase
  10.1007/s10509-016-2662-y} {\bibfield  {journal} {\bibinfo  {journal}
  {Astrophys. Space Sci.}\ }\textbf {\bibinfo {volume} {361}},\ \bibinfo {eid}
  {68} (\bibinfo {year} {2016}{\natexlab{a}})},\ \Eprint
  {http://arxiv.org/abs/1507.07377} {arXiv:1507.07377 [gr-qc]} \BibitemShut
  {NoStop}%
\bibitem [{\citenamefont {Chen}\ \emph {et~al.}(2015)\citenamefont {Chen},
  \citenamefont {Wu},\ and\ \citenamefont {Wei}}]{Chen2}%
  \BibitemOpen
  \bibfield  {author} {\bibinfo {author} {\bibfnamefont {Zu-Cheng}\
  \bibnamefont {Chen}}, \bibinfo {author} {\bibfnamefont {You}\ \bibnamefont
  {Wu}}, \ and\ \bibinfo {author} {\bibfnamefont {Hao}\ \bibnamefont {Wei}},\
  }\bibfield  {title} {\enquote {\bibinfo {title} {{Post-Newtonian
  Approximation of Teleparallel Gravity Coupled with a Scalar Field}},}\ }\href
  {\doibase 10.1016/j.nuclphysb.2015.03.012} {\bibfield  {journal} {\bibinfo
  {journal} {Nucl. Phys.}\ ,\ \bibinfo {pages} {422--438}} (\bibinfo {year}
  {2015})},\ \Eprint {http://arxiv.org/abs/1410.7715} {arXiv:1410.7715 [gr-qc]}
  \BibitemShut {NoStop}%
\bibitem [{\citenamefont {{Nashed}}(2002)}]{Nashed1}%
  \BibitemOpen
  \bibfield  {author} {\bibinfo {author} {\bibfnamefont {G.~G.~L.}\
  \bibnamefont {{Nashed}}},\ }\bibfield  {title} {\enquote {\bibinfo {title}
  {{Vacuum nonsingular black hole in tetrad theory of gravitation}},}\
  }\href@noop {} {\bibfield  {journal} {\bibinfo  {journal} {Nuovo Cimento Soc.
  Ital. Fis.}\ ,\ \bibinfo {pages} {521}} (\bibinfo {year} {2002})},\ \Eprint
  {http://arxiv.org/abs/gr-qc/0109017} {arXiv:gr-qc/0109017 [gr-qc]}
  \BibitemShut {NoStop}%
\bibitem [{\citenamefont {Kr\v{s}\v{s}\'{a}k}(2015)}]{Martin3}%
  \BibitemOpen
  \bibfield  {author} {\bibinfo {author} {\bibfnamefont {Martin}\ \bibnamefont
  {Kr\v{s}\v{s}\'{a}k}},\ }\bibfield  {title} {\enquote {\bibinfo {title}
  {{Holographic Renormalization in Teleparallel Gravity}},}\ }\href@noop {} {\
  (\bibinfo {year} {2015})},\ \Eprint {http://arxiv.org/abs/1510.06676}
  {arXiv:1510.06676 [gr-qc]} \BibitemShut {NoStop}%
\bibitem [{\citenamefont {{Shirafuji}}\ and\ \citenamefont
  {{Nashed}}(1997)}]{Nashed3}%
  \BibitemOpen
  \bibfield  {author} {\bibinfo {author} {\bibfnamefont {T.}~\bibnamefont
  {{Shirafuji}}}\ and\ \bibinfo {author} {\bibfnamefont {G.~G.}\ \bibnamefont
  {{Nashed}}},\ }\bibfield  {title} {\enquote {\bibinfo {title} {{Energy and
  Momentum in the Tetrad Theory of Gravitation}},}\ }\href {\doibase
  10.1143/PTP.98.1355} {\bibfield  {journal} {\bibinfo  {journal} {Prog. Theor.
  Phys.}\ }\textbf {\bibinfo {volume} {98}},\ \bibinfo {pages} {1355--1370}
  (\bibinfo {year} {1997})},\ \Eprint {http://arxiv.org/abs/gr-qc/9711010}
  {arXiv:gr-qc/9711010 [gr-qc]} \BibitemShut {NoStop}%
\bibitem [{\citenamefont {{Nashed}}(2010)}]{Nashed4}%
  \BibitemOpen
  \bibfield  {author} {\bibinfo {author} {\bibfnamefont {G.~G.~L.}\
  \bibnamefont {{Nashed}}},\ }\bibfield  {title} {\enquote {\bibinfo {title}
  {{Stationary axisymmetric solutions and their energy contents in teleparallel
  equivalent of Einstein theory}},}\ }\href {\doibase
  10.1007/s10509-010-0375-1} {\bibfield  {journal} {\bibinfo  {journal}
  {Astrophys. Space Sci.}\ }\textbf {\bibinfo {volume} {330}},\ \bibinfo
  {pages} {173--181} (\bibinfo {year} {2010})},\ \Eprint
  {http://arxiv.org/abs/1503.01379} {arXiv:1503.01379 [gr-qc]} \BibitemShut
  {NoStop}%
\bibitem [{\citenamefont {Capozziello}\ \emph {et~al.}(2015)\citenamefont
  {Capozziello}, \citenamefont {Luongo},\ and\ \citenamefont
  {Saridakis}}]{Saridakis2}%
  \BibitemOpen
  \bibfield  {author} {\bibinfo {author} {\bibfnamefont {Salvatore}\
  \bibnamefont {Capozziello}}, \bibinfo {author} {\bibfnamefont {Orlando}\
  \bibnamefont {Luongo}}, \ and\ \bibinfo {author} {\bibfnamefont
  {Emmanuel~N.}\ \bibnamefont {Saridakis}},\ }\bibfield  {title} {\enquote
  {\bibinfo {title} {{Transition redshift in $f(T)$ cosmology and observational
  constraints}},}\ }\href {\doibase 10.1103/PhysRevD.91.124037} {\bibfield
  {journal} {\bibinfo  {journal} {Phys. Rev. D}\ }\textbf {\bibinfo {volume}
  {91}},\ \bibinfo {pages} {124037} (\bibinfo {year} {2015})},\ \Eprint
  {http://arxiv.org/abs/1503.02832} {arXiv:1503.02832 [gr-qc]} \BibitemShut
  {NoStop}%
\bibitem [{\citenamefont {Nashed}\ and\ \citenamefont
  {El~Hanafy}(2014)}]{Waleed1}%
  \BibitemOpen
  \bibfield  {author} {\bibinfo {author} {\bibfnamefont {G.~G.~L.}\
  \bibnamefont {Nashed}}\ and\ \bibinfo {author} {\bibfnamefont
  {W.}~\bibnamefont {El~Hanafy}},\ }\bibfield  {title} {\enquote {\bibinfo
  {title} {{A Built-in Inflation in the $f(T)$-Cosmology}},}\ }\href {\doibase
  10.1140/epjc/s10052-014-3099-5} {\bibfield  {journal} {\bibinfo  {journal}
  {Eur. Phys. J. C}\ }\textbf {\bibinfo {volume} {74}},\ \bibinfo {pages}
  {3099} (\bibinfo {year} {2014})},\ \Eprint {http://arxiv.org/abs/1403.0913}
  {arXiv:1403.0913 [gr-qc]} \BibitemShut {NoStop}%
\bibitem [{\citenamefont {Rezazadeh}\ \emph {et~al.}(2016)\citenamefont
  {Rezazadeh}, \citenamefont {Abdolmaleki},\ and\ \citenamefont
  {Karami}}]{Karami}%
  \BibitemOpen
  \bibfield  {author} {\bibinfo {author} {\bibfnamefont {K.}~\bibnamefont
  {Rezazadeh}}, \bibinfo {author} {\bibfnamefont {A.}~\bibnamefont
  {Abdolmaleki}}, \ and\ \bibinfo {author} {\bibfnamefont {K.}~\bibnamefont
  {Karami}},\ }\bibfield  {title} {\enquote {\bibinfo {title} {{Power-law and
  intermediate inflationary models in $f(T)$-gravity}},}\ }\href {\doibase
  10.1007/JHEP01(2016)131} {\bibfield  {journal} {\bibinfo  {journal} {J. High
  Energy Phys.}\ }\textbf {\bibinfo {volume} {01}},\ \bibinfo {pages} {131}
  (\bibinfo {year} {2016})},\ \Eprint {http://arxiv.org/abs/1509.08769}
  {arXiv:1509.08769 [gr-qc]} \BibitemShut {NoStop}%
\bibitem [{\citenamefont {Wu}\ \emph {et~al.}(2015)\citenamefont {Wu},
  \citenamefont {Chen}, \citenamefont {Wang},\ and\ \citenamefont
  {Wei}}]{Chen1}%
  \BibitemOpen
  \bibfield  {author} {\bibinfo {author} {\bibfnamefont {You}\ \bibnamefont
  {Wu}}, \bibinfo {author} {\bibfnamefont {Zu-Cheng}\ \bibnamefont {Chen}},
  \bibinfo {author} {\bibfnamefont {Jiaxin}\ \bibnamefont {Wang}}, \ and\
  \bibinfo {author} {\bibfnamefont {Hao}\ \bibnamefont {Wei}},\ }\bibfield
  {title} {\enquote {\bibinfo {title} {{$f(T)$ non-linear massive gravity and
  the cosmic acceleration}},}\ }\href {\doibase 10.1088/0253-6102/63/6/701}
  {\bibfield  {journal} {\bibinfo  {journal} {Commun. Theor. Phys.}\ }\textbf
  {\bibinfo {volume} {63}},\ \bibinfo {pages} {701--708} (\bibinfo {year}
  {2015})},\ \Eprint {http://arxiv.org/abs/1503.05281} {arXiv:1503.05281
  [gr-qc]} \BibitemShut {NoStop}%
\bibitem [{\citenamefont {{El Hanafy}}\ and\ \citenamefont
  {{Nashed}}(2015)}]{Waleed2}%
  \BibitemOpen
  \bibfield  {author} {\bibinfo {author} {\bibfnamefont {W.}~\bibnamefont {{El
  Hanafy}}}\ and\ \bibinfo {author} {\bibfnamefont {G.~G.~L.}\ \bibnamefont
  {{Nashed}}},\ }\bibfield  {title} {\enquote {\bibinfo {title} {{The hidden
  flat like universe}},}\ }\href {\doibase 10.1140/epjc/s10052-015-3501-y}
  {\bibfield  {journal} {\bibinfo  {journal} {Eur. Phys. J. C}\ }\textbf
  {\bibinfo {volume} {75}},\ \bibinfo {eid} {279} (\bibinfo {year} {2015})},\
  \Eprint {http://arxiv.org/abs/1409.7199} {arXiv:1409.7199 [hep-th]}
  \BibitemShut {NoStop}%
\bibitem [{\citenamefont {{Guo}}\ \emph {et~al.}(2016)\citenamefont {{Guo}},
  \citenamefont {{Fu}}, \citenamefont {{Zhang}},\ and\ \citenamefont
  {{Liu}}}]{cosm1}%
  \BibitemOpen
  \bibfield  {author} {\bibinfo {author} {\bibfnamefont {W.-D.}\ \bibnamefont
  {{Guo}}}, \bibinfo {author} {\bibfnamefont {Q.-M.}\ \bibnamefont {{Fu}}},
  \bibinfo {author} {\bibfnamefont {Y.-P.}\ \bibnamefont {{Zhang}}}, \ and\
  \bibinfo {author} {\bibfnamefont {Y.-X.}\ \bibnamefont {{Liu}}},\ }\bibfield
  {title} {\enquote {\bibinfo {title} {{Tensor perturbations of f (T )
  branes}},}\ }\href {\doibase 10.1103/PhysRevD.93.044002} {\bibfield
  {journal} {\bibinfo  {journal} {Phys. Rev. D}\ }\textbf {\bibinfo {volume}
  {93}},\ \bibinfo {eid} {044002} (\bibinfo {year} {2016})},\ \Eprint
  {http://arxiv.org/abs/1511.07143} {arXiv:1511.07143 [hep-th]} \BibitemShut
  {NoStop}%
\bibitem [{\citenamefont {{Myrzakulov}}\ \emph {et~al.}(2015)\citenamefont
  {{Myrzakulov}}, \citenamefont {{S{\'a}ez-G{\'o}mez}},\ and\ \citenamefont
  {{Tsyba}}}]{cosm2}%
  \BibitemOpen
  \bibfield  {author} {\bibinfo {author} {\bibfnamefont {R.}~\bibnamefont
  {{Myrzakulov}}}, \bibinfo {author} {\bibfnamefont {D.}~\bibnamefont
  {{S{\'a}ez-G{\'o}mez}}}, \ and\ \bibinfo {author} {\bibfnamefont
  {P.}~\bibnamefont {{Tsyba}}},\ }\bibfield  {title} {\enquote {\bibinfo
  {title} {{Cosmological solutions in $F(T)$ gravity with the presence of
  spinor fields}},}\ }\href {\doibase 10.1142/S0219887815500231} {\bibfield
  {journal} {\bibinfo  {journal} {Int. J. Geom. Meth. Mod. Phys.}\ }\textbf
  {\bibinfo {volume} {12}},\ \bibinfo {pages} {1550023} (\bibinfo {year}
  {2015})},\ \Eprint {http://arxiv.org/abs/1311.5261} {arXiv:1311.5261 [gr-qc]}
  \BibitemShut {NoStop}%
\bibitem [{\citenamefont {{Sharif}}\ and\ \citenamefont
  {{Nazir}}(2015)}]{cosm3}%
  \BibitemOpen
  \bibfield  {author} {\bibinfo {author} {\bibfnamefont {M.}~\bibnamefont
  {{Sharif}}}\ and\ \bibinfo {author} {\bibfnamefont {K.}~\bibnamefont
  {{Nazir}}},\ }\bibfield  {title} {\enquote {\bibinfo {title} {{Cosmological
  evolution of generalized ghost pilgrim dark energy in $f(T)$ gravity}},}\
  }\href {\doibase 10.1007/s10509-015-2572-4} {\bibfield  {journal} {\bibinfo
  {journal} {Astrophys. Space Sci.}\ }\textbf {\bibinfo {volume} {360}},\
  \bibinfo {eid} {25} (\bibinfo {year} {2015})}\BibitemShut {NoStop}%
\bibitem [{\citenamefont {{Odintsov}}\ \emph {et~al.}(2015)\citenamefont
  {{Odintsov}}, \citenamefont {{Oikonomou}},\ and\ \citenamefont
  {{Saridakis}}}]{cosm4}%
  \BibitemOpen
  \bibfield  {author} {\bibinfo {author} {\bibfnamefont {S.~D.}\ \bibnamefont
  {{Odintsov}}}, \bibinfo {author} {\bibfnamefont {V.~K.}\ \bibnamefont
  {{Oikonomou}}}, \ and\ \bibinfo {author} {\bibfnamefont {E.~N.}\ \bibnamefont
  {{Saridakis}}},\ }\bibfield  {title} {\enquote {\bibinfo {title}
  {{Superbounce and loop quantum ekpyrotic cosmologies from modified gravity:
  $F(R)$ , $F(G)$ and $F(T)$ theories}},}\ }\href {\doibase
  10.1016/j.aop.2015.08.021} {\bibfield  {journal} {\bibinfo  {journal} {Ann.
  Phys. (Amsterdam)}\ }\textbf {\bibinfo {volume} {363}},\ \bibinfo {pages}
  {141--163} (\bibinfo {year} {2015})},\ \Eprint
  {http://arxiv.org/abs/1501.06591} {arXiv:1501.06591 [gr-qc]} \BibitemShut
  {NoStop}%
\bibitem [{\citenamefont {{Feng}}\ \emph {et~al.}(2015)\citenamefont {{Feng}},
  \citenamefont {{Ge}}, \citenamefont {{Li}}, \citenamefont {{Lin}},\ and\
  \citenamefont {{Zhai}}}]{cosm5}%
  \BibitemOpen
  \bibfield  {author} {\bibinfo {author} {\bibfnamefont {C.-j.}\ \bibnamefont
  {{Feng}}}, \bibinfo {author} {\bibfnamefont {F.-f.}\ \bibnamefont {{Ge}}},
  \bibinfo {author} {\bibfnamefont {X.-z.}\ \bibnamefont {{Li}}}, \bibinfo
  {author} {\bibfnamefont {R.-h.}\ \bibnamefont {{Lin}}}, \ and\ \bibinfo
  {author} {\bibfnamefont {X.-h.}\ \bibnamefont {{Zhai}}},\ }\bibfield  {title}
  {\enquote {\bibinfo {title} {{Towards realistic f (T ) models with nonminimal
  torsion-matter coupling extension}},}\ }\href {\doibase
  10.1103/PhysRevD.92.104038} {\bibfield  {journal} {\bibinfo  {journal} {Phys.
  Rev. D}\ }\textbf {\bibinfo {volume} {92}},\ \bibinfo {eid} {104038}
  (\bibinfo {year} {2015})},\ \Eprint {http://arxiv.org/abs/1511.07935}
  {arXiv:1511.07935 [gr-qc]} \BibitemShut {NoStop}%
\bibitem [{\citenamefont {{Sharif}}\ and\ \citenamefont
  {{Saima}}(2015)}]{cosm6}%
  \BibitemOpen
  \bibfield  {author} {\bibinfo {author} {\bibfnamefont {M.}~\bibnamefont
  {{Sharif}}}\ and\ \bibinfo {author} {\bibfnamefont {J.}~\bibnamefont
  {{Saima}}},\ }\bibfield  {title} {\enquote {\bibinfo {title} {{Phase Space
  Analysis and Anisotropic Universe Model in $f(T)$ Gravity}},}\ }\href
  {\doibase 10.1088/0253-6102/63/2/09} {\bibfield  {journal} {\bibinfo
  {journal} {Commun. Theor. Phys.}\ }\textbf {\bibinfo {volume} {63}},\
  \bibinfo {eid} {168-180} (\bibinfo {year} {2015})}\BibitemShut {NoStop}%
\bibitem [{\citenamefont {Iorio}\ \emph {et~al.}(2015)\citenamefont {Iorio},
  \citenamefont {Radicella},\ and\ \citenamefont {Ruggiero}}]{Iorio}%
  \BibitemOpen
  \bibfield  {author} {\bibinfo {author} {\bibfnamefont {Lorenzo}\ \bibnamefont
  {Iorio}}, \bibinfo {author} {\bibfnamefont {Ninfa}\ \bibnamefont
  {Radicella}}, \ and\ \bibinfo {author} {\bibfnamefont {Matteo~Luca}\
  \bibnamefont {Ruggiero}},\ }\bibfield  {title} {\enquote {\bibinfo {title}
  {{Constraining $f(T)$ gravity in the Solar System}},}\ }\href {\doibase
  10.1088/1475-7516/2015/08/021} {\bibfield  {journal} {\bibinfo  {journal} {J.
  Cosmol. Astropart. Phys.}\ }\textbf {\bibinfo {volume} {1508}},\ \bibinfo
  {pages} {021} (\bibinfo {year} {2015})},\ \Eprint
  {http://arxiv.org/abs/1505.06996} {arXiv:1505.06996 [gr-qc]} \BibitemShut
  {NoStop}%
\bibitem [{\citenamefont {Ruggiero}(2016)}]{Ruggiero}%
  \BibitemOpen
  \bibfield  {author} {\bibinfo {author} {\bibfnamefont {Matteo~Luca}\
  \bibnamefont {Ruggiero}},\ }\bibfield  {title} {\enquote {\bibinfo {title}
  {{Light bending in $f(T)$ gravity}},}\ }\href {\doibase
  10.1142/S0218271816500735} {\bibfield  {journal} {\bibinfo  {journal} {Int.
  J. Mod. Phys.}\ ,\ \bibinfo {pages} {1650073}} (\bibinfo {year} {2016})},\
  \Eprint {http://arxiv.org/abs/1601.00588} {arXiv:1601.00588 [gr-qc]}
  \BibitemShut {NoStop}%
\bibitem [{\citenamefont {Junior}\ \emph
  {et~al.}(2015{\natexlab{a}})\citenamefont {Junior}, \citenamefont
  {Rodrigues},\ and\ \citenamefont {Houndjo}}]{Rod1}%
  \BibitemOpen
  \bibfield  {author} {\bibinfo {author} {\bibfnamefont {Ednaldo L.~B.}\
  \bibnamefont {Junior}}, \bibinfo {author} {\bibfnamefont {Manuel~E.}\
  \bibnamefont {Rodrigues}}, \ and\ \bibinfo {author} {\bibfnamefont {Mahouton
  J.~S.}\ \bibnamefont {Houndjo}},\ }\bibfield  {title} {\enquote {\bibinfo
  {title} {{Regular black holes in $f(T)$ Gravity through a nonlinear
  electrodynamics source}},}\ }\href {\doibase 10.1088/1475-7516/2015/10/060}
  {\bibfield  {journal} {\bibinfo  {journal} {J. Cosmol. Astropart. Phys.}\
  }\textbf {\bibinfo {volume} {1510}},\ \bibinfo {pages} {060} (\bibinfo {year}
  {2015}{\natexlab{a}})},\ \Eprint {http://arxiv.org/abs/1503.07857}
  {arXiv:1503.07857 [gr-qc]} \BibitemShut {NoStop}%
\bibitem [{\citenamefont {Junior}\ \emph
  {et~al.}(2015{\natexlab{b}})\citenamefont {Junior}, \citenamefont
  {Rodrigues},\ and\ \citenamefont {Houndjo}}]{Rod2}%
  \BibitemOpen
  \bibfield  {author} {\bibinfo {author} {\bibfnamefont {Ednaldo L.~B.}\
  \bibnamefont {Junior}}, \bibinfo {author} {\bibfnamefont {Manuel~E.}\
  \bibnamefont {Rodrigues}}, \ and\ \bibinfo {author} {\bibfnamefont {Mahouton
  J.~S.}\ \bibnamefont {Houndjo}},\ }\bibfield  {title} {\enquote {\bibinfo
  {title} {{Born-Infeld and Charged Black Holes with non-linear source in
  $f(T)$ Gravity}},}\ }\href {\doibase 10.1088/1475-7516/2015/06/037}
  {\bibfield  {journal} {\bibinfo  {journal} {J. Cosmol. Astropart. Phys.}\
  }\textbf {\bibinfo {volume} {1506}},\ \bibinfo {pages} {037} (\bibinfo {year}
  {2015}{\natexlab{b}})},\ \Eprint {http://arxiv.org/abs/1503.07427}
  {arXiv:1503.07427 [gr-qc]} \BibitemShut {NoStop}%
\bibitem [{\citenamefont {{Paliathanasis}}\ \emph {et~al.}(2014)\citenamefont
  {{Paliathanasis}}, \citenamefont {{Basilakos}}, \citenamefont {{Saridakis}},
  \citenamefont {{Capozziello}}, \citenamefont {{Atazadeh}}, \citenamefont
  {{Darabi}},\ and\ \citenamefont {{Tsamparlis}}}]{astro1}%
  \BibitemOpen
  \bibfield  {author} {\bibinfo {author} {\bibfnamefont {A.}~\bibnamefont
  {{Paliathanasis}}}, \bibinfo {author} {\bibfnamefont {S.}~\bibnamefont
  {{Basilakos}}}, \bibinfo {author} {\bibfnamefont {E.~N.}\ \bibnamefont
  {{Saridakis}}}, \bibinfo {author} {\bibfnamefont {S.}~\bibnamefont
  {{Capozziello}}}, \bibinfo {author} {\bibfnamefont {K.}~\bibnamefont
  {{Atazadeh}}}, \bibinfo {author} {\bibfnamefont {F.}~\bibnamefont
  {{Darabi}}}, \ and\ \bibinfo {author} {\bibfnamefont {M.}~\bibnamefont
  {{Tsamparlis}}},\ }\bibfield  {title} {\enquote {\bibinfo {title} {{New
  Schwarzschild-like solutions in $f(T)$ gravity through Noether
  symmetries}},}\ }\href {\doibase 10.1103/PhysRevD.89.104042} {\bibfield
  {journal} {\bibinfo  {journal} {Phys. Rev. D}\ }\textbf {\bibinfo {volume}
  {89}},\ \bibinfo {eid} {104042} (\bibinfo {year} {2014})},\ \Eprint
  {http://arxiv.org/abs/1402.5935} {arXiv:1402.5935 [gr-qc]} \BibitemShut
  {NoStop}%
\bibitem [{\citenamefont {{Ruggiero}}\ and\ \citenamefont
  {{Radicella}}(2015)}]{astro2}%
  \BibitemOpen
  \bibfield  {author} {\bibinfo {author} {\bibfnamefont {M.~L.}\ \bibnamefont
  {{Ruggiero}}}\ and\ \bibinfo {author} {\bibfnamefont {N.}~\bibnamefont
  {{Radicella}}},\ }\bibfield  {title} {\enquote {\bibinfo {title} {{Weak-field
  spherically symmetric solutions in $f(T)$ gravity}},}\ }\href {\doibase
  10.1103/PhysRevD.91.104014} {\bibfield  {journal} {\bibinfo  {journal} {Phys.
  Rev. D}\ }\textbf {\bibinfo {volume} {91}},\ \bibinfo {eid} {104014}
  (\bibinfo {year} {2015})},\ \Eprint {http://arxiv.org/abs/1501.02198}
  {arXiv:1501.02198 [gr-qc]} \BibitemShut {NoStop}%
\bibitem [{\citenamefont {{Darabi}}\ \emph {et~al.}(2015)\citenamefont
  {{Darabi}}, \citenamefont {{Mousavi}},\ and\ \citenamefont
  {{Atazadeh}}}]{astro3}%
  \BibitemOpen
  \bibfield  {author} {\bibinfo {author} {\bibfnamefont {F.}~\bibnamefont
  {{Darabi}}}, \bibinfo {author} {\bibfnamefont {M.}~\bibnamefont {{Mousavi}}},
  \ and\ \bibinfo {author} {\bibfnamefont {K.}~\bibnamefont {{Atazadeh}}},\
  }\bibfield  {title} {\enquote {\bibinfo {title} {{Geodesic deviation equation
  in $f(T)$ gravity}},}\ }\href {\doibase 10.1103/PhysRevD.91.084023}
  {\bibfield  {journal} {\bibinfo  {journal} {Phys. Rev. D}\ }\textbf {\bibinfo
  {volume} {91}},\ \bibinfo {eid} {084023} (\bibinfo {year} {2015})},\ \Eprint
  {http://arxiv.org/abs/1501.00103} {arXiv:1501.00103 [gr-qc]} \BibitemShut
  {NoStop}%
\bibitem [{\citenamefont {Awad}(2013)}]{awad2013}%
  \BibitemOpen
  \bibfield  {author} {\bibinfo {author} {\bibfnamefont {Adel}\ \bibnamefont
  {Awad}},\ }\bibfield  {title} {\enquote {\bibinfo {title} {{Fixed points and
  FLRW cosmologies: Flat case}},}\ }\href {\doibase 10.1103/PhysRevD.87.109902,
  10.1103/PhysRevD.87.103001} {\bibfield  {journal} {\bibinfo  {journal} {Phys.
  Rev. D}\ }\textbf {\bibinfo {volume} {87}},\ \bibinfo {pages} {103001}
  (\bibinfo {year} {2013})},\ \bibinfo {note} {[Erratum: Phys.
  Rev.D87,no.10,109902(2013)]},\ \Eprint {http://arxiv.org/abs/1303.2014}
  {arXiv:1303.2014 [gr-qc]} \BibitemShut {NoStop}%
\bibitem [{\citenamefont {{Nojiri}}\ \emph {et~al.}(2005)\citenamefont
  {{Nojiri}}, \citenamefont {{Odintsov}},\ and\ \citenamefont
  {{Tsujikawa}}}]{Nojiri:2005prd}%
  \BibitemOpen
  \bibfield  {author} {\bibinfo {author} {\bibfnamefont {Shin'ichi}\
  \bibnamefont {{Nojiri}}}, \bibinfo {author} {\bibfnamefont {Sergei~D.}\
  \bibnamefont {{Odintsov}}}, \ and\ \bibinfo {author} {\bibfnamefont {Shinji}\
  \bibnamefont {{Tsujikawa}}},\ }\bibfield  {title} {\enquote {\bibinfo {title}
  {Properties of singularities in the (phantom) dark energy universe},}\ }\href
  {\doibase 10.1103/PhysRevD.71.063004} {\bibfield  {journal} {\bibinfo
  {journal} {Phys. Rev. D}\ }\textbf {\bibinfo {volume} {71}},\ \bibinfo
  {pages} {063004} (\bibinfo {year} {2005})},\ \Eprint
  {http://arxiv.org/abs/hep-th/0501025} {arXiv:hep-th/0501025 [hep-th]}
  \BibitemShut {NoStop}%
\bibitem [{\citenamefont {{Awad}}\ \emph {et~al.}(2016)\citenamefont {{Awad}},
  \citenamefont {{El Hanafy}}, \citenamefont {{Nashed}},\ and\ \citenamefont
  {{Saridakis}}}]{HNS:2016}%
  \BibitemOpen
  \bibfield  {author} {\bibinfo {author} {\bibfnamefont {A.}~\bibnamefont
  {{Awad}}}, \bibinfo {author} {\bibfnamefont {W.}~\bibnamefont {{El Hanafy}}},
  \bibinfo {author} {\bibfnamefont {G.~G.~L.}\ \bibnamefont {{Nashed}}}, \ and\
  \bibinfo {author} {\bibfnamefont {E.~N.}\ \bibnamefont {{Saridakis}}},\
  }\href@noop {} {\enquote {\bibinfo {title} {{Phase Portraits of $f(T)$
  Cosmology}},}\ } (\bibinfo {year} {2016}),\ \bibinfo {note} {(to be
  published)}\BibitemShut {NoStop}%
\bibitem [{\citenamefont {Oikonomou}(2016)}]{Oikonomou:2016ass}%
  \BibitemOpen
  \bibfield  {author} {\bibinfo {author} {\bibfnamefont {V.~K.}\ \bibnamefont
  {Oikonomou}},\ }\bibfield  {title} {\enquote {\bibinfo {title} {Gauss-bonnet
  cosmology unifying late and early-time acceleration eras with intermediate
  eras},}\ }\href {\doibase 10.1007/s10509-016-2800-6} {\bibfield  {journal}
  {\bibinfo  {journal} {Astrophys. Space Sci.}\ }\textbf {\bibinfo {volume}
  {361}},\ \bibinfo {pages} {211} (\bibinfo {year} {2016})},\ \Eprint
  {http://arxiv.org/abs/1606.02164} {arXiv:1606.02164 [gr-qc]} \BibitemShut
  {NoStop}%
\bibitem [{\citenamefont {{Wanas}}(2009)}]{Wanas2}%
  \BibitemOpen
  \bibfield  {author} {\bibinfo {author} {\bibfnamefont {M.~I.}\ \bibnamefont
  {{Wanas}}},\ }\bibfield  {title} {\enquote {\bibinfo {title} {{An
  Ap-Structure with Finslerian Flavor I:. the Principal Idea}},}\ }\href
  {\doibase 10.1142/S0217732309030412} {\bibfield  {journal} {\bibinfo
  {journal} {Modern Phys. Lett. A}\ }\textbf {\bibinfo {volume} {24}},\
  \bibinfo {pages} {1749--1762} (\bibinfo {year} {2009})},\ \Eprint
  {http://arxiv.org/abs/0801.1132} {arXiv:0801.1132 [gr-qc]} \BibitemShut
  {NoStop}%
\bibitem [{\citenamefont {{Youssef}}\ \emph {et~al.}(2009)\citenamefont
  {{Youssef}}, \citenamefont {{Abed}},\ and\ \citenamefont
  {{Soleiman}}}]{Nabil1}%
  \BibitemOpen
  \bibfield  {author} {\bibinfo {author} {\bibfnamefont {N.~L.}\ \bibnamefont
  {{Youssef}}}, \bibinfo {author} {\bibfnamefont {S.~H.}\ \bibnamefont
  {{Abed}}}, \ and\ \bibinfo {author} {\bibfnamefont {A.}~\bibnamefont
  {{Soleiman}}},\ }\bibfield  {title} {\enquote {\bibinfo {title} {{A Global
  Approach to the Theory of Connections in Finsler Geometry}},}\ }\href@noop {}
  {\bibfield  {journal} {\bibinfo  {journal} {Tensor (Japan)}\ }\textbf
  {\bibinfo {volume} {71}},\ \bibinfo {pages} {187--208} (\bibinfo {year}
  {2009})},\ \Eprint {http://arxiv.org/abs/0801.3220} {arXiv:0801.3220
  [math.DG]} \BibitemShut {NoStop}%
\bibitem [{\citenamefont {{Youssef}}\ \emph {et~al.}(2008)\citenamefont
  {{Youssef}}, \citenamefont {{Abed}},\ and\ \citenamefont
  {{Soleiman}}}]{Nabil2}%
  \BibitemOpen
  \bibfield  {author} {\bibinfo {author} {\bibfnamefont {N.~L.}\ \bibnamefont
  {{Youssef}}}, \bibinfo {author} {\bibfnamefont {S.~H.}\ \bibnamefont
  {{Abed}}}, \ and\ \bibinfo {author} {\bibfnamefont {A.}~\bibnamefont
  {{Soleiman}}},\ }\bibfield  {title} {\enquote {\bibinfo {title} {{A Global
  Theory of Conformal Finsler Geometry}},}\ }\href@noop {} {\bibfield
  {journal} {\bibinfo  {journal} {Tensor (Japan)}\ }\textbf {\bibinfo {volume}
  {69}},\ \bibinfo {pages} {155--178} (\bibinfo {year} {2008})},\ \Eprint
  {http://arxiv.org/abs/math/0610052} {arXiv:math/0610052 [math.DG]}
  \BibitemShut {NoStop}%
\bibitem [{\citenamefont {{Tamim}}\ and\ \citenamefont
  {{Youssef}}(1999)}]{Nabil3}%
  \BibitemOpen
  \bibfield  {author} {\bibinfo {author} {\bibfnamefont {A.~A.}\ \bibnamefont
  {{Tamim}}}\ and\ \bibinfo {author} {\bibfnamefont {N.~L.}\ \bibnamefont
  {{Youssef}}},\ }\bibfield  {title} {\enquote {\bibinfo {title} {{On
  Generalized Randers Manifolds}},}\ }\href@noop {} {\bibfield  {journal}
  {\bibinfo  {journal} {Alg. Groups Geom.}\ }\textbf {\bibinfo {volume} {16}},\
  \bibinfo {pages} {115--126} (\bibinfo {year} {1999})},\ \Eprint
  {http://arxiv.org/abs/math/0607572} {arXiv:math/0607572 [math.DG]}
  \BibitemShut {NoStop}%
\bibitem [{\citenamefont {{Wanas}}(2007)}]{Wanas1}%
  \BibitemOpen
  \bibfield  {author} {\bibinfo {author} {\bibfnamefont {M.~I.}\ \bibnamefont
  {{Wanas}}},\ }\bibfield  {title} {\enquote {\bibinfo {title} {{On the
  Relation Between Mass and Charge: a Pure Geometric Approach}},}\ }\href
  {\doibase 10.1142/S0219887807002144} {\bibfield  {journal} {\bibinfo
  {journal} {Int. J. Geom. Methods Mod. Phys.}\ }\textbf {\bibinfo {volume}
  {04}},\ \bibinfo {pages} {373} (\bibinfo {year} {2007})},\ \Eprint
  {http://arxiv.org/abs/gr-qc/0703036} {arXiv:gr-qc/0703036 [gr-qc]}
  \BibitemShut {NoStop}%
\bibitem [{\citenamefont {{Mikhail}}\ \emph {et~al.}(1995)\citenamefont
  {{Mikhail}}, \citenamefont {{Wanas}},\ and\ \citenamefont {{Eid}}}]{Wanas3}%
  \BibitemOpen
  \bibfield  {author} {\bibinfo {author} {\bibfnamefont {F.~I.}\ \bibnamefont
  {{Mikhail}}}, \bibinfo {author} {\bibfnamefont {M.~I.}\ \bibnamefont
  {{Wanas}}}, \ and\ \bibinfo {author} {\bibfnamefont {A.~M.}\ \bibnamefont
  {{Eid}}},\ }\bibfield  {title} {\enquote {\bibinfo {title} {{Theoretical
  Interpretation of Cosmic Magnetic Fields}},}\ }\href {\doibase
  10.1007/BF00984977} {\bibfield  {journal} {\bibinfo  {journal} {Astrophys.
  Space Sci.}\ }\textbf {\bibinfo {volume} {228}},\ \bibinfo {pages} {221--237}
  (\bibinfo {year} {1995})}\BibitemShut {NoStop}%
\bibitem [{\citenamefont {{Wanas}}(1986)}]{Wanas4}%
  \BibitemOpen
  \bibfield  {author} {\bibinfo {author} {\bibfnamefont {M.~I.}\ \bibnamefont
  {{Wanas}}},\ }\bibfield  {title} {\enquote {\bibinfo {title} {{Geometrical
  structures for cosmological applications}},}\ }\href {\doibase
  10.1007/BF00637758} {\bibfield  {journal} {\bibinfo  {journal} {Astrophys.
  Space Sci.}\ }\textbf {\bibinfo {volume} {127}},\ \bibinfo {pages} {21--25}
  (\bibinfo {year} {1986})}\BibitemShut {NoStop}%
\bibitem [{\citenamefont {{Wanas}}\ \emph {et~al.}(2014)\citenamefont
  {{Wanas}}, \citenamefont {{Youssef}},\ and\ \citenamefont {{El
  Hanafy}}}]{Waleed4}%
  \BibitemOpen
  \bibfield  {author} {\bibinfo {author} {\bibfnamefont {M.~I.}\ \bibnamefont
  {{Wanas}}}, \bibinfo {author} {\bibfnamefont {N.~L.}\ \bibnamefont
  {{Youssef}}}, \ and\ \bibinfo {author} {\bibfnamefont {W.}~\bibnamefont {{El
  Hanafy}}},\ }\bibfield  {title} {\enquote {\bibinfo {title} {{Pure Geometric
  Field Theory: Description of Gravity and Material Distribution}},}\
  }\href@noop {} {\  (\bibinfo {year} {2014})},\ \bibinfo {note} {submitted to
  Grav. Cosm.},\ \Eprint {http://arxiv.org/abs/1404.2485} {arXiv:1404.2485
  [gr-qc]} \BibitemShut {NoStop}%
\bibitem [{\citenamefont {{Maluf}}(2013)}]{M2013}%
  \BibitemOpen
  \bibfield  {author} {\bibinfo {author} {\bibfnamefont {J.~W.}\ \bibnamefont
  {{Maluf}}},\ }\bibfield  {title} {\enquote {\bibinfo {title} {{The
  teleparallel equivalent of general relativity}},}\ }\href {\doibase
  10.1002/andp.201200272} {\bibfield  {journal} {\bibinfo  {journal} {Ann.
  Phys. (Berlin)}\ }\textbf {\bibinfo {volume} {525}},\ \bibinfo {pages}
  {339--357} (\bibinfo {year} {2013})},\ \Eprint
  {http://arxiv.org/abs/1303.3897} {arXiv:1303.3897 [gr-qc]} \BibitemShut
  {NoStop}%
\bibitem [{\citenamefont {{Ferraro}}\ and\ \citenamefont
  {{Fiorini}}(2008)}]{FF08}%
  \BibitemOpen
  \bibfield  {author} {\bibinfo {author} {\bibfnamefont {R.}~\bibnamefont
  {{Ferraro}}}\ and\ \bibinfo {author} {\bibfnamefont {F.}~\bibnamefont
  {{Fiorini}}},\ }\bibfield  {title} {\enquote {\bibinfo {title} {{Born-Infeld
  gravity in Weitzenb{\"o}ck spacetime}},}\ }\href {\doibase
  10.1103/PhysRevD.78.124019} {\bibfield  {journal} {\bibinfo  {journal} {Phys.
  Rev. D}\ }\textbf {\bibinfo {volume} {78}},\ \bibinfo {eid} {124019}
  (\bibinfo {year} {2008})},\ \Eprint {http://arxiv.org/abs/0812.1981}
  {arXiv:0812.1981 [gr-qc]} \BibitemShut {NoStop}%
\bibitem [{\citenamefont {{Fiorini}}(2013)}]{F13}%
  \BibitemOpen
  \bibfield  {author} {\bibinfo {author} {\bibfnamefont {F.}~\bibnamefont
  {{Fiorini}}},\ }\bibfield  {title} {\enquote {\bibinfo {title} {{Nonsingular
  Promises from Born-Infeld Gravity}},}\ }\href@noop {} {\bibfield  {journal}
  {\bibinfo  {journal} {Phys. Rev. Lett.}\ }\textbf {\bibinfo {volume} {111}},\
  \bibinfo {pages} {041104} (\bibinfo {year} {2013})},\ \Eprint
  {http://arxiv.org/abs/1306.4392} {arXiv:1306.4392 [gr-qc]} \BibitemShut
  {NoStop}%
\bibitem [{\citenamefont {{Kofinas}}\ and\ \citenamefont
  {{Saridakis}}(2014{\natexlab{a}})}]{KS114}%
  \BibitemOpen
  \bibfield  {author} {\bibinfo {author} {\bibfnamefont {G.}~\bibnamefont
  {{Kofinas}}}\ and\ \bibinfo {author} {\bibfnamefont {E.N.}\ \bibnamefont
  {{Saridakis}}},\ }\bibfield  {title} {\enquote {\bibinfo {title}
  {{Teleparallel equivalent of Gauss-Bonnet gravity and its modifications}},}\
  }\href {\doibase 10.1103/PhysRevD.90.084044} {\bibfield  {journal} {\bibinfo
  {journal} {Phys. Rev. D}\ }\textbf {\bibinfo {volume} {90}},\ \bibinfo {eid}
  {084044} (\bibinfo {year} {2014}{\natexlab{a}})},\ \Eprint
  {http://arxiv.org/abs/1404.2249} {arXiv:1404.2249 [gr-qc]} \BibitemShut
  {NoStop}%
\bibitem [{\citenamefont {{Kofinas}}\ and\ \citenamefont
  {{Saridakis}}(2014{\natexlab{b}})}]{KS214}%
  \BibitemOpen
  \bibfield  {author} {\bibinfo {author} {\bibfnamefont {G.}~\bibnamefont
  {{Kofinas}}}\ and\ \bibinfo {author} {\bibfnamefont {E.N.}\ \bibnamefont
  {{Saridakis}}},\ }\bibfield  {title} {\enquote {\bibinfo {title}
  {{Cosmological applications of $f(T,T_{G})$ gravity}},}\ }\href {\doibase
  10.1103/PhysRevD.90.084045} {\bibfield  {journal} {\bibinfo  {journal} {Phys.
  Rev. D}\ }\textbf {\bibinfo {volume} {90}},\ \bibinfo {eid} {084045}
  (\bibinfo {year} {2014}{\natexlab{b}})},\ \Eprint
  {http://arxiv.org/abs/1408.0107} {arXiv:1408.0107 [gr-qc]} \BibitemShut
  {NoStop}%
\bibitem [{\citenamefont {{Kofinas}}\ \emph {et~al.}(2014)\citenamefont
  {{Kofinas}}, \citenamefont {{Leon}},\ and\ \citenamefont
  {{Saridakis}}}]{KS314}%
  \BibitemOpen
  \bibfield  {author} {\bibinfo {author} {\bibfnamefont {G.}~\bibnamefont
  {{Kofinas}}}, \bibinfo {author} {\bibfnamefont {G.}~\bibnamefont {{Leon}}}, \
  and\ \bibinfo {author} {\bibfnamefont {E.N.}\ \bibnamefont {{Saridakis}}},\
  }\bibfield  {title} {\enquote {\bibinfo {title} {{Dynamical behavior in
  $f(T,T_{G})$ cosmology}},}\ }\href {\doibase 10.1088/0264-9381/31/17/175011}
  {\bibfield  {journal} {\bibinfo  {journal} {Classical and Quantum Gravity}\
  }\textbf {\bibinfo {volume} {31}},\ \bibinfo {eid} {175011} (\bibinfo {year}
  {2014})},\ \Eprint {http://arxiv.org/abs/1404.7100} {arXiv:1404.7100 [gr-qc]}
  \BibitemShut {NoStop}%
\bibitem [{\citenamefont {{Bengochea}}\ and\ \citenamefont
  {{Ferraro}}(2009)}]{BF09}%
  \BibitemOpen
  \bibfield  {author} {\bibinfo {author} {\bibfnamefont {G.~R.}\ \bibnamefont
  {{Bengochea}}}\ and\ \bibinfo {author} {\bibfnamefont {R.}~\bibnamefont
  {{Ferraro}}},\ }\bibfield  {title} {\enquote {\bibinfo {title} {{Dark torsion
  as the cosmic speed-up}},}\ }\href {\doibase 10.1103/PhysRevD.79.124019}
  {\bibfield  {journal} {\bibinfo  {journal} {Phys. Rev. D}\ }\textbf {\bibinfo
  {volume} {79}},\ \bibinfo {eid} {124019} (\bibinfo {year} {2009})},\ \Eprint
  {http://arxiv.org/abs/0812.1205} {arXiv:0812.1205 [astro-ph]} \BibitemShut
  {NoStop}%
\bibitem [{\citenamefont {{Linder}}(2010)}]{L10}%
  \BibitemOpen
  \bibfield  {author} {\bibinfo {author} {\bibfnamefont {E.~V.}\ \bibnamefont
  {{Linder}}},\ }\bibfield  {title} {\enquote {\bibinfo {title} {{Einstein's
  other gravity and the acceleration of the Universe}},}\ }\href {\doibase
  10.1103/PhysRevD.81.127301} {\bibfield  {journal} {\bibinfo  {journal} {Phys.
  Rev. D}\ }\textbf {\bibinfo {volume} {81}},\ \bibinfo {eid} {127301}
  (\bibinfo {year} {2010})},\ \Eprint {http://arxiv.org/abs/1005.3039}
  {arXiv:1005.3039 [astro-ph]} \BibitemShut {NoStop}%
\bibitem [{\citenamefont {{Bamba}}\ \emph {et~al.}(2010)\citenamefont
  {{Bamba}}, \citenamefont {{Geng}},\ and\ \citenamefont {{Lee}}}]{1008.4036}%
  \BibitemOpen
  \bibfield  {author} {\bibinfo {author} {\bibfnamefont {K.}~\bibnamefont
  {{Bamba}}}, \bibinfo {author} {\bibfnamefont {C.-Q.}\ \bibnamefont {{Geng}}},
  \ and\ \bibinfo {author} {\bibfnamefont {C.-C.}\ \bibnamefont {{Lee}}},\
  }\bibfield  {title} {\enquote {\bibinfo {title} {{Comment on ''Einstein's
  Other Gravity and the Acceleration of the Universe''}},}\ }\href@noop {}
  {\bibfield  {journal} {\bibinfo  {journal} {ArXiv e-prints}\ } (\bibinfo
  {year} {2010})},\ \Eprint {http://arxiv.org/abs/1008.4036} {arXiv:1008.4036
  [astro-ph]} \BibitemShut {NoStop}%
\bibitem [{\citenamefont {{Bamba}}\ \emph {et~al.}(2011)\citenamefont
  {{Bamba}}, \citenamefont {{Geng}}, \citenamefont {{Lee}},\ and\ \citenamefont
  {{Luo}}}]{1011.0508}%
  \BibitemOpen
  \bibfield  {author} {\bibinfo {author} {\bibfnamefont {K.}~\bibnamefont
  {{Bamba}}}, \bibinfo {author} {\bibfnamefont {C.-Q.}\ \bibnamefont {{Geng}}},
  \bibinfo {author} {\bibfnamefont {C.-C.}\ \bibnamefont {{Lee}}}, \ and\
  \bibinfo {author} {\bibfnamefont {L.-W.}\ \bibnamefont {{Luo}}},\ }\bibfield
  {title} {\enquote {\bibinfo {title} {{Equation of state for dark energy in
  $f(T)$ gravity}},}\ }\href {\doibase 10.1088/1475-7516/2011/01/021}
  {\bibfield  {journal} {\bibinfo  {journal} {J. Cosmol. Astropart. Phys.}\
  }\textbf {\bibinfo {volume} {1}},\ \bibinfo {eid} {021} (\bibinfo {year}
  {2011})},\ \Eprint {http://arxiv.org/abs/1011.0508} {arXiv:1011.0508
  [astro-ph]} \BibitemShut {NoStop}%
\bibitem [{\citenamefont {{Ferraro}}\ and\ \citenamefont
  {{Fiorini}}(2007)}]{FF07}%
  \BibitemOpen
  \bibfield  {author} {\bibinfo {author} {\bibfnamefont {R.}~\bibnamefont
  {{Ferraro}}}\ and\ \bibinfo {author} {\bibfnamefont {F.}~\bibnamefont
  {{Fiorini}}},\ }\bibfield  {title} {\enquote {\bibinfo {title} {{Modified
  teleparallel gravity: Inflation without an inflaton}},}\ }\href {\doibase
  10.1103/PhysRevD.75.084031} {\bibfield  {journal} {\bibinfo  {journal} {Phys.
  Rev. D}\ }\textbf {\bibinfo {volume} {75}},\ \bibinfo {eid} {084031}
  (\bibinfo {year} {2007})},\ \Eprint {http://arxiv.org/abs/gr-qc/0610067}
  {arXiv:gr-qc/0610067 [gr-qc]} \BibitemShut {NoStop}%
\bibitem [{\citenamefont {{Ferraro}}\ and\ \citenamefont
  {{Fiorini}}(2011)}]{FF11}%
  \BibitemOpen
  \bibfield  {author} {\bibinfo {author} {\bibfnamefont {R.}~\bibnamefont
  {{Ferraro}}}\ and\ \bibinfo {author} {\bibfnamefont {F.}~\bibnamefont
  {{Fiorini}}},\ }\bibfield  {title} {\enquote {\bibinfo {title} {{Non-trivial
  frames for $f(T)$ theories of gravity and beyond}},}\ }\href {\doibase
  10.1016/j.physletb.2011.06.049} {\bibfield  {journal} {\bibinfo  {journal}
  {Phys. Lett.}\ }\textbf {\bibinfo {volume} {702B}},\ \bibinfo {pages}
  {75--80} (\bibinfo {year} {2011})},\ \Eprint {http://arxiv.org/abs/1103.0824}
  {arXiv:1103.0824 [gr-qc]} \BibitemShut {NoStop}%
\bibitem [{\citenamefont {{El Hanafy}}\ and\ \citenamefont
  {{Nashed}}(2016{\natexlab{b}})}]{Waleed5}%
  \BibitemOpen
  \bibfield  {author} {\bibinfo {author} {\bibfnamefont {W.}~\bibnamefont {{El
  Hanafy}}}\ and\ \bibinfo {author} {\bibfnamefont {G.~G.~L.}\ \bibnamefont
  {{Nashed}}},\ }\bibfield  {title} {\enquote {\bibinfo {title}
  {{Reconstruction of $f(T)$-gravity in the absence of matter}},}\ }\href
  {\doibase 10.1007/s10509-016-2786-0} {\bibfield  {journal} {\bibinfo
  {journal} {Astrophys. Space Sci.}\ }\textbf {\bibinfo {volume} {361}},\
  \bibinfo {pages} {1--16} (\bibinfo {year} {2016}{\natexlab{b}})},\ \Eprint
  {http://arxiv.org/abs/1410.2467} {arXiv:1410.2467 [hep-th]} \BibitemShut
  {NoStop}%
\bibitem [{\citenamefont {{Bamba}}\ \emph {et~al.}(2014)\citenamefont
  {{Bamba}}, \citenamefont {{Nojiri}},\ and\ \citenamefont
  {{Odintsov}}}]{BNO14}%
  \BibitemOpen
  \bibfield  {author} {\bibinfo {author} {\bibfnamefont {K.}~\bibnamefont
  {{Bamba}}}, \bibinfo {author} {\bibfnamefont {S.}~\bibnamefont {{Nojiri}}}, \
  and\ \bibinfo {author} {\bibfnamefont {S.~D.}\ \bibnamefont {{Odintsov}}},\
  }\bibfield  {title} {\enquote {\bibinfo {title} {{Trace-anomaly driven
  inflation in $f(T)$ gravity and in minimal massive bigravity}},}\ }\href
  {\doibase 10.1016/j.physletb.2014.02.041} {\bibfield  {journal} {\bibinfo
  {journal} {Phys. Lett.}\ }\textbf {\bibinfo {volume} {731B}},\ \bibinfo
  {pages} {257--264} (\bibinfo {year} {2014})},\ \Eprint
  {http://arxiv.org/abs/1401.7378} {arXiv:1401.7378 [gr-qc]} \BibitemShut
  {NoStop}%
\bibitem [{\citenamefont {{Bamba}}\ and\ \citenamefont
  {{Odintsov}}(2014)}]{BO14}%
  \BibitemOpen
  \bibfield  {author} {\bibinfo {author} {\bibfnamefont {K.}~\bibnamefont
  {{Bamba}}}\ and\ \bibinfo {author} {\bibfnamefont {S.~D.}\ \bibnamefont
  {{Odintsov}}},\ }\bibfield  {title} {\enquote {\bibinfo {title} {{Universe
  acceleration in modified gravities: $f(R)$ and $f(T)$ cases}},}\ }\href@noop
  {} {\bibfield  {journal} {\bibinfo  {journal} {Proc. Sci.,}\ }\textbf
  {\bibinfo {volume} {KMI2013}},\ \bibinfo {pages} {023} (\bibinfo {year}
  {2014})},\ \Eprint {http://arxiv.org/abs/1402.7114} {arXiv:1402.7114
  [hep-th]} \BibitemShut {NoStop}%
\bibitem [{\citenamefont {{El Hanafy}}\ and\ \citenamefont
  {{Nashed}}(2016{\natexlab{c}})}]{Waleed6}%
  \BibitemOpen
  \bibfield  {author} {\bibinfo {author} {\bibfnamefont {W.}~\bibnamefont {{El
  Hanafy}}}\ and\ \bibinfo {author} {\bibfnamefont {G.~G.~L.}\ \bibnamefont
  {{Nashed}}},\ }\bibfield  {title} {\enquote {\bibinfo {title} {{The hidden
  flat like universe II}},}\ }\href {\doibase 10.1007/s10509-016-2853-6}
  {\bibfield  {journal} {\bibinfo  {journal} {Astrophys. Space Sci.}\ }\textbf
  {\bibinfo {volume} {361}},\ \bibinfo {pages} {266} (\bibinfo {year}
  {2016}{\natexlab{c}})},\ \Eprint {http://arxiv.org/abs/1510.02337}
  {arXiv:1510.02337 [gr-qc]} \BibitemShut {NoStop}%
\bibitem [{\citenamefont {{Jamil}}\ \emph {et~al.}(2015)\citenamefont
  {{Jamil}}, \citenamefont {{Momeni}},\ and\ \citenamefont
  {{Myrzakulov}}}]{JMM14}%
  \BibitemOpen
  \bibfield  {author} {\bibinfo {author} {\bibfnamefont {M.}~\bibnamefont
  {{Jamil}}}, \bibinfo {author} {\bibfnamefont {D.}~\bibnamefont {{Momeni}}}, \
  and\ \bibinfo {author} {\bibfnamefont {R.}~\bibnamefont {{Myrzakulov}}},\
  }\bibfield  {title} {\enquote {\bibinfo {title} {{Warm Intermediate Inflation
  in $f(T)$ Gravity}},}\ }\href {\doibase 10.1007/s10773-014-2303-6} {\bibfield
   {journal} {\bibinfo  {journal} {Int. J. Theor. Phys.}\ }\textbf {\bibinfo
  {volume} {54}},\ \bibinfo {pages} {1098--1112} (\bibinfo {year} {2015})},\
  \Eprint {http://arxiv.org/abs/1309.3269} {arXiv:1309.3269 [gr-qc]}
  \BibitemShut {NoStop}%
\bibitem [{\citenamefont {{Harko}}\ \emph {et~al.}(2014)\citenamefont
  {{Harko}}, \citenamefont {{Lobo}}, \citenamefont {{Otalora}},\ and\
  \citenamefont {{Saridakis}}}]{HLOS14}%
  \BibitemOpen
  \bibfield  {author} {\bibinfo {author} {\bibfnamefont {T.}~\bibnamefont
  {{Harko}}}, \bibinfo {author} {\bibfnamefont {F.~S.~N.}\ \bibnamefont
  {{Lobo}}}, \bibinfo {author} {\bibfnamefont {G.}~\bibnamefont {{Otalora}}}, \
  and\ \bibinfo {author} {\bibfnamefont {E.~N.}\ \bibnamefont {{Saridakis}}},\
  }\bibfield  {title} {\enquote {\bibinfo {title} {{Nonminimal torsion-matter
  coupling extension of $f(T)$ gravity}},}\ }\href {\doibase
  10.1103/PhysRevD.89.124036} {\bibfield  {journal} {\bibinfo  {journal} {Phys.
  Rev. D}\ }\textbf {\bibinfo {volume} {89}},\ \bibinfo {eid} {124036}
  (\bibinfo {year} {2014})},\ \Eprint {http://arxiv.org/abs/1404.6212}
  {arXiv:1404.6212 [gr-qc]} \BibitemShut {NoStop}%
\bibitem [{\citenamefont {{Nashed}}(2015{\natexlab{a}})}]{Nd15}%
  \BibitemOpen
  \bibfield  {author} {\bibinfo {author} {\bibfnamefont {G.~L.}\ \bibnamefont
  {{Nashed}}},\ }\bibfield  {title} {\enquote {\bibinfo {title} {{FRW in
  quadratic form of $f(T)$ gravitational theories}},}\ }\href {\doibase
  10.1007/s10714-015-1917-1} {\bibfield  {journal} {\bibinfo  {journal} {Gen.
  Relativ. Gravit.}\ }\textbf {\bibinfo {volume} {47}},\ \bibinfo {pages} {75}
  (\bibinfo {year} {2015}{\natexlab{a}})},\ \Eprint
  {http://arxiv.org/abs/1506.08695} {arXiv:1506.08695 [gr-qc]} \BibitemShut
  {NoStop}%
\bibitem [{\citenamefont {{Nashed}}(2015{\natexlab{b}})}]{Nd215}%
  \BibitemOpen
  \bibfield  {author} {\bibinfo {author} {\bibfnamefont {G.~G.~L.}\
  \bibnamefont {{Nashed}}},\ }\bibfield  {title} {\enquote {\bibinfo {title}
  {{Anisotropic models with two fluids in linear and quadratic forms of $f(T)$
  gravitational theories}},}\ }\href {\doibase 10.1007/s10509-015-2339-y}
  {\bibfield  {journal} {\bibinfo  {journal} {Astrophys. Space Sci.}\ }\textbf
  {\bibinfo {volume} {357}},\ \bibinfo {eid} {111} (\bibinfo {year}
  {2015}{\natexlab{b}})}\BibitemShut {NoStop}%
\bibitem [{\citenamefont {{Li}}\ \emph
  {et~al.}(2011{\natexlab{a}})\citenamefont {{Li}}, \citenamefont
  {{Sotiriou}},\ and\ \citenamefont {{Barrow}}}]{1010.1041}%
  \BibitemOpen
  \bibfield  {author} {\bibinfo {author} {\bibfnamefont {B.}~\bibnamefont
  {{Li}}}, \bibinfo {author} {\bibfnamefont {T.~P.}\ \bibnamefont
  {{Sotiriou}}}, \ and\ \bibinfo {author} {\bibfnamefont {J.~D.}\ \bibnamefont
  {{Barrow}}},\ }\bibfield  {title} {\enquote {\bibinfo {title} {{$f(T)$
  gravity and local Lorentz invariance}},}\ }\href {\doibase
  10.1103/PhysRevD.83.064035} {\bibfield  {journal} {\bibinfo  {journal} {Phys.
  Rev. D}\ }\textbf {\bibinfo {volume} {83}},\ \bibinfo {eid} {064035}
  (\bibinfo {year} {2011}{\natexlab{a}})},\ \Eprint
  {http://arxiv.org/abs/1010.1041} {arXiv:1010.1041 [gr-qc]} \BibitemShut
  {NoStop}%
\bibitem [{\citenamefont {{Sotiriou}}\ \emph {et~al.}(2011)\citenamefont
  {{Sotiriou}}, \citenamefont {{Li}},\ and\ \citenamefont
  {{Barrow}}}]{1012.4039}%
  \BibitemOpen
  \bibfield  {author} {\bibinfo {author} {\bibfnamefont {T.~P.}\ \bibnamefont
  {{Sotiriou}}}, \bibinfo {author} {\bibfnamefont {B.}~\bibnamefont {{Li}}}, \
  and\ \bibinfo {author} {\bibfnamefont {J.~D.}\ \bibnamefont {{Barrow}}},\
  }\bibfield  {title} {\enquote {\bibinfo {title} {{Generalizations of
  teleparallel gravity and local Lorentz symmetry}},}\ }\href {\doibase
  10.1103/PhysRevD.83.104030} {\bibfield  {journal} {\bibinfo  {journal} {Phys.
  Rev. D}\ }\textbf {\bibinfo {volume} {83}},\ \bibinfo {eid} {104030}
  (\bibinfo {year} {2011})},\ \Eprint {http://arxiv.org/abs/1012.4039}
  {arXiv:1012.4039 [gr-qc]} \BibitemShut {NoStop}%
\bibitem [{\citenamefont {{Ferraro}}\ and\ \citenamefont
  {{Fiorini}}(2015)}]{LLT1}%
  \BibitemOpen
  \bibfield  {author} {\bibinfo {author} {\bibfnamefont {R.}~\bibnamefont
  {{Ferraro}}}\ and\ \bibinfo {author} {\bibfnamefont {F.}~\bibnamefont
  {{Fiorini}}},\ }\bibfield  {title} {\enquote {\bibinfo {title} {{Remnant
  group of local Lorentz transformations in $f(T)$ theories}},}\ }\href
  {\doibase 10.1103/PhysRevD.91.064019} {\bibfield  {journal} {\bibinfo
  {journal} {Phys. Rev. D}\ }\textbf {\bibinfo {volume} {91}},\ \bibinfo {eid}
  {064019} (\bibinfo {year} {2015})},\ \Eprint {http://arxiv.org/abs/1412.3424}
  {arXiv:1412.3424 [gr-qc]} \BibitemShut {NoStop}%
\bibitem [{\citenamefont {{Li}}\ \emph
  {et~al.}(2011{\natexlab{b}})\citenamefont {{Li}}, \citenamefont {{Miao}},\
  and\ \citenamefont {{Miao}}}]{Li:2011rn}%
  \BibitemOpen
  \bibfield  {author} {\bibinfo {author} {\bibfnamefont {M.}~\bibnamefont
  {{Li}}}, \bibinfo {author} {\bibfnamefont {R.-X.}\ \bibnamefont {{Miao}}}, \
  and\ \bibinfo {author} {\bibfnamefont {Y.-G.}\ \bibnamefont {{Miao}}},\
  }\bibfield  {title} {\enquote {\bibinfo {title} {{Degrees of freedom of
  $f(T)$ gravity}},}\ }\href {\doibase 10.1007/JHEP07(2011)108} {\bibfield
  {journal} {\bibinfo  {journal} {J. High Energy Phys.}\ }\textbf {\bibinfo
  {volume} {7}},\ \bibinfo {eid} {108} (\bibinfo {year}
  {2011}{\natexlab{b}})},\ \Eprint {http://arxiv.org/abs/1105.5934}
  {arXiv:1105.5934 [hep-th]} \BibitemShut {NoStop}%
\bibitem [{\citenamefont {{Ong}}\ \emph {et~al.}(2013)\citenamefont {{Ong}},
  \citenamefont {{Izumi}}, \citenamefont {{Nester}},\ and\ \citenamefont
  {{Chen}}}]{Ong:2013qja}%
  \BibitemOpen
  \bibfield  {author} {\bibinfo {author} {\bibfnamefont {Y.~C.}\ \bibnamefont
  {{Ong}}}, \bibinfo {author} {\bibfnamefont {K.}~\bibnamefont {{Izumi}}},
  \bibinfo {author} {\bibfnamefont {J.~M.}\ \bibnamefont {{Nester}}}, \ and\
  \bibinfo {author} {\bibfnamefont {P.}~\bibnamefont {{Chen}}},\ }\bibfield
  {title} {\enquote {\bibinfo {title} {{Problems with propagation and time
  evolution in $f(T)$ gravity}},}\ }\href {\doibase 10.1103/PhysRevD.88.024019}
  {\bibfield  {journal} {\bibinfo  {journal} {Phys. Rev. D}\ }\textbf {\bibinfo
  {volume} {88}},\ \bibinfo {eid} {024019} (\bibinfo {year} {2013})},\ \Eprint
  {http://arxiv.org/abs/1303.0993} {arXiv:1303.0993 [gr-qc]} \BibitemShut
  {NoStop}%
\bibitem [{\citenamefont {{Bamba}}\ \emph
  {et~al.}(2013{\natexlab{a}})\citenamefont {{Bamba}}, \citenamefont
  {{Odintsov}},\ and\ \citenamefont {{S{\'a}ez-G{\'o}mez}}}]{Bamba:2013jqa}%
  \BibitemOpen
  \bibfield  {author} {\bibinfo {author} {\bibfnamefont {K.}~\bibnamefont
  {{Bamba}}}, \bibinfo {author} {\bibfnamefont {S.~D.}\ \bibnamefont
  {{Odintsov}}}, \ and\ \bibinfo {author} {\bibfnamefont {D.}~\bibnamefont
  {{S{\'a}ez-G{\'o}mez}}},\ }\bibfield  {title} {\enquote {\bibinfo {title}
  {{Conformal symmetry and accelerating cosmology in teleparallel gravity}},}\
  }\href {\doibase 10.1103/PhysRevD.88.084042} {\bibfield  {journal} {\bibinfo
  {journal} {Phys. Rev. D}\ }\textbf {\bibinfo {volume} {88}},\ \bibinfo {eid}
  {084042} (\bibinfo {year} {2013}{\natexlab{a}})},\ \Eprint
  {http://arxiv.org/abs/1308.5789} {arXiv:1308.5789 [gr-qc]} \BibitemShut
  {NoStop}%
\bibitem [{\citenamefont {{Bamba}}\ \emph
  {et~al.}(2013{\natexlab{b}})\citenamefont {{Bamba}}, \citenamefont
  {{Capozziello}}, \citenamefont {{De Laurentis}}, \citenamefont {{Nojiri}},\
  and\ \citenamefont {{S{\'a}ez-G{\'o}mez}}}]{Bamba:2013ooa}%
  \BibitemOpen
  \bibfield  {author} {\bibinfo {author} {\bibfnamefont {K.}~\bibnamefont
  {{Bamba}}}, \bibinfo {author} {\bibfnamefont {S.}~\bibnamefont
  {{Capozziello}}}, \bibinfo {author} {\bibfnamefont {M.}~\bibnamefont {{De
  Laurentis}}}, \bibinfo {author} {\bibfnamefont {S.}~\bibnamefont {{Nojiri}}},
  \ and\ \bibinfo {author} {\bibfnamefont {D.}~\bibnamefont
  {{S{\'a}ez-G{\'o}mez}}},\ }\bibfield  {title} {\enquote {\bibinfo {title}
  {{No further gravitational wave modes in $F(T)$ gravity}},}\ }\href {\doibase
  10.1016/j.physletb.2013.10.022} {\bibfield  {journal} {\bibinfo  {journal}
  {Phys. Lett.}\ }\textbf {\bibinfo {volume} {727B}},\ \bibinfo {pages}
  {194--198} (\bibinfo {year} {2013}{\natexlab{b}})},\ \Eprint
  {http://arxiv.org/abs/1309.2698} {arXiv:1309.2698 [gr-qc]} \BibitemShut
  {NoStop}%
\bibitem [{\citenamefont {{Izumi}}\ \emph {et~al.}(2014)\citenamefont
  {{Izumi}}, \citenamefont {{Gu}},\ and\ \citenamefont
  {{Ong}}}]{Izumi:2013dca}%
  \BibitemOpen
  \bibfield  {author} {\bibinfo {author} {\bibfnamefont {K.}~\bibnamefont
  {{Izumi}}}, \bibinfo {author} {\bibfnamefont {J.-A.}\ \bibnamefont {{Gu}}}, \
  and\ \bibinfo {author} {\bibfnamefont {Y.~C.}\ \bibnamefont {{Ong}}},\
  }\bibfield  {title} {\enquote {\bibinfo {title} {{Acausality and nonunique
  evolution in generalized teleparallel gravity}},}\ }\href {\doibase
  10.1103/PhysRevD.89.084025} {\bibfield  {journal} {\bibinfo  {journal} {Phys.
  Rev. D}\ }\textbf {\bibinfo {volume} {89}},\ \bibinfo {eid} {084025}
  (\bibinfo {year} {2014})},\ \Eprint {http://arxiv.org/abs/1309.6461}
  {arXiv:1309.6461 [gr-qc]} \BibitemShut {NoStop}%
\bibitem [{\citenamefont {{Chen}}\ \emph {et~al.}(2015)\citenamefont {{Chen}},
  \citenamefont {{Izumi}}, \citenamefont {{Nester}},\ and\ \citenamefont
  {{Ong}}}]{Chen:2014qtl}%
  \BibitemOpen
  \bibfield  {author} {\bibinfo {author} {\bibfnamefont {P.}~\bibnamefont
  {{Chen}}}, \bibinfo {author} {\bibfnamefont {K.}~\bibnamefont {{Izumi}}},
  \bibinfo {author} {\bibfnamefont {J.~M.}\ \bibnamefont {{Nester}}}, \ and\
  \bibinfo {author} {\bibfnamefont {Y.~C.}\ \bibnamefont {{Ong}}},\ }\bibfield
  {title} {\enquote {\bibinfo {title} {{Remnant symmetry, propagation, and
  evolution in f (T ) gravity}},}\ }\href {\doibase 10.1103/PhysRevD.91.064003}
  {\bibfield  {journal} {\bibinfo  {journal} {Phys. Rev. D}\ }\textbf {\bibinfo
  {volume} {91}},\ \bibinfo {eid} {064003} (\bibinfo {year} {2015})},\ \Eprint
  {http://arxiv.org/abs/1412.8383} {arXiv:1412.8383 [gr-qc]} \BibitemShut
  {NoStop}%
\bibitem [{\citenamefont {{Bahamonde}}\ \emph {et~al.}(2015)\citenamefont
  {{Bahamonde}}, \citenamefont {{B{\"o}hmer}},\ and\ \citenamefont
  {{Wright}}}]{Bahamonde:2015zma}%
  \BibitemOpen
  \bibfield  {author} {\bibinfo {author} {\bibfnamefont {S.}~\bibnamefont
  {{Bahamonde}}}, \bibinfo {author} {\bibfnamefont {C.~G.}\ \bibnamefont
  {{B{\"o}hmer}}}, \ and\ \bibinfo {author} {\bibfnamefont {M.}~\bibnamefont
  {{Wright}}},\ }\bibfield  {title} {\enquote {\bibinfo {title} {{Modified
  teleparallel theories of gravity}},}\ }\href {\doibase
  10.1103/PhysRevD.92.104042} {\bibfield  {journal} {\bibinfo  {journal} {Phys.
  Rev. D}\ }\textbf {\bibinfo {volume} {92}},\ \bibinfo {eid} {104042}
  (\bibinfo {year} {2015})},\ \Eprint {http://arxiv.org/abs/1508.05120}
  {arXiv:1508.05120 [gr-qc]} \BibitemShut {NoStop}%
\bibitem [{\citenamefont {{Kr\v{s}\v{s}\'{a}k}}\ and\ \citenamefont
  {{Saridakis}}(2016)}]{Martin2}%
  \BibitemOpen
  \bibfield  {author} {\bibinfo {author} {\bibfnamefont {M.}~\bibnamefont
  {{Kr\v{s}\v{s}\'{a}k}}}\ and\ \bibinfo {author} {\bibfnamefont {E.~N.}\
  \bibnamefont {{Saridakis}}},\ }\bibfield  {title} {\enquote {\bibinfo {title}
  {{The covariant formulation of $f(T)$ gravity}},}\ }\href {\doibase
  10.1088/0264-9381/33/11/115009} {\bibfield  {journal} {\bibinfo  {journal}
  {Classical and Quantum Gravity}\ }\textbf {\bibinfo {volume} {33}},\ \bibinfo
  {eid} {115009} (\bibinfo {year} {2016})},\ \Eprint
  {http://arxiv.org/abs/1510.08432} {arXiv:1510.08432 [gr-qc]} \BibitemShut
  {NoStop}%
\bibitem [{\citenamefont {Cai}\ \emph {et~al.}(2016{\natexlab{b}})\citenamefont
  {Cai}, \citenamefont {Capozziello}, \citenamefont {De~Laurentis},\ and\
  \citenamefont {Saridakis}}]{Saridakis1}%
  \BibitemOpen
  \bibfield  {author} {\bibinfo {author} {\bibfnamefont {Yi-Fu}\ \bibnamefont
  {Cai}}, \bibinfo {author} {\bibfnamefont {Salvatore}\ \bibnamefont
  {Capozziello}}, \bibinfo {author} {\bibfnamefont {Mariafelicia}\ \bibnamefont
  {De~Laurentis}}, \ and\ \bibinfo {author} {\bibfnamefont {Emmanuel~N.}\
  \bibnamefont {Saridakis}},\ }\bibfield  {title} {\enquote {\bibinfo {title}
  {{$f(T)$ teleparallel gravity and cosmology}},}\ }\href {\doibase
  10.1088/0034-4885/79/10/106901} {\bibfield  {journal} {\bibinfo  {journal}
  {Rept. Prog. Phys.}\ }\textbf {\bibinfo {volume} {79}},\ \bibinfo {pages}
  {106901} (\bibinfo {year} {2016}{\natexlab{b}})},\ \Eprint
  {http://arxiv.org/abs/1511.07586} {arXiv:1511.07586 [gr-qc]} \BibitemShut
  {NoStop}%
\bibitem [{\citenamefont {Weinberg}(1972)}]{Wberg1972}%
  \BibitemOpen
  \bibfield  {author} {\bibinfo {author} {\bibfnamefont {S.}~\bibnamefont
  {Weinberg}},\ }\href@noop {} {\emph {\bibinfo {title} {Gravitation and
  cosmology: principles and applications of the general theory of
  relativity}}}\ (\bibinfo  {publisher} {Wiley, New York},\ \bibinfo {year}
  {1972})\ p.\ \bibinfo {pages} {688}\BibitemShut {NoStop}%
\bibitem [{\citenamefont {{Ade}}\ \emph
  {et~al.}(2016{\natexlab{a}})\citenamefont {{Ade}}, \citenamefont {{Aghanim}},
  \citenamefont {{Arnaud}}, \citenamefont {{Ashdown}}, \citenamefont
  {{Aumont}}, \citenamefont {{Baccigalupi}}, \citenamefont {{Banday}},
  \citenamefont {{Barreiro}}, \citenamefont {{Bartlett}},\ and\ \citenamefont
  {et~al. (Planck~Collaboration)}}]{Planck:2015xua}%
  \BibitemOpen
  \bibfield  {author} {\bibinfo {author} {\bibfnamefont {P.~A.~R.}\
  \bibnamefont {{Ade}}}, \bibinfo {author} {\bibfnamefont {N.}~\bibnamefont
  {{Aghanim}}}, \bibinfo {author} {\bibfnamefont {M.}~\bibnamefont {{Arnaud}}},
  \bibinfo {author} {\bibfnamefont {M.}~\bibnamefont {{Ashdown}}}, \bibinfo
  {author} {\bibfnamefont {J.}~\bibnamefont {{Aumont}}}, \bibinfo {author}
  {\bibfnamefont {C.}~\bibnamefont {{Baccigalupi}}}, \bibinfo {author}
  {\bibfnamefont {A.~J.}\ \bibnamefont {{Banday}}}, \bibinfo {author}
  {\bibfnamefont {R.~B.}\ \bibnamefont {{Barreiro}}}, \bibinfo {author}
  {\bibfnamefont {J.~G.}\ \bibnamefont {{Bartlett}}}, \ and\ \bibinfo {author}
  {\bibnamefont {et~al. (Planck~Collaboration)}},\ }\bibfield  {title}
  {\enquote {\bibinfo {title} {{Planck 2015 results. XIII. Cosmological
  parameters}},}\ }\href {\doibase 10.1051/0004-6361/201525830} {\bibfield
  {journal} {\bibinfo  {journal} {Astron. Astrophys.}\ }\textbf {\bibinfo
  {volume} {594}} (\bibinfo {year} {2016}{\natexlab{a}}),\
  10.1051/0004-6361/201525830},\ \Eprint {http://arxiv.org/abs/1502.01589}
  {arXiv:1502.01589 [astro-ph]} \BibitemShut {NoStop}%
\bibitem [{\citenamefont {{Ade}}\ \emph
  {et~al.}(2016{\natexlab{b}})\citenamefont {{Ade}}, \citenamefont {{Aghanim}},
  \citenamefont {{Arnaud}}, \citenamefont {{Arroja}}, \citenamefont
  {{Ashdown}}, \citenamefont {{Aumont}}, \citenamefont {{Baccigalupi}},
  \citenamefont {{Ballardini}}, \citenamefont {{Banday}},\ and\ \citenamefont
  {et~al. (Planck~Collaboration)}}]{Ade:2015lrj}%
  \BibitemOpen
  \bibfield  {author} {\bibinfo {author} {\bibfnamefont {P.~A.~R.}\
  \bibnamefont {{Ade}}}, \bibinfo {author} {\bibfnamefont {N.}~\bibnamefont
  {{Aghanim}}}, \bibinfo {author} {\bibfnamefont {M.}~\bibnamefont {{Arnaud}}},
  \bibinfo {author} {\bibfnamefont {F.}~\bibnamefont {{Arroja}}}, \bibinfo
  {author} {\bibfnamefont {M.}~\bibnamefont {{Ashdown}}}, \bibinfo {author}
  {\bibfnamefont {J.}~\bibnamefont {{Aumont}}}, \bibinfo {author}
  {\bibfnamefont {C.}~\bibnamefont {{Baccigalupi}}}, \bibinfo {author}
  {\bibfnamefont {M.}~\bibnamefont {{Ballardini}}}, \bibinfo {author}
  {\bibfnamefont {A.~J.}\ \bibnamefont {{Banday}}}, \ and\ \bibinfo {author}
  {\bibnamefont {et~al. (Planck~Collaboration)}},\ }\bibfield  {title}
  {\enquote {\bibinfo {title} {{Planck 2015 results. XX. Constraints on
  inflation}},}\ }\href {\doibase 10.1051/0004-6361/201525898} {\bibfield
  {journal} {\bibinfo  {journal} {Astron. Astrophys.}\ }\textbf {\bibinfo
  {volume} {594}} (\bibinfo {year} {2016}{\natexlab{b}}),\
  10.1051/0004-6361/201525898},\ \Eprint {http://arxiv.org/abs/1502.02114}
  {arXiv:1502.02114 [astro-ph]} \BibitemShut {NoStop}%
\bibitem [{\citenamefont {{Ade}}\ \emph {et~al.}(2015)\citenamefont {{Ade}},
  \citenamefont {{Aghanim}}, \citenamefont {{Ahmed}}, \citenamefont {{Aikin}},
  \citenamefont {{Alexander}}, \citenamefont {{Arnaud}}, \citenamefont
  {{Aumont}}, \citenamefont {{Baccigalupi}}, \citenamefont {{Banday}},
  \citenamefont {et~al. (BICEP2/Keck},\ and\ \citenamefont
  {Collaborations)}}]{Ade:2015tva}%
  \BibitemOpen
  \bibfield  {author} {\bibinfo {author} {\bibfnamefont {P.~A.~R.}\
  \bibnamefont {{Ade}}}, \bibinfo {author} {\bibfnamefont {N.}~\bibnamefont
  {{Aghanim}}}, \bibinfo {author} {\bibfnamefont {Z.}~\bibnamefont {{Ahmed}}},
  \bibinfo {author} {\bibfnamefont {R.~W.}\ \bibnamefont {{Aikin}}}, \bibinfo
  {author} {\bibfnamefont {K.~D.}\ \bibnamefont {{Alexander}}}, \bibinfo
  {author} {\bibfnamefont {M.}~\bibnamefont {{Arnaud}}}, \bibinfo {author}
  {\bibfnamefont {J.}~\bibnamefont {{Aumont}}}, \bibinfo {author}
  {\bibfnamefont {C.}~\bibnamefont {{Baccigalupi}}}, \bibinfo {author}
  {\bibfnamefont {A.~J.}\ \bibnamefont {{Banday}}}, \bibinfo {author}
  {\bibnamefont {et~al. (BICEP2/Keck}}, \ and\ \bibinfo {author} {\bibfnamefont
  {Planck}\ \bibnamefont {Collaborations)}},\ }\bibfield  {title} {\enquote
  {\bibinfo {title} {{Joint Analysis of BICEP2/Keck Array and Planck Data}},}\
  }\href {\doibase 10.1103/PhysRevLett.114.101301} {\bibfield  {journal}
  {\bibinfo  {journal} {Phys. Rev. Lett.}\ }\textbf {\bibinfo {volume} {114}},\
  \bibinfo {eid} {101301} (\bibinfo {year} {2015})},\ \Eprint
  {http://arxiv.org/abs/1502.00612} {arXiv:1502.00612 [astro-ph]} \BibitemShut
  {NoStop}%
\bibitem [{\citenamefont {{Ade}}\ \emph
  {et~al.}(2016{\natexlab{c}})\citenamefont {{Ade}}, \citenamefont {{Ahmed}},
  \citenamefont {{Aikin}}, \citenamefont {{Alexander}}, \citenamefont
  {{Barkats}}, \citenamefont {{Benton}}, \citenamefont {{Bischoff}},
  \citenamefont {{Bock}}, \citenamefont {et. al.},\ and\ \citenamefont
  {({BICEP2 Collaboration and Keck Array Collaboration})}}]{Array:2015xqh}%
  \BibitemOpen
  \bibfield  {author} {\bibinfo {author} {\bibfnamefont {P.~A.~R.}\
  \bibnamefont {{Ade}}}, \bibinfo {author} {\bibfnamefont {Z.}~\bibnamefont
  {{Ahmed}}}, \bibinfo {author} {\bibfnamefont {R.~W.}\ \bibnamefont
  {{Aikin}}}, \bibinfo {author} {\bibfnamefont {K.~D.}\ \bibnamefont
  {{Alexander}}}, \bibinfo {author} {\bibfnamefont {D.}~\bibnamefont
  {{Barkats}}}, \bibinfo {author} {\bibfnamefont {S.~J.}\ \bibnamefont
  {{Benton}}}, \bibinfo {author} {\bibfnamefont {C.~A.}\ \bibnamefont
  {{Bischoff}}}, \bibinfo {author} {\bibfnamefont {J.~J.}\ \bibnamefont
  {{Bock}}}, \bibinfo {author} {\bibnamefont {et. al.}}, \ and\ \bibinfo
  {author} {\bibnamefont {({BICEP2 Collaboration and Keck Array
  Collaboration})}},\ }\bibfield  {title} {\enquote {\bibinfo {title}
  {{Improved Constraints on Cosmology and Foregrounds from BICEP2 and Keck
  Array Cosmic Microwave Background Data with Inclusion of 95 GHz Band}},}\
  }\href {\doibase 10.1103/PhysRevLett.116.031302} {\bibfield  {journal}
  {\bibinfo  {journal} {Phys. Rev. Lett.}\ }\textbf {\bibinfo {volume} {116}},\
  \bibinfo {eid} {031302} (\bibinfo {year} {2016}{\natexlab{c}})},\ \Eprint
  {http://arxiv.org/abs/1510.09217} {arXiv:1510.09217 [astro-ph]} \BibitemShut
  {NoStop}%
\bibitem [{\citenamefont {{Mukhanov}}\ \emph {et~al.}(1992)\citenamefont
  {{Mukhanov}}, \citenamefont {{Feldman}},\ and\ \citenamefont
  {{Brandenberger}}}]{Mukhanov:1992}%
  \BibitemOpen
  \bibfield  {author} {\bibinfo {author} {\bibfnamefont {V.~F.}\ \bibnamefont
  {{Mukhanov}}}, \bibinfo {author} {\bibfnamefont {H.~A.}\ \bibnamefont
  {{Feldman}}}, \ and\ \bibinfo {author} {\bibfnamefont {R.~H.}\ \bibnamefont
  {{Brandenberger}}},\ }\bibfield  {title} {\enquote {\bibinfo {title} {{Theory
  of cosmological perturbations}},}\ }\href {\doibase
  10.1016/0370-1573(92)90044-Z} {\bibfield  {journal} {\bibinfo  {journal}
  {Phys. Rep.}\ }\textbf {\bibinfo {volume} {215}},\ \bibinfo {pages}
  {203--333} (\bibinfo {year} {1992})}\BibitemShut {NoStop}%
\bibitem [{\citenamefont {Bunch}\ and\ \citenamefont
  {Davies}(1978)}]{Bunch:1978yq}%
  \BibitemOpen
  \bibfield  {author} {\bibinfo {author} {\bibfnamefont {T.~S.}\ \bibnamefont
  {Bunch}}\ and\ \bibinfo {author} {\bibfnamefont {P.~C.~W.}\ \bibnamefont
  {Davies}},\ }\bibfield  {title} {\enquote {\bibinfo {title} {{Quantum Field
  Theory in de Sitter Space: Renormalization by Point Splitting}},}\ }\href
  {\doibase 10.1098/rspa.1978.0060} {\bibfield  {journal} {\bibinfo  {journal}
  {Proc. Roy. Soc. Lond.}\ ,\ \bibinfo {pages} {117--134}} (\bibinfo {year}
  {1978})}\BibitemShut {NoStop}%
\bibitem [{\citenamefont {{Chen}}\ \emph {et~al.}(2011)\citenamefont {{Chen}},
  \citenamefont {{Dent}}, \citenamefont {{Dutta}},\ and\ \citenamefont
  {{Saridakis}}}]{Chen:2010va}%
  \BibitemOpen
  \bibfield  {author} {\bibinfo {author} {\bibfnamefont {S.-H.}\ \bibnamefont
  {{Chen}}}, \bibinfo {author} {\bibfnamefont {J.~B.}\ \bibnamefont {{Dent}}},
  \bibinfo {author} {\bibfnamefont {S.}~\bibnamefont {{Dutta}}}, \ and\
  \bibinfo {author} {\bibfnamefont {E.~N.}\ \bibnamefont {{Saridakis}}},\
  }\bibfield  {title} {\enquote {\bibinfo {title} {{Cosmological perturbations
  in $f(T)$ gravity}},}\ }\href {\doibase 10.1103/PhysRevD.83.023508}
  {\bibfield  {journal} {\bibinfo  {journal} {Phys. Rev. D}\ }\textbf {\bibinfo
  {volume} {83}},\ \bibinfo {eid} {023508} (\bibinfo {year} {2011})},\ \Eprint
  {http://arxiv.org/abs/1008.1250} {arXiv:1008.1250 [astro-ph]} \BibitemShut
  {NoStop}%
\end{thebibliography}
%
\end{document}